\documentclass[5p,twocolumn,authoryear,times]{elsarticle}

\usepackage{lineno,hyperref}
\modulolinenumbers[5]
\hypersetup{colorlinks,citecolor=blue,linkcolor=blue,urlcolor=blue}

\journal{arXiv}

\usepackage{graphicx}	
\usepackage{amsmath}	
\usepackage{amssymb}	
\usepackage{bm}

\usepackage{natbib}
\usepackage[usenames,dvipsnames]{color}
\usepackage{verbatim}
\usepackage{multirow}
\usepackage{booktabs}
\usepackage{subfigure}
\usepackage{color, colortbl}

\usepackage[T1]{fontenc}
\usepackage[utf8]{inputenc}
\usepackage[tracking=smallcaps]{microtype}
\SetTracking[no ligatures = f]{encoding = *,shape = sc}{0}

\usepackage{listings}
\usepackage[normalem]{ulem}
\usepackage[font=footnotesize,labelfont=bf]{caption}
\usepackage{footnote}
\makesavenoteenv{tabular}
\makesavenoteenv{table}
\usepackage{supertabular}
\usepackage{threeparttable}
\usepackage{longtable}
\usepackage{rotating}
\usepackage{lscape}

\usepackage{lineno}

\usepackage{bold-extra}
\usepackage{slantsc}
\usepackage{times}

\usepackage{blindtext}
\makeatletter
\newcommand{\removelatexerror}{\let\@latex@error\@gobble}
\makeatother

\usepackage[]{algorithm2e}
\usepackage{siunitx}

\newcommand{\mic}{~\si{\micro\metre}}

\newcommand{\caesar}{\textsc{caesar}}

\newcommand{\scorpio}{\textsc{Scorpio}}
 
\definecolor{babypink}{rgb}{0.96, 0.76, 0.76}
\definecolor{beaublue}{rgb}{0.74, 0.83, 0.9}
\definecolor{mediumturquoise}{rgb}{0.28, 0.82, 0.8}
\definecolor{mossgreen}{rgb}{0.68, 0.87, 0.68}
\definecolor{mustard}{rgb}{1.0, 0.86, 0.35}
\definecolor{olivine}{rgb}{0.6, 0.73, 0.45}
\definecolor{orchid}{rgb}{0.85, 0.44, 0.84}
\definecolor{palecerulean}{rgb}{0.61, 0.77, 0.89}
\definecolor{palegold}{rgb}{0.9, 0.75, 0.54}
\definecolor{paleplum}{rgb}{0.8, 0.6, 0.8}
\definecolor{palespringbud}{rgb}{0.93, 0.92, 0.74}
\definecolor{pastelgray}{rgb}{0.81, 0.81, 0.77}
\definecolor{pastelviolet}{rgb}{0.8, 0.6, 0.79}
\definecolor{pastelred}{rgb}{1.0, 0.41, 0.38}
\definecolor{pearl}{rgb}{0.94, 0.92, 0.84}
\definecolor{pistachio}{rgb}{0.58, 0.77, 0.45}	
\definecolor{teal}{rgb}{0.0, 0.5, 0.5}
\definecolor{tiffanyblue}{rgb}{0.04, 0.73, 0.71}
\definecolor{turquoise}{rgb}{0.19, 0.84, 0.78}
\definecolor{verdigris}{rgb}{0.26, 0.7, 0.68}
	
\definecolor{lightgray}{gray}{0.9}

\newcommand{\hii}{H\textsc{ii}}

 {\pagebreak[4]\global\pdfpageattr\expandafter{\the\pdfpageattr/Rotate 90}}%
 {\pagebreak[4]\global\pdfpageattr\expandafter{\the\pdfpageattr/Rotate 0}}%
 
\usepackage{listings}
\lstnewenvironment{codebkg}{%
  \lstset{
    escapechar=\&,
    basicstyle=\ttfamily\scriptsize,
    backgroundcolor=\color{lightgray},
    frame=single,
    breaklines=true,    
    framerule=0pt,
    columns=fullflexible
  }
}
{}

\begin{document}
\begin{frontmatter}

\title{Classification of compact radio sources in the Galactic plane\\with supervised machine learning}

\author[1]{S. Riggi\corref{cor}}%
\ead{simone.riggi@inaf.it}
\author[1]{G. Umana}
\author[1]{C. Trigilio}
\author[1]{C. Bordiu}
\author[1]{F. Bufano}
\author[1]{A. Ingallinera}
\author[1]{F. Cavallaro}
\author[2]{Y. Gordon}
\author[3,4]{R.P. Norris}
\author[5,6]{G. G\"urkan}
\author[1]{P. Leto}
\author[1]{C. Buemi}
\author[1]{S. Loru}
\author[7]{A.M. Hopkins}
\author[3]{M.D. Filipovi\'c}
\author[1,8]{T. Cecconello}

\cortext[cor]{Corresponding author}%
\address[1]{INAF - Osservatorio Astrofisico di Catania, Via Santa Sofia 78, 95123 Catania, Italy}%
\address[2]{Department of Physics, University of Wisconsin-Madison, 1150 University Ave, Madison, WI 53706, USA}
\address[3]{Western Sydney University, Locked Bag 1797, Penrith South DC, NSW 2751, Australia}%
\address[4]{CSIRO Space \& Astronomy, P.O. Box 76, Epping, NSW 1710, Australia}%
\address[5]{Th\"uringer Landessternwarte Tautenburg (TLS), Sternwarte 5, D-07778 Tautenburg, Germany}%
\address[6]{CSIRO Space \& Astronomy, ATNF, PO Box 1130, Bentley WA 6102, Australia}
\address[7]{Australian Astronomical Optics, Macquarie University, 105 Delhi Rd, North Ryde, NSW 2113, Australia}%
\address[8]{Department of Electrical, Electronic and Computer Engineering, University of Catania, Viale Andrea Doria 6, 95125 Catania, Italy}

\begin{abstract}
Generation of science-ready data from processed data products is one of the major challenges in next-generation radio continuum surveys with the Square Kilometre Array (SKA) and its precursors, due to the expected data volume and the need to achieve a high degree of automated processing. 
Source extraction, characterization, and classification are the major stages involved in this process.\\
In this work we focus on the classification of compact radio sources in the Galactic plane using both radio and infrared images as inputs. To this aim, we produced a curated dataset of $\sim$20,000 images of compact sources of different astronomical classes, obtained from past radio and infrared surveys, and novel radio data from pilot surveys carried out with the Australian SKA Pathfinder (ASKAP). Radio spectral index information was also obtained for a subset of the data. We then trained two different classifiers on the produced dataset. The first model uses gradient-boosted decision trees and is trained on a set of pre-computed features derived from the data, which include radio-infrared colour indices and the radio spectral index. The second model is trained directly on multi-channel images, employing convolutional neural networks.\\Using a completely supervised procedure, we obtained a high classification accuracy (F1-score>90\%) for separating Galactic objects from the extragalactic background. Individual class discrimination performances, ranging from 60\% to 75\%, increased by 10\% when adding far-infrared and spectral index information, with extragalactic objects, PNe and \hii{} regions identified with higher accuracies.\\
The implemented tools and trained models were publicly released, and made available to the radioastronomical community for future application on new radio data. 
\end{abstract}


\end{frontmatter}



\section{Introduction}
\label{sec:intro}
The Square Kilometre Array (SKA) \citep{SKADesignDoc} will open a golden era in radio astronomy due to its anticipated sensitivity, frequency coverage and angular resolution. While the SKA is currently in the construction phase, SKA precursor telescopes have already started their planned survey programs, delivering valuable scientific results even during the commissioning phase. 
Among them, the Evolutionary Map of the Universe (EMU) program \citep{Norris2011} of the Australian SKA Pathfinder (ASKAP, \citealt{Johnston2008,ASKAPSystemDesign}) will survey $\sim$75\% of the sky at $\sim$940 MHz with an angular resolution of 10" and a target rms of 15 $\mu$Jy/beam. As EMU is expected to detect $\sim$50 million sources, the cataloguing process will require a significant degree of automation and knowledge extraction compared to previous surveys. Source finding is a major stage involved in such post-processing of observations.

In the last years, several developments were made within the SKA precursor community, and new tools were produced to improve compact source extraction and measurement capabilities (e.g. completeness, reliability, positional and flux density accuracy) and processing speed, also employing parallel computing methodologies (e.g. see \citealt{Riggi2019} and references therein).
Fewer efforts, however, has been spent on source classification, particularly for Galactic science targets, as almost all source finders do not provide any information (e.g. labels or tags) on the extracted source class identity. The implication for Galactic plane observations is that, after taking out source classifications made through automated cross-matching to previously classified objects (e.g. $\sim$5\% in the Scorpio field in \citealt{Riggi2021a}), the vast majority of the catalogued sources are unclassified. Of these, more than 90\% are typically single-island and single-component sources\footnote{By "island" we denote a group of 4-connected pixels in the analysed map having brightness above a threshold (typically 2.5$-$3.0 $\sigma_{rms}$), located around a seed pixel with brightness above a detection threshold (typically 5 $\sigma_{rms}$). An island can include multiple source "components", each typically modelled with a 2D Gaussian distribution.}. From the number of previously classified objects, it is reasonable to expect that the majority of unknown sources are extragalactic (radio galaxies, quasars), and \hii{} regions, with a smaller fraction\footnote{PNe and pulsars are, respectively, $\sim$60 and 30\% less numerous than \hii{} regions according to existing catalogue counts (see Section~\ref{subsec:compact-sources} and catalogue references therein).} of Planetary Nebulae (PNe) and pulsars, and an even smaller fraction ($<$10\%) of radio stars of different types and evolution stage (e.g. including evolved massive stars like Luminous Blue Variables or Wolf-Rayet), or even completely new or unexpected classes of objects. Classification tools could therefore significantly increase the number of known sources in the Galaxy, or at least guide science groups in proposing follow-up multi-wavelength observations for selected source samples. Machine learning, in general, and specifically deep learning techniques, have proven to be very powerful for this kind of analysis. We summarize here the developments made on radio source classification in recent works.\\
Most of the contributions focused on galaxy morphology classification for extragalactic science cases. 
For example, \cite{Aniyan2017} employed convolutional neural networks (CNNs) for classification of Fanaroff–Riley (FR) type I and type II \citep{FanaroffRiley1974}, and bent-tailed radio galaxies, using images from the Very Large Array (VLA) FIRST\footnote{FIRST: Faint Images of the Radio Sky at Twenty-cm \citep{Becker1995}} survey.
Similar analysis were conducted using CNNs \citep{Lukic2018,Wu2019,Maslej2021,Rustige2022} or capsule networks \citep{Lukic2019} on FIRST\footnote{The Radio Galaxy Zoo (RGZ) DR1 dataset \citep{Banfield2015} is $\sim$99\% made up by FIRST survey images.} and LOFAR (Low Frequency Array) images \citep{Alegre2022}. \cite{Sadeghi2021} studied morphological-based classification of FRI/FRII radio galaxies with Support Vector Machine (SVM) \citep{Cortes1995} models, using computed Zernike moments of source images from the FIRST survey. 
Radio galaxy morphology was also studied using semi-supervised \citep{Slijepcevic2022} and unsupervised learning methods, for example employing Kohonen maps \citep{Polsterer2016,Gupta2022,Galvin2020} or K-means clustering algorithm applied to compressed features learnt by convolutional autoencoders and Self-Organising Maps (SOMs) \citep{Ralph2019}.\\
Various works used ML techniques to target Galactic science objectives, such as the identification of Galactic objects or selected object classes from the dominant background of extragalactic sources, or the discovery of anomalous/unexpected objects. Among them, \cite{Akras2019} employed decision trees for classifying PNe against mimics (\hii{} regions, stars, YSO) using near- and mid-infrared colours. \cite{Iskandar2020} tested several deep network architectures to identify PNe from rejected PNe listed in the HASH\footnote{HASH: Hong Kong/AAO/Strasbourg H-alpha \citep{Parker2016}} and Pan-STARRS\footnote{Pan-STARRS: Panoramic Survey Telescope and Rapid Response System \citep{Flewelling2020}} databases, using infrared (WISE\footnote{WISE: Wide-Field Infrared Survey Explorer \citep{Wright2010}}) and optical (IPHAS\footnote{IPHAS: INT Photometric H$\alpha$ Survey of the Northern Galactic Plane \citep{Drew2005}}, VPHAS+\footnote{VPHAS+: VST/OmegaCAM Photometric H$\alpha$ Survey \citep{Drew2014}}, SHS\footnote{SHS: SuperCOSMOS Halpha Survey \citep{Parker2005}}/SSS\footnote{SSS: SuperCOSMOS Sky Survey \citep{Hambly2001}}) images. \cite{Anderson2012} considered mid- and far-infrared colours, providing diagnostic selection criteria for discriminating PNe and \hii{} regions. \cite{Morello2018} also considered near-infrared (2MASS\footnote{IRAC: Infrared Array Camera \citep{irac}}) colours to identify new Wolf-Rayet star candidates from other stellar populations contaminants (Young Stellar Objects (YSOs), asymptotic giant branch (AGB) candidates, Be/M$-$S type stars), using variants of the k-nearest neighbours algorithm. None of these studies, however, included radio data in their analysis or had the radio domain as their primary target. In this context, various ML applications were instead primarily developed for classification of radio sources in the Galactic plane. Among them, \cite{Liu2019} used radio data from different surveys (MGPS\footnote{MGPS: Molonglo Galactic Plane Survey \citep{mgps}}, MAGPIS\footnote{MAGPIS: Multi-Array Galactic Plane Imaging Survey \citep{MAGPIS}}, NVSS\footnote{NVSS: NRAO VLA Sky Survey \citep{Condon1998}}, CGPS\footnote{CGPS: Canadian Galactic Plane Survey \citep{cgps}}) to train a deep CNN to identify Supernova Remnants (SNRs) from non-SNRs (e.g. regions surrounding the SNRs in their analysis). Several studies \citep{Bates2012,Lyon2016,Tan2018} employed machine learning methodologies to classify pulsars from non-pulsars or to filter pulsar candidates. We also recently provided some contributions in this field. In \cite{Riggi2023} we have applied the Mask R-CNN object detection framework to detect and classify compact point-source, extended radio galaxies, and imaging artefacts, making use of radio data from the FIRST, Scorpio ATCA\footnote{ATCA: Australia Telescope Compact Array}\citep{Umana2015a} and ASKAP EMU pilot surveys. In \cite{Riggi2021a} we have trained a decision tree to identify Galactic-like sources from extragalactic ones on the basis of their radio-infrared colours. The classifier was also applied to a set of 284 unclassified sources selected in the ASKAP \scorpio{} survey field, highlighting potential Galactic objects for future studies. This analysis was however limited by the size and reliability of the dataset used for model training, mostly based on past low-resolution Galactic plane surveys.\\In this work, we made significant steps further, building a much larger and curated dataset of different Galactic and extragalactic compact objects, including previous and newest radio data in the Galactic plane, combined with mid- and far-infrared data, and measuring the radio spectral index for a portion of them. 
Such a dataset will be used as a reference for performing classification studies with different machine learning methodologies in a series of planned papers. 
The scope of this first paper, besides presenting the dataset, is firstly to 
explore and select suitable parameters for source classification, from more traditional science-aware features (e.g. radio-infrared colours, spectral indices), to more abstract features automatically learnt in convolutional neural network architectures. Secondly, we would like to build and test a supervised learning model 
able to predict a classification label for an input set of unknown sources, from the considered set of class categories, providing also the relative membership score. As a final goal, we aim to deliver the trained model and the classification tool/service to end users, supporting SKA and precursor science projects planned in the Galactic plane (e.g. production of added-value catalogues from pipeline catalogue products, source selection for follow-up analysis, and so forth). In future papers we will focus on testing unsupervised techniques for cluster search and anomaly detection on the same dataset.\\ 
This paper is organized as follows. In Section~\ref{sec:observations} we describe the observational radio data and supplementary surveys used to create our compact source image dataset. The source classes considered for the analysis, the methodology followed to prepare the dataset, and summary dataset information, are reported in Section~\ref{sec:dataset}. In Section~\ref{sec:feat-extraction} we describe the techniques explored to extract a set of sensitive features for classification from the produced dataset. The results of our classification analysis are reported in Section~\ref{sec:classification-analysis}. Details on the analysis pipeline and the implemented tool are provided in Section~\ref{sec:method}. In Section~\ref{sec:summary}, we summarized our findings, and highlighted future steps and analysis that are planned to be done with the produced dataset.
 
\begin{table}[htb]
\centering%
\footnotesize%
\caption{Centres of the ASKAP EMU pilot phase 2 images used in this work. Each image covers an area of $\sim$40 deg$^{2}$. Column (1) indicates the observation scheduling blocks.}
\begin{tabular}{ccc}
\hline%
\hline%
SB & Right ascension (J2000) & Declination (J2000)\\%
& \textit{(hh:mm:ss.ss)} & \textit{(dd:mm:ss.ss)}\\%
\hline%
28280 & 16:50:44.24 & $-$41:47:22.93 \\%
32043 & 16:45:03.28 & $-$46:28:26.90 \\
32145 & 17:18:08.79 & $-$41:52:51.98 \\%
32526 & 17:15:03.15 & $-$46:28:53.52 \\%
33284 & 15:53:57.03 & $-$55:41:52.77 \\%
\hline%
\hline%
\end{tabular}
\label{tab:askap-observations}
\end{table}

\section{Observational data}
\label{sec:observations}

\subsection{ASKAP radio surveys}
We searched for sources of different classes in ASKAP pilot survey observations, carried out both far and towards the Galactic plane. Details are reported in the following sections.

\subsubsection{ASKAP EMU pilot survey data}
\label{subsec:askap-data}
The ASKAP-EMU survey \citep{Norris2011} started observations at the end of 2022. This work makes use of different radio continuum maps that were produced with the ASKAP telescope during the commissioning and science preparation activities for EMU:

\begin{itemize}
\item \emph{Early Science data}: The \textsc{Scorpio} field was the only region observed in the Galactic plane by ASKAP at multiple radio frequencies during the Early Science and pilot 1 phase. First observations, done in 2018 with 15 antennas at 912 MHz, and covering $\sim\!40$ square degrees centred on ($l$,$b$)=(343.5$^{\circ}$, 0.75$^{\circ}$), were described in \cite{Umana2021} along with data reduction, while scientific results on compact sources were presented in \cite{Riggi2021a}.\\As the array was nearly completed, new observations of the same region were carried out with 30 antennas in band 1 (900 MHz), 2 (1250 MHz), and 3 (1550 MHz), each with a 288 MHz bandwidth, thus providing a much better sensitivity and an almost full frequency coverage from 0.75 to 1.7 GHz when combining all observations. Observation configuration and data reduction were described in more detail in \cite{Ingallinera2022}. Final data products\footnote{These data still have a parametrized primary beam correction in the three bands, affecting flux density measurement by $\sim$10\% \citep{Riggi2021a}, as precise measurements of the beam shape became available afterwards at pilot 1 phase \citep{Norris2021}.} include a total intensity map at the reference frequency of 1243 MHz and 5 sub-band maps (reference frequencies: 871 MHz, 1015 MHz, 1356 MHz, 1480 MHz, 1615 MHz).\\The synthesized 
beam of the total intensity maps is 9.4"$\times$7.7" at a position angle of 84$^{\circ}$. The background rms noise in regions far from the Galactic plane and point-sources was found of the order of 50 $\mu$Jy/beam.
\item \emph{Pilot data}: When the array was completed, a pilot survey program was undertaken 
within EMU. In Phase 1, the survey covered $\sim$270 deg$^{2}$ of the Dark Energy Survey area, reaching an angular resolution of 11"$-$18" and $\sim$30 $\mu$Jy/beam noise rms at 944 MHz \citep{Norris2021}. Observations towards the Galactic plane were carried out in Phase 2. They consist of 5 tiles, each covering $\sim$40 deg$^{2}$. Their coordinate centers and corresponding observation scheduling blocks are reported in Table~\ref{tab:askap-observations}.
The achieved angular resolution of the total intensity maps varies from 14" to 20", and the noise rms is of the order of 200 $\mu$Jy/beam far from the Galactic plane and from regions of diffuse emission.
\end{itemize}

\subsubsection{The Rapid ASKAP Continuum Survey}
\label{subsec:askap-racs}
The RACS survey \citep{McConnell2020} is the first large area survey carried out at 887.5 MHz with ASKAP. It reached an angular resolution of 15"$-$25", a rms sensitivity of 0.2$-$0.4 mJy/beam, and source positional accuracy better than 1", delivering a catalogue of 2,123,638 sources, 95\% complete above 3 mJy \citep{Hale2021}\footnote{Image products are publicly available through the CSIRO ASKAP Science Data Archive (CASDA) at \scriptsize{\url{https://data.csiro.au/domain/casdaObservation}}}.

\subsection{Previous radio surveys}
We also searched for sources in the following previous radio surveys carried out between 1.4 and 5 GHz. Some of them cover a large portion of the Galactic plane in the first quadrant. Details are reported below:
\begin{itemize}

\item \emph{The HI/OH/Recombination line survey
of the Milky Way}: THOR \citep{thor} is a Galactic plane survey (14.0$^{\circ}$<$l$<67.4$^{\circ}$, $|b|$<1.25$^{\circ}$) carried out with the VLA in C-configuration at 1.42 GHz. Observations achieved an angular resolution of 10"$-$25" with a noise rms of 0.3$-$1.0 mJy/beam\footnote{THOR data products are available at \scriptsize{\url{https://www2.mpia-hd.mpg.de/thor/DATA/www/}}}.
\item \emph{The Global view on Star formation in the Milky Way}: GLOSTAR \citep{GLOSTAR1,GLOSTAR2} is a Galactic plane survey (28$^{\circ}$<$l$<36$^{\circ}$, $|b|$<1$^{\circ}$) carried out with the VLA in B and D configuration at 4$-$8 GHz. The integrated map has a resolution of 18" and a sensitivity of
$\sim$60$-$150 $\mu$Jy/beam at the effective frequency of 5.8 GHz\footnote{GLOSTAR data products are available at \scriptsize{\url{https://glostar.mpifr-bonn.mpg.de/glostar/image_server}}}.
\item \emph{Multi-Array Galactic Plane Imaging Survey}: MAGPIS \citep{MAGPIS} is a Galactic plane survey (5$^{\circ}$<$l$<48.5$^{\circ}$, $|b|$<0.8$^{\circ}$) carried out with the VLA in B, C, and D configurations at 1.4 GHz. Observations achieved an angular resolution of 6" with a noise rms of 0.3 mJy/beam\footnote{MAGPIS data products can be downloaded from the public cutout web interface \scriptsize{\url{https://third.ucllnl.org/cgi-bin/gpscutout}}}.
\item \emph{The Coordinated Radio and Infrared Survey for High-Mass Star Formation}: CORNISH \citep{CORNISH1,CORNISH2} is a Galactic plane survey (10$^{\circ}<l<$65$^{\circ}$, $|b|<$1.1$^{\circ}$) carried out with the VLA in B and BnA configurations at 5 GHz. Observations achieved an angular resolution of 1.5" with a noise rms of 0.4 mJy/beam\footnote{CORNISH data products can be retrieved from the public cutout web interface \scriptsize{\url{https://cornish.leeds.ac.uk/public/img_server.php}}}.
\item \emph{Faint Images of the Radio Sky at Twenty cm (FIRST) survey}: The FIRST survey \citep{Becker1995} is a large area ($\sim$10,500 deg$^{2}$, $\sim$80\% covering the north Galactic cap) carried out at 1.4 GHz with the NRAO VLA. It reached an angular resolution of $\sim$5.4", a rms sensitivity of 0.15 mJy/beam, and source positional accuracy better than 1", delivering a catalogue of 946,432 sources in its latest version \citep{Helfand2015}, 95\% complete above 2 mJy\footnote{FIRST data products are publicly available at \scriptsize{\url{ftp://archive.stsci.edu/pub/vla_first/data}} or can be retrieved from the cutout web service interface \scriptsize{\url{https://third.ucllnl.org/cgi-bin/firstcutout}}}. 
\end{itemize}

\begin{table}[htb]
\centering%
\footnotesize%
\caption{Reference catalogues considered for searching radio stars in our dataset.}
\begin{threeparttable}
\begin{tabular}{lcl}
\hline%
\hline%
Reference & $N_{entries}$ & Type\\%
\hline%
\cite{Wendker1995} & 1128 & mostly M-type stars\\%
\hline%
\cite{Benaglia2010} & 65 & O$-$B2 stars with wind radio\\%
& & emission\\%
\hline%
\cite{Kimball2009}\tnote{(a)} & 112 & $\sim$75\% late-type stars (K$-$M)\\%
\hline%
\cite{Rosslowe2015}\tnote{(b)} & 667 & Wolf-Rayet (WR) stars\\%
\hline%
\cite{Richardson2018} & 88\tnote{(c)} & Luminous Blue Variables\\%
\hline%
\cite{Wachter2010} & 71 & Spitzer massive stars with \\%
& & circumstellar shells, including \\%
& & early and late-type stars, \\%
& & 16 LBVs candidates and 6 WRs\\%
\hline%
\cite{Leto2021} & 50 & magnetic chemically \\%
\cite{Shultz2022} & & peculiar (MCP) radio stars\\%
\hline%
\cite{Liu2007} & 187 & Low-mass X-ray binaries\\%
& & in the Galaxy, LMC, and SMC,\\%
& & 4th ed.\tnote{(d)}\\%
\hline%
\cite{Liu2006} & 114 & High-mass X-ray binaries \\%
& & in the Galaxy, 4th ed. \tnote{(e)}\\%
\hline%
\hline%
\end{tabular}
\begin{tablenotes}\footnotesize
\item[(a)] \footnotesize{As the authors stated, this sample is expected to be contaminated by optically faint radio quasars, with only few tens of candidates expected to be truly radio stars.} 
\item[(b)] \scriptsize{\url{http://pacrowther.staff.shef.ac.uk/WRcat/index.php}}
\item[(c)] \footnotesize{Counts include 60 Galactic LBVs and extragalactic LBVs from the Local Group (LMC, SMC).}
\item[(d)] \scriptsize{\url{https://heasarc.gsfc.nasa.gov/W3Browse/all/lmxbcat.html}}
\item[(r)] \scriptsize{\url{https://heasarc.gsfc.nasa.gov/w3browse/all/hmxbcat.html}}
\end{tablenotes}
\end{threeparttable}

\label{tab:radio-star-catalogue}
\end{table}

\subsection{Supplementary infrared data}
\label{subsec:infrared-surveys}
In this study, we complemented the radio observations with mid- and far-infrared data from the following surveys:
\begin{itemize}
\item \emph{AllWISE} \citep{Cutri2013}: This WISE survey is fully covering the Galactic plane in four bands at 3.4$\mic$ (W1), 4.6$\mic$ (W2), 12$\mic$ (W3), and 22$\mic$ (W4). The angular resolutions are 6.1", 6.4", 6.5" and 12" and the 5$\sigma$ flux sensitivities for point sources are 0.08 mJy, 0.11 mJy, 1 mJy and 6 mJy, respectively.
\item \emph{GLIMPSE} (Galactic Legacy Infrared MidPlane Survey Extraordinaire) 8.0$\mic$ surveys \citep{Churchwell2009} of the \emph{Spitzer Space Telescope} \citep{Werner2004}: 
GLIMPSE (I, II) fully covers this Galactic coordinate range: 0$^{\circ}$<$l$<65$^{\circ}$, 295$^{\circ}$<$l$<360$^{\circ}$, $|b|\le$1\footnote{The exact sky coverage of all GLIPSE surveys (including GLIMPSE-3D) is summarized at \scriptsize{\url{https://irsa.ipac.caltech.edu/data/SPITZER/GLIMPSE/overview.html}}.}.
The angular resolution is 2" and the 5$\sigma$ flux sensitivity $\sim$0.4 mJy.
\item \emph{Hi-GAL} (Herschel infrared Galactic plane Survey) 70$\mic$ survey \citep{Molinari2016} of the \emph{Herschel Space Observatory} \citep{Pilbratt2010}: 
The survey covers $|l|\le$60$^{\circ}$, $|b|\le$1$^{\circ}$, with an angular resolution of $\sim$8.5" and a 1$\sigma$ flux sensitivity $\sim$20 MJy/sr.
\end{itemize}

\begin{table*}[htb]
\centering%
\footnotesize%
\caption{Summary information on the compact source data extracted from previous radio surveys (FIRST, THOR, GLOSTAR, MAGPIS, CORNISH). Columns (3) and (4) are the average radio source angular size and its standard deviation in arcsec. Column (5) lists the number of sources from previous radio surveys for each considered class or sub-class in columns (1) and (2) with available Mid-Infrared data (3.4$\mic$, 4.6$\mic$, 12$\mic$, 22$\mic$) from AllWISE survey. Column (6) lists the number of radio sources with available Mid-Infrared data (3.4$\mic$, 4.6$\mic$, 8$\mic$, 12$\mic$, 22$\mic$) from AllWISE and GLIMPSE surveys, and Far-Infrared data (70$\mic$) from Hi-GAL survey.  Column (7) reports the number of radio sources with average spectral index information available (see text). Columns (8) and (9) reports how many sources listed in columns (5) and (6), respectively, also have a measured spectral index.}
\begin{tabular}{llccccccc}
\hline%
\hline%
\rowcolor{palegold}%
\multicolumn{1}{c}{\footnotesize{\textsc{class}}} & \multicolumn{1}{c}{\footnotesize{\textsc{subclass}}} & \multicolumn{1}{c}{\footnotesize{$\langle\textsc{r}\rangle$ \tiny{(")}}} & 
\multicolumn{1}{c}{\footnotesize{$\sigma_{\textsc{r}}$ \tiny{(")}}} & 
\footnotesize{$n_{\textsc{mir}}$} & \footnotesize{$n_{\textsc{mir}+\textsc{fir}}$} & \footnotesize{$n_{\alpha}$} & \footnotesize{$n_{\alpha+\textsc{mir}}$} & \footnotesize{$n_{\alpha+\textsc{mir}+\textsc{fir}}$}\\%
\rowcolor{palegold}%
\multicolumn{1}{c}{\tiny{(1)}} & \multicolumn{1}{c}{\tiny{(2)}} 
& \multicolumn{1}{c}{\tiny{(3)}} & \multicolumn{1}{c}{\tiny{(4)}} & \multicolumn{1}{c}{\tiny{(5)}} & \multicolumn{1}{c}{\tiny{(6)}} & \multicolumn{1}{c}{\tiny{(7)}} & \multicolumn{1}{c}{\tiny{(8)}} & \multicolumn{1}{c}{\tiny{(9)}}\\%
\hline%
\textsc{\hii{}} & & 22.1 & 12.6 & 2295 & 2257 & 1214 & 1178 & 1168\\%
\hline%
\textsc{pn} & & 18.2 & 7.1 & 1411 & 1214 & 783 & 782 & 718\\%
\hline%
\textsc{pulsar} & & 10.8 & 2.9 & 645 & 534 & 221 & 221 & 57\\%
\hline%
\textsc{yso} & & 10.6 & 4.1 & 552 & 501 & 215 & 204 & 71\\%
\hline%
\multirow{5}{*}{\textsc{star}} & \textsc{wr} & 10.1 & 3.9 & 51 & 47 & 7 & 7 & 6\\%
 & \textsc{lbv} & 20.7 & 11.3 & 25 & 24 & 11 & 10 & 10\\%
\textsc{star} & \textsc{xb} & 12.5 & 4.9 & 14 & 9 & 7 & 7 & 6\\%
 & \textsc{other} & 9.7 & 6.1 & 348 & 217 & 89 & 88 & 74 \\%
\cmidrule(lr){2-9}%
 & \textsc{all} & & & 438 & 297 & 114 & 112 & 96\\%
\hline%
\textsc{rg} & & 15.2 & 4.4 & 6920 & 264 & 1285 & 1285 & 134\\%
\textsc{qso} & & 18.6 & 4.8 & 5136 & 0 & 1045 & 1045 & 0\\%
\hline%
\rowcolor{lightgray}%
\textsc{all} & & & & 17397 & 5013 & 4877 & 4827 & 2244\\%
\hline%
\hline%
\end{tabular}
\label{tab:dataset-train}
\end{table*}

\begin{table*}[htb]
\centering%
\footnotesize%
\caption{Summary information on the compact source data extracted from different ASKAP radio surveys (RACS, EMU pilot). See Table~\ref{tab:dataset-train} caption for column description.}
\begin{tabular}{llccccccc}
\hline%
\hline%
\rowcolor{pistachio}%
\multicolumn{1}{c}{\footnotesize{\textsc{class}}} & \multicolumn{1}{c}{\footnotesize{\textsc{subclass}}} & \multicolumn{1}{c}{\footnotesize{$\langle\textsc{r}\rangle$ \tiny{(")}}} & 
\multicolumn{1}{c}{\footnotesize{$\sigma_{\textsc{r}}$ \tiny{(")}}} &
\footnotesize{$n_{\textsc{mir}}$} & \footnotesize{$n_{\textsc{mir}+\textsc{fir}}$} & \footnotesize{$n_{\alpha}$} & \footnotesize{$n_{\alpha+\textsc{mir}}$} & \footnotesize{$n_{\alpha+\textsc{mir}+\textsc{fir}}$}\\%
\rowcolor{pistachio}%
\multicolumn{1}{c}{\tiny{(1)}} & \multicolumn{1}{c}{\tiny{(2)}}  & \multicolumn{1}{c}{\tiny{(3)}} & \multicolumn{1}{c}{\tiny{(4)}} & \multicolumn{1}{c}{\tiny{(5)}} & \multicolumn{1}{c}{\tiny{(6)}} & \multicolumn{1}{c}{\tiny{(7)}} & \multicolumn{1}{c}{\tiny{(8)}} & \multicolumn{1}{c}{\tiny{(9)}}\\%
\hline%
\textsc{\hii{}} & & 26.8 & 13.9 & 369 & 364 & 57 & 56 & 55\\%
\hline%
\textsc{pn} & & 27.0 & 8.7 & 174 & 45 & 33 & 33 & 10\\%
\hline%
\textsc{pulsar} & & 18.5 & 6.6 & 166 & 72 & 11 & 11 & 9\\%
\hline%
\textsc{yso} & & 12.4 & 4.1 & 37 & 30 & 17 & 16 & 15\\%
\hline%
 \multirow{5}{*}{\textsc{star}} & \textsc{wr} & 14.5 & 3.9 & 11 & 7 & 1 & 1 & 1\\%
 & \textsc{lbv}  & 23.1 & 10.4 & 3 & 3 & 0 & 0 & 0\\%
\textsc{star} & \textsc{xb}  & 12.5 & 1.7 & 2 & 0 & 0 & 0 & 0\\%
 & \textsc{other}  & 14.2 & 3.6 & 7 & 1 & 2 & 2 & 0 \\%
\cmidrule(lr){2-9}%
  & \textsc{all} & & & 23 & 11 & 3 & 3 & 1\\%
\hline%
\textsc{rg} &  & 22.3 & 6.0 & 1892 & 0 & 0 & 0 & 0\\%
\textsc{qso} &  & 26.1 & 7.7 & 1343 & 0 & 0 & 0 & 0\\%
\hline%
\rowcolor{lightgray}%
\textsc{all} & & & & 4004 & 522 & 121 & 119 & 90\\%
\hline%
\hline%
\end{tabular}
\label{tab:dataset-test}
\end{table*}

\section{Generating training/testing datasets}
\label{sec:dataset}

\begin{figure*}[htb]
\centering%
\subtable[3.4 $\mic$]{\includegraphics[scale=0.24]{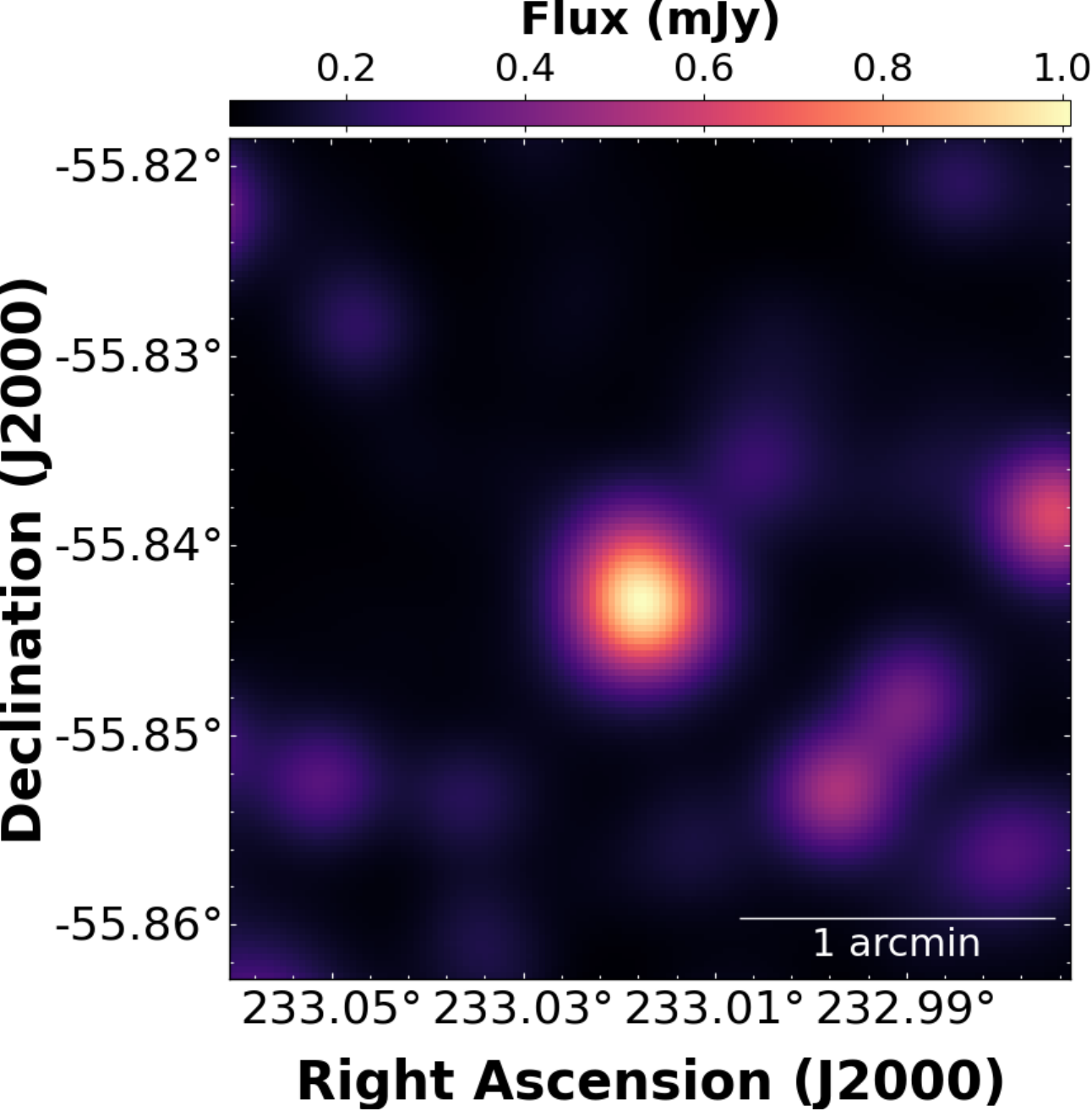}}%
\subtable[4.6 $\mic$]{\includegraphics[scale=0.24]{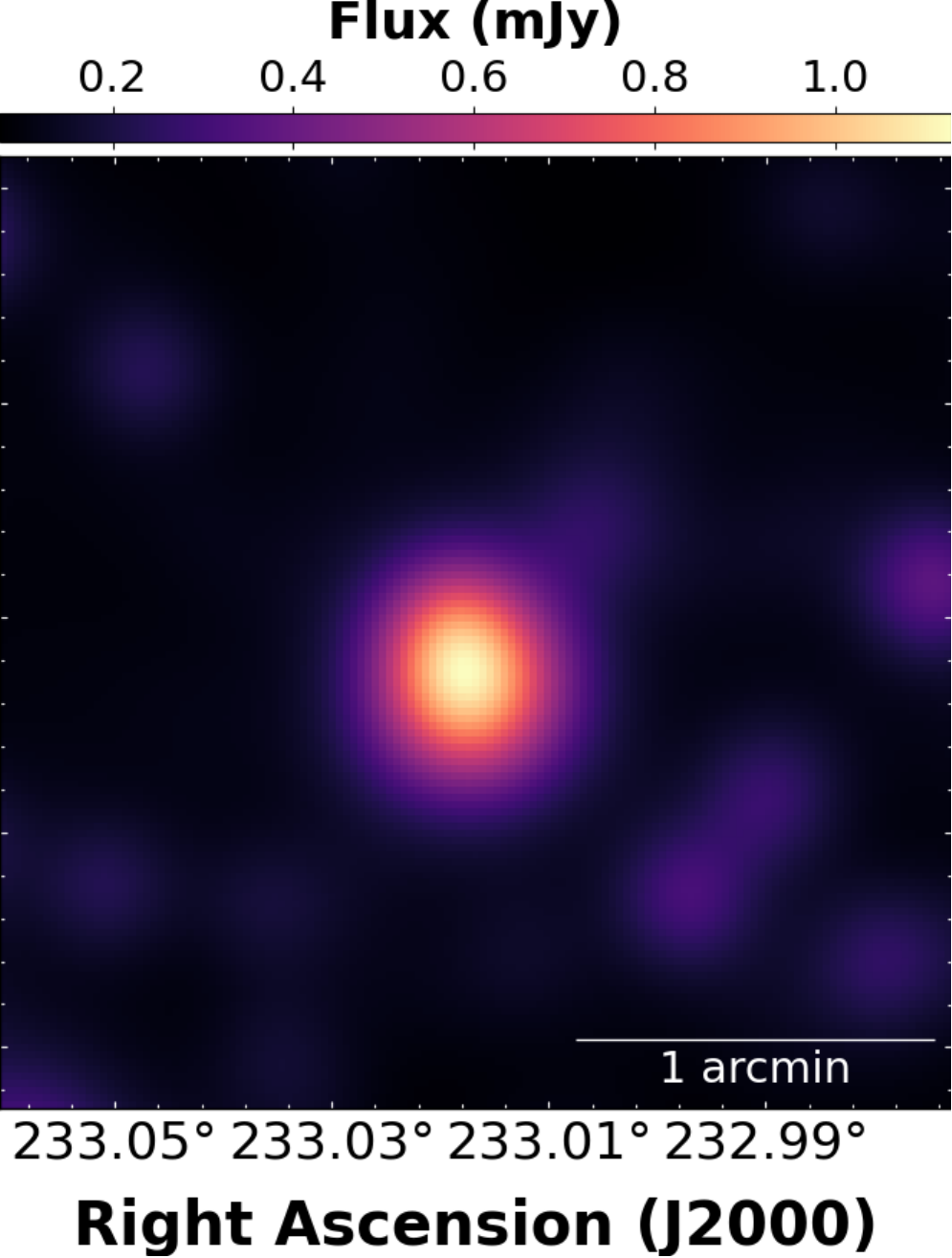}}%
\hspace{0.1cm}%
\subtable[8 $\mic$]{\includegraphics[scale=0.24]{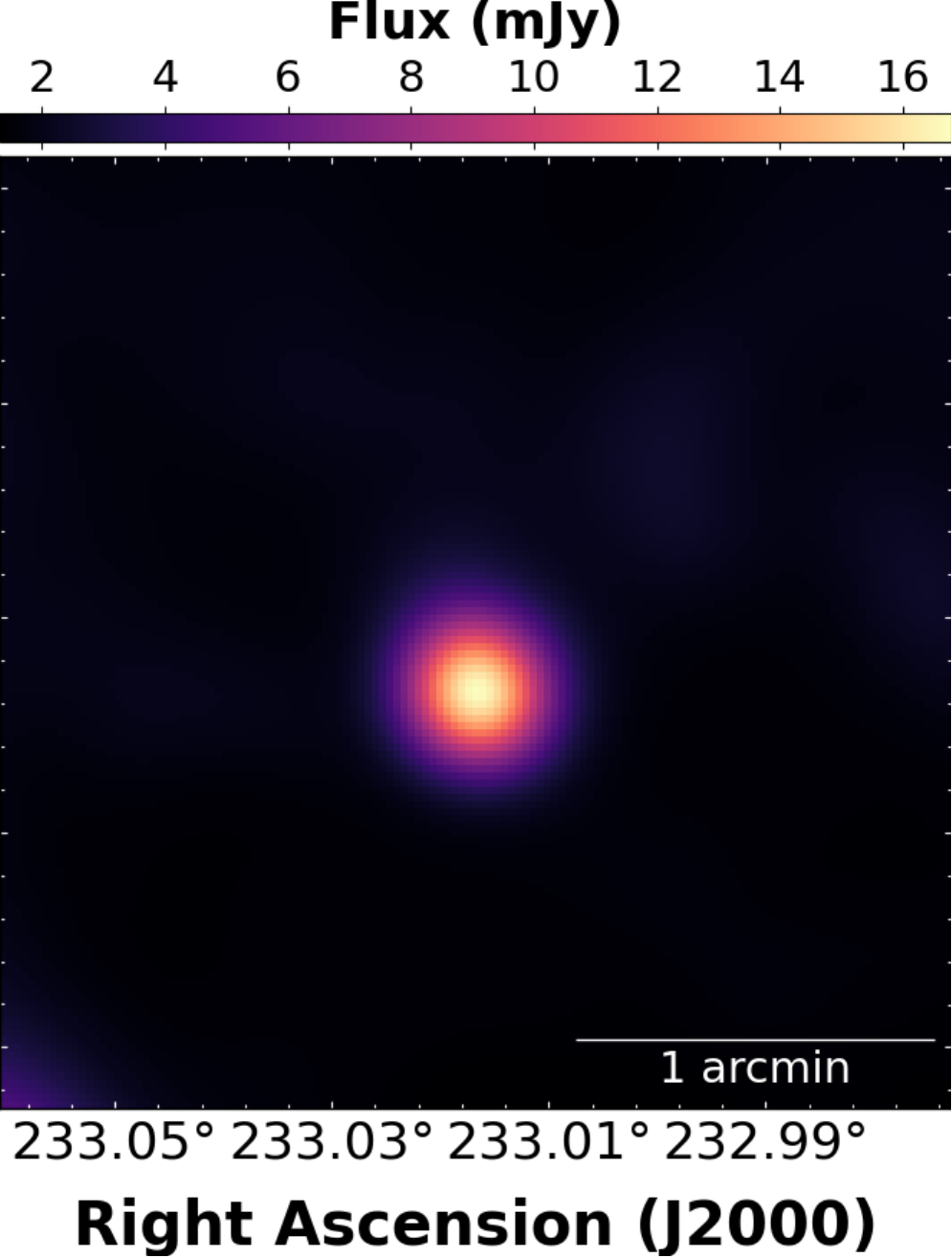}}
\subtable[12 $\mic$]{\includegraphics[scale=0.24]{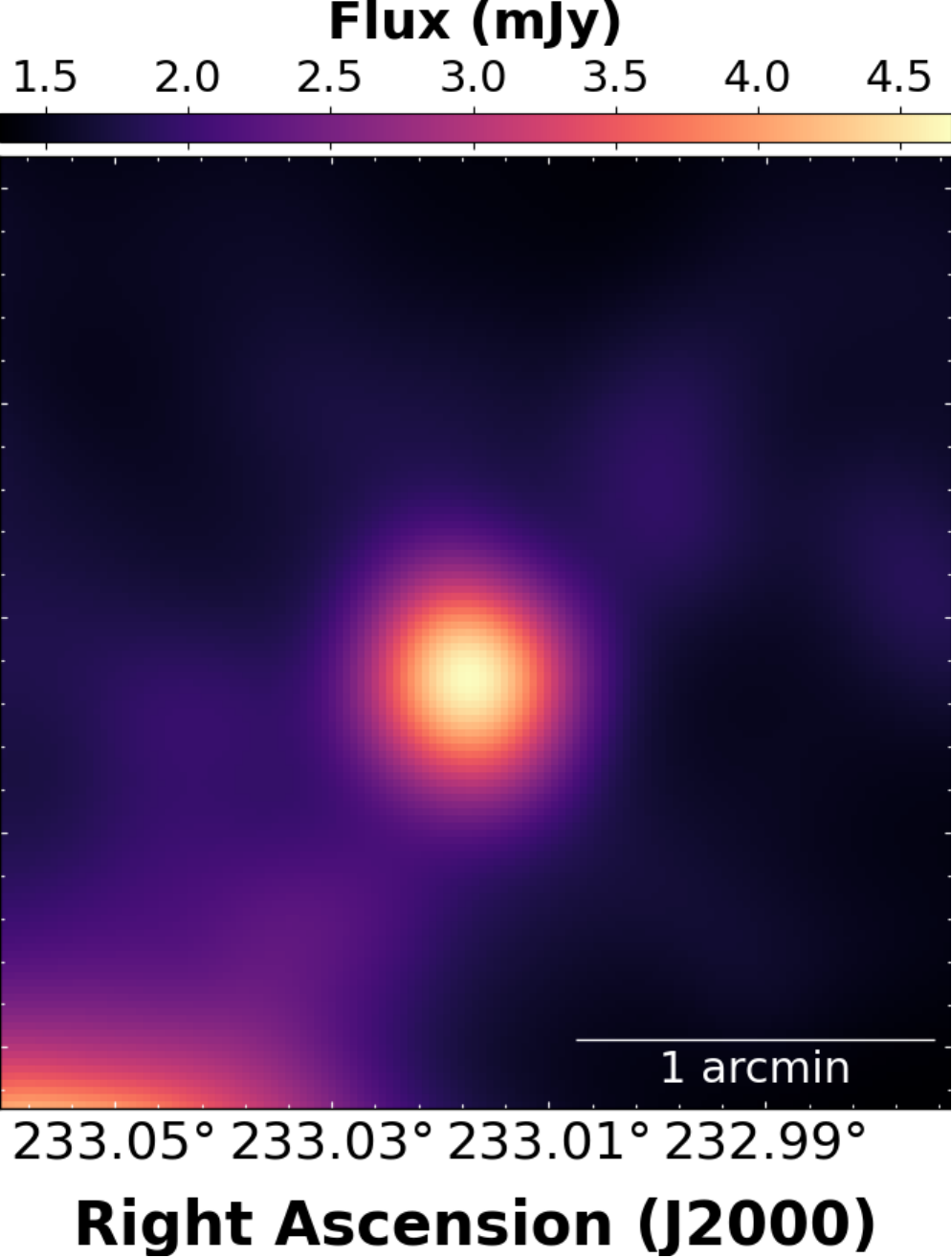}}\\%
\hspace{0.1cm}%
\subtable[22 $\mic$]{\includegraphics[scale=0.24]{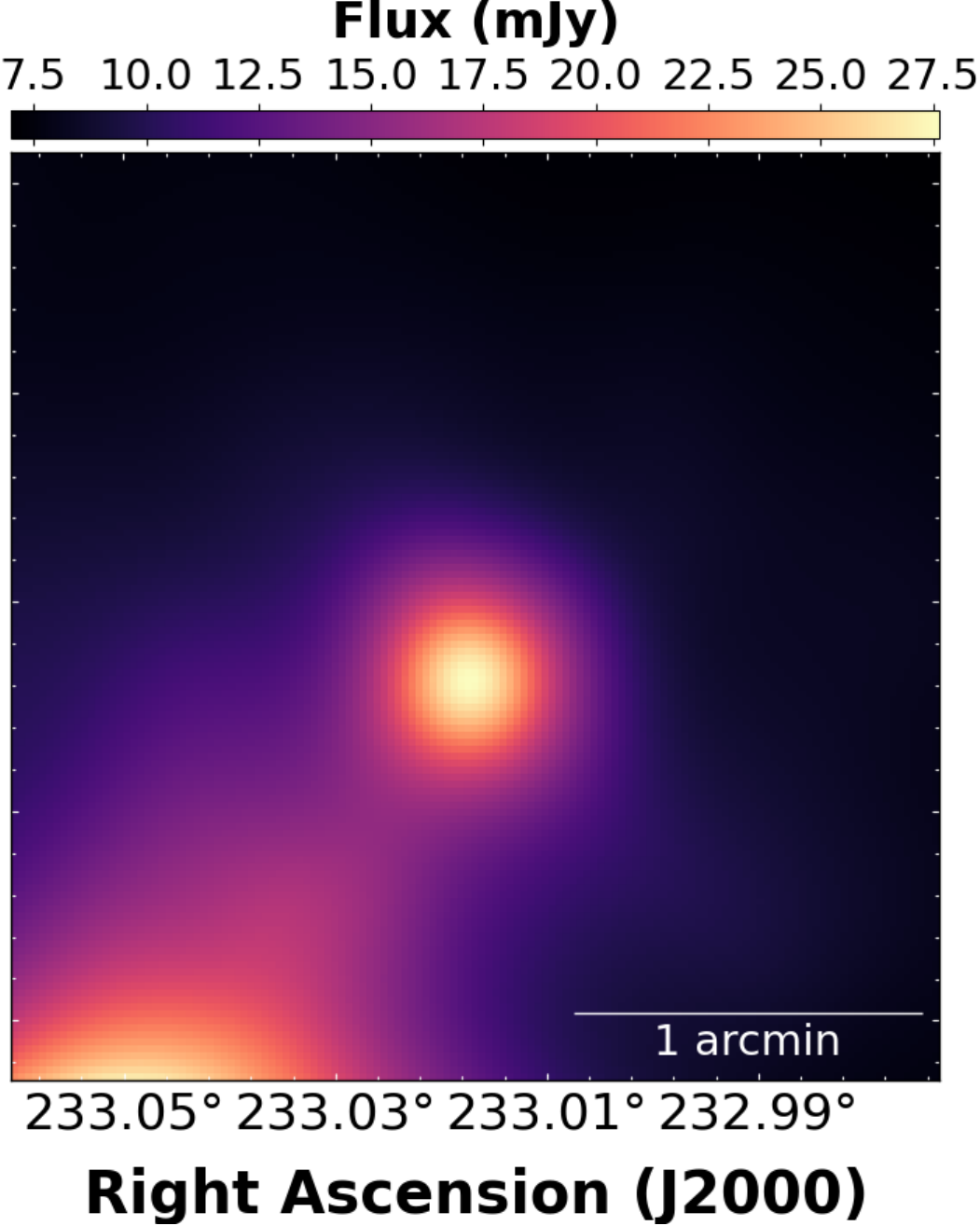}}%
\subtable[70 $\mic$]{\includegraphics[scale=0.24]{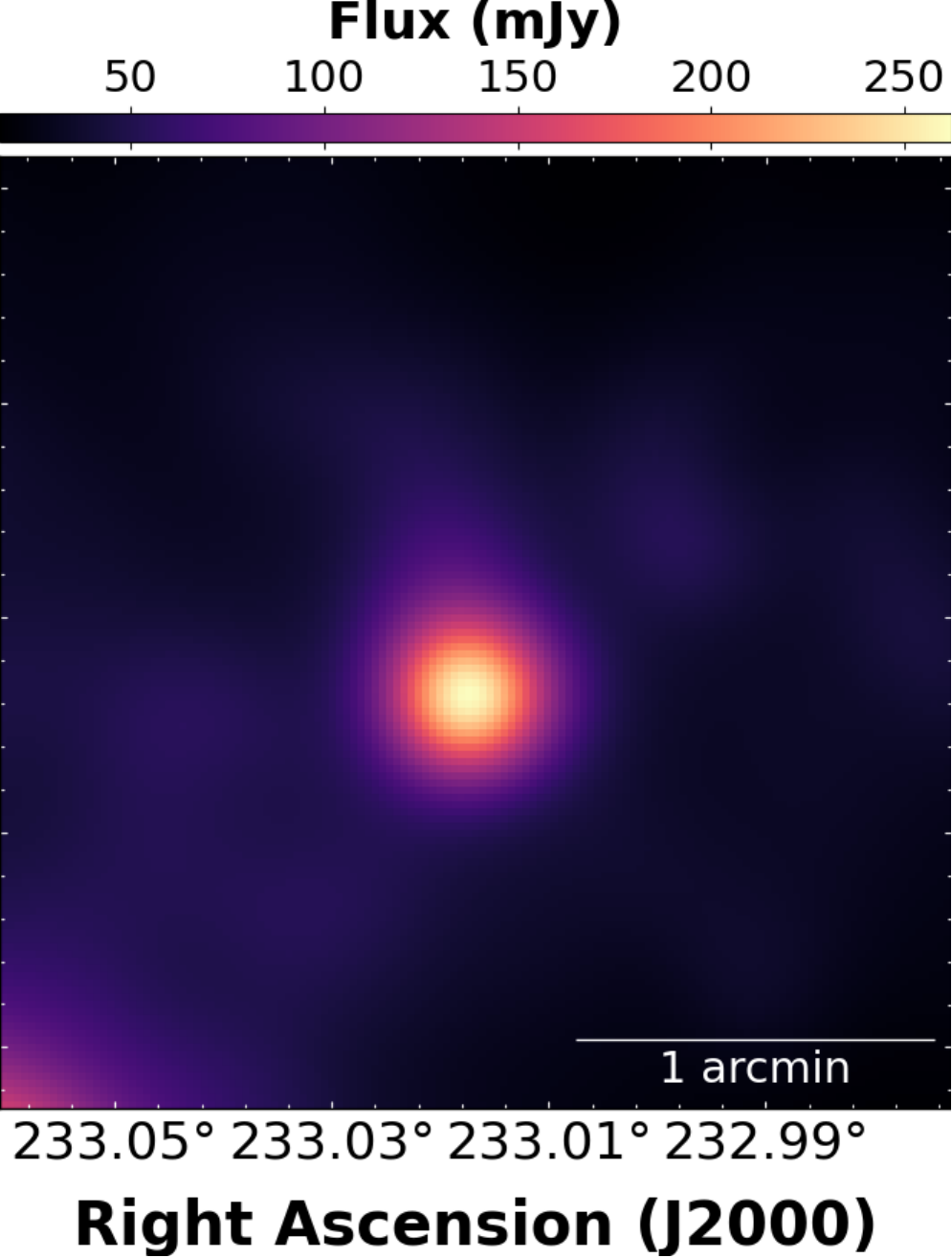}}%
\hspace{0.1cm}%
\subtable[radio]{\includegraphics[scale=0.24]{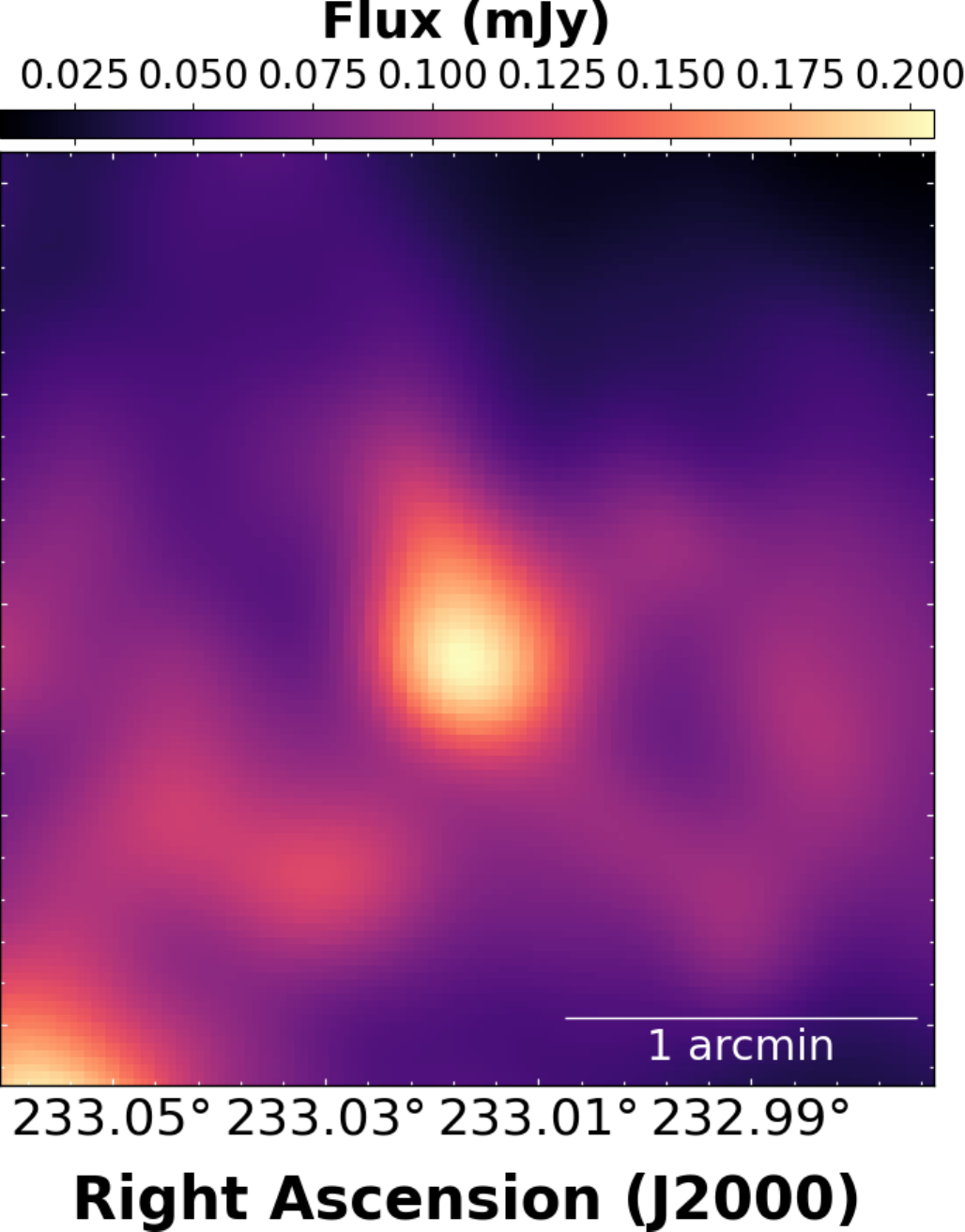}}%
\caption{Template source (G324.161+00.264, \hii{} region) from the dataset, observed in 7-bands (3.4$\mic$, 4.6$\mic$, 8$\mic$, 12$\mic$, 22$\mic$, 70$\mic$, and ASKAP radio 944 MHz), shown in left to right panels, respectively.}%
\label{fig:sample-source}
\end{figure*}

\subsection{Compact source sample}
\label{subsec:compact-sources}
To build our dataset, we searched for compact sources in the available radio data (Section~\ref{sec:observations}), using the following selection criteria:

\begin{enumerate}
\item Isolated single-island point-sources or slightly resolved sources. We assumed an upper threshold of 10$\times$ synthesized beam size\footnote{The average size of sources (e.g. \hii{} regions or PNe) in the dataset is $\sim$2.5$\times$ synthesized beam size.};
\item No diffuse, extended or complex radio morphologies, e.g. no child point-sources or inner filaments inside source contour;
\item Source position cross-matching to known or candidate objects in reference catalogues, within a match radius equal to the synthesized beam size;
\item Source island clearly distinguishable from the background, e.g. peak flux larger than 3$\sigma$ and number of island pixels larger than 6;
\item Source island not located at radio map borders.
\end{enumerate}

We considered possible associations to these classes of astrophysical objects (Galactic or Extragalactic), having a compact radio morphology (as defined above), in most of the cases (e.g. pulsars or radio stars), or in a considerable fraction of cases (e.g. \hii{} regions, PNe) compared to more extended morphologies:

\begin{itemize}
\item \emph{Radio stars}: We included in this class stars of different spectral types and evolution stages, including late stages, like Wolf-Rayet (WR) stars or Luminous Blue Variables (LBVs), and X-ray binaries (hereafter abbreviated as XBs for brevity). The sensitivity of existing telescopes has been the major limitation in radio star searches, as the emission is rather faint, often below the mJy level. Furthermore, a limited angular resolution, e.g. above 1" \citep{Helfand1999}, makes cross-matching with densely populated optical catalogues ineffective. In fact, the number of reported radio stars is rather low, and no comprehensive catalogue, including all possible stellar types, is currently available. To build a sufficiently large dataset, we considered different reference catalogues of known and candidate radio stars, to be cross-matched with available radio data. References are reported in Table~\ref{tab:radio-star-catalogue}.
\item \emph{\hii{} regions}: We have used the WISE Catalogue of Galactic \hii{} regions \citep{Anderson2014,Makai2017}, as a reference for searching \hii{} region associations in our radio data. The catalogue is actively updated online\footnote{\scriptsize{\url{http://astro.phys.wvu.edu/wise/}}}. The version used for this work (v2.2) contains 8412 entries, $\sim$10\% of them with measured radio flux information reported at 20$-$21 cm.
\item \emph{Planetary Nebulae} (PNe): We have used the Hong Kong/AAO/Strasbourg H-alpha (HASH) Planetary Nebula Database \citep{Parker2016}, representing the largest compilation to date, as a reference for searching PNe in our radio data. The HASH catalogue is actively updated online\footnote{\scriptsize{\url{http://202.189.117.101:8999/gpne/dbMainPage.php}}}. The version used for this work contains 5591 entries, $\sim$24\% of them with measured radio flux density reported at 20 cm or 36 cm.
\item \emph{Young Stellar Objects} (YSOs): We carried out a search for possible associations to confirmed YSOs in our radio data, using the SIMBAD database \footnote{\scriptsize{\url{https://simbad.unistra.fr/simbad/}}} as a reference. No distinction is made among possible evolution or mass classes of YSOs. In the search, we discarded all matches found to compact radio sources, previously labelled as \hii{} regions and PNe.
\item \emph{Pulsars}: We have searched for pulsar matches in our radio data, using the ATNF Pulsar Catalogue\footnote{\scriptsize{\url{https://www.atnf.csiro.au/research/pulsar/psrcat/}}} \citep{Manchester2005} as a reference. The version used (version 1.63) for this work contains 2800 entries, 67\% of them with measured radio flux density reported at 21 cm.
\item \emph{Active Galactic Nuclei}: For our analysis, we considered a catalogue of radio galaxies and quasars obtained by \cite{Kimball2008} through cross-matching of different radio surveys (FIRST, primarily) with optical data from the Sloan Digital Sky Survey (SDSS) \citep{York2000}, providing source spectroscopic classification ("GALAXY", "QSO") (see \citealt{Bolton2012} for details). After applying the criteria given by \cite{Kimball2008} to select compact and unresolved sources, we selected 7967 radio galaxies (RG), and 5994 QSOs. By visual inspection, we removed residual extended sources passing the selection cuts, and sources found with incorrect/unclear position reported in the catalogue, as compared to FIRST images. The final selected sample includes: 6646 radio galaxies, and 5213 QSOs.
\end{itemize}

A brief description of physical properties for each of these source classes is reported in Appendix~\ref{appendix:source-classes}. As we expect these types to be the most abundant classes of compact sources found in Galactic plane observations, we did not consider other rarer classes. Actually, star-forming galaxies (SFG) are expected to become dominant over AGNs at sub-mJy flux levels (<100 $\mu$Jy) \citep{Mancuso2017} but their counts should be very small in FIRST/ASKAP-RACS surveys, given their sensitivities. This is, however, not the case for future ASKAP-EMU observations, so future studies should aim to incorporate SFGs in our dataset, once reference labelled catalogues become available within EMU.\\
Sources detected in our considered radio maps are reported in Tables~\ref{tab:dataset-train} and \ref{tab:dataset-test}. The resulting dataset is not expected to be completely free of spurious associations, due to the cross-matching procedure and to possible object misclassifications affecting the reference catalogues. Indeed, one of the goal of this and future studies is to make these unlikely classifications discoverable by means of both supervised and unsupervised techniques. The uncertainty associated with the automated cross-matching procedure was evaluated on the ASKAP data by comparing the observed \hii{} regions matches (i.e. the most densely populated reference catalogue) against the expected number of matches purely arising by chance. Following \cite{Riggi2021a,Mauch2007,Ching2017}, the latter was estimated by averaging the number of matches found with multiple random catalogues in which the measured source positions were uniformly randomized inside the radio map. We found that less than 3\% of the selected matches are spurious. For each class, the obtained matches were all validated by visual inspection to reduce the number of spurious associations.

\subsection{Image dataset preparation}
\label{subsec:data-preparation}
Using the \emph{scutout} tool\footnote{\scriptsize{\url{https://github.com/SKA-INAF/scutout}}}, we extracted postage-stamp images around each compact source detected in reference radio maps listed in Section~\ref{sec:observations}. Additionally, 
source cutouts were extracted from the supplementary infrared survey maps described in Section~\ref{subsec:infrared-surveys}. Cutout raw size was set to 10$\times$ the source radius $r_{s}$\footnote{$r_{s}$ was computed as the radius of the circle circumscribed to the source bounding box obtained from source segmentation mask.}. The image cutout set for each source, including the radio plus a configurable number of infrared bands (3.4$\mic$, 4.6$\mic$, 8$\mic$, 12$\mic$, 22$\mic$, 70$\mic$), were all re-processed (e.g. re-gridding/re-projection, re-scaling, cropping) to bring them to the same pixel size, sky coordinate system, resolution, flux density units (Jy/pixel), and final image size (2.5$\times r_{s}$). Final images have a different size in pixels, depending on the source size radius $r_{s}$. In the analysis reported in Section~\ref{subsec:auto-feature-analysis}, all source images will be resized to a common size in pixels.\\
As the 8$\mic$ and far-infrared surveys only cover the Galactic plane, in contrast to the full WISE sky coverage, we considered two possible radio-infrared combinations when making the image cutouts, denoted throughout the paper as follows:
\begin{itemize}
\item 5-bands (or radio+MIR\footnote{Mid-Infrared}) dataset: comprising radio, 3.4$\mic$, 4.6$\mic$, 12$\mic$, and 22$\mic$ images; 
\item 7-bands (or radio+MIR+FIR\footnote{Far-Infrared}) dataset: comprising radio, 3.4$\mic$, 4.6$\mic$, 8$\mic$, 12$\mic$, 22$\mic$, and 70$\mic$ images.
\end{itemize}
In Fig.~\ref{fig:sample-source} we report the image data for a sample source (\texttt{G324.161+00.264} \hii{} region) detected in 7 different channels. Infrared (3.4$\mic$, 4.6$\mic$, 8$\mic$, 12$\mic$, 22$\mic$, 70$\mic$) and radio (ASKAP) data are shown in left to right panels, respectively.\\The number of available images finally selected in previous radio surveys (FIRST, THOR, GLOSTAR, MAGPIS, CORNISH) and in ASKAP surveys is reported for each source class in Tables~\ref{tab:dataset-train} and \ref{tab:dataset-test}, respectively. Columns (5) and (6) reports the number of sources detected in radio, for which MIR and FIR images are available\footnote{Availability of MIR or FIR images does not imply that the source is actually detected in that infrared bands.}. Overall, $\sim$17400 radio sources are available in the first dataset with MIR (5-bands) information, $\sim$30\% of them with also FIR information (7-bands). Extragalactic sources are almost completely missing in our 7-bands dataset, due to the limited coverage of far-infrared surveys. A major consequence is that, unfortunately, galactic-extragalactic source separation studies can be carried out only with the 5-bands dataset. On the other hand, this is, to the best of our knowledge, the largest radio data compilation simultaneously including different classes of Galactic and extragalactic compact objects, suitable for machine-learning and other algorithmic studies.

\begin{table}[htb]
\centering%
\footnotesize%
\caption{Summary of extracted color features used for classification analysis. See Section~\ref{subsec:color-indices} for details.}
\begin{tabular}{lll}
\hline%
\hline%
Dataset & Feature ID & Description\\%
\hline%
5-band & F$_{1}$, \dots, F$_{4}$ & c$_{\text{radio,j}}$ ($j$=12, 22, 3.4, 4.6 \mic)\\%
(radio+MIR) & F$_{5}$, \dots, F$_{8}$ & IoU$_{\text{radio,j}}$ ($j$=12, 22, 3.4, 4.6 \mic)\\%
& F$_{9}$, \dots, F$_{12}$ & SSIM$_{\text{radio,j}}$ ($j$=12, 22, 3.4, 4.6 \mic)\\%
\hline%
7-band & F$_{1}$, \dots, F$_{6}$ & c$_{\text{radio,j}}$ ($j$=12, 22, 3.4, 4.6, 8, 70 \mic)\\%
(radio+MIR+FIR) & F$_{7}$, \dots, F$_{12}$ & IoU$_{\text{radio,j}}$ ($j$=12, 22, 3.4, 4.6, 8, 70 \mic)\\%
& F$_{13}$, \dots, F$_{18}$ & SSIM$_{\text{radio,j}}$ ($j$=12, 22, 3.4, 4.6, 8, 70 \mic)\\%
\hline%
\hline%
\end{tabular}
\label{tab:color-features}
\end{table}

\begin{table}[htb]
\centering%
\scriptsize%
\caption{Percentage of radio sources potentially detected (e.g. IoU>0) in each infrared band.}
\begin{tabular}{lcccccc}
\hline%
\hline%
\multicolumn{1}{c}{\footnotesize{Class}} & \footnotesize{3.4$\mic$} & \footnotesize{4.6$\mic$} & \footnotesize{8$\mic$} & \footnotesize{12$\mic$} & \footnotesize{22$\mic$} & \footnotesize{70$\mic$}\\%
& (\%) & (\%) & (\%) & (\%) & (\%) & (\%)\\%
\hline%
\hii{} & 37.6 & 49.2 & 63.0 & 58.1 & 64.5 & 67.7 \\%
PN & 17.0 & 21.5 & 28.5 & 45.1 & 55.3 & 47.5 \\%
PULSAR & 4.8 & 4.2 & 1.7 & 0.7 & 0.4 & 0.5 \\%
YSO & 19.5 & 26.8 & 36.4 & 30.4 & 32.9 & 38.9 \\%
STAR & 35.6 & 34.9 & 26.3 & 18.2 & 19.1 & 14.0 \\%
RG & 86.5 & 83.6 & 7.6 & 37.1 & 21.1 & 0.0 \\%
QSO & 69.9 & 75.6 & $-$ & 60.3 & 36.4 & $-$  \\%
\hline%
\rowcolor{lightgray}%
ALL & 64.2 & 66.6 & 41.4 & 45.4 & 33.2 & 47.0 \\%
\hline%
\hline%
\end{tabular}
\label{tab:dataset-match-stats}
\end{table}

\section{Feature extraction and data exploration}
\label{sec:feat-extraction}
In this section, we describe the methods used to process our dataset and extract parameters suitable for data inspection and source classification.

\begin{figure}[htb]
\centering%
\includegraphics[scale=0.37]{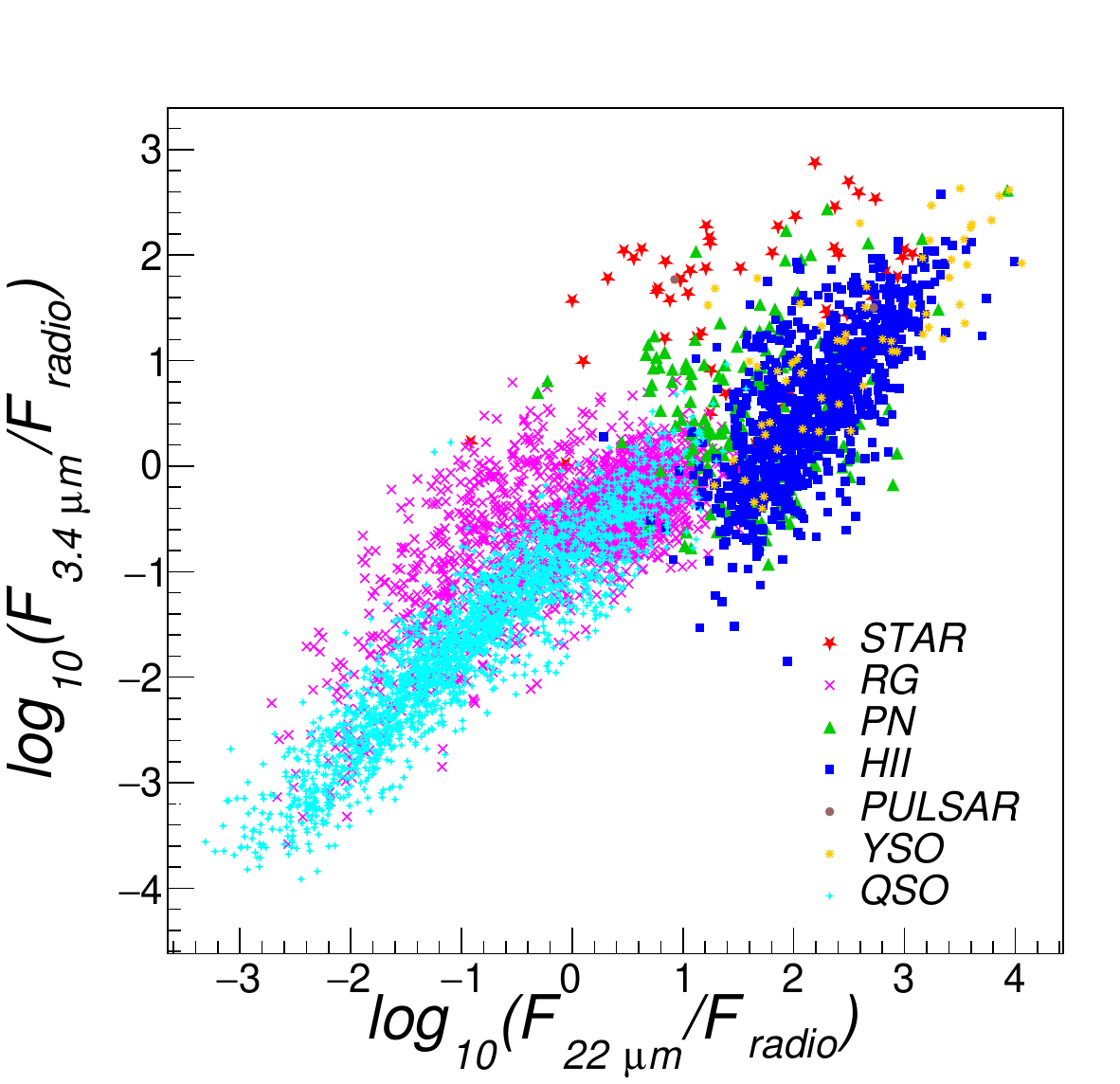}
\vspace{-0.2cm}
\includegraphics[scale=0.37]{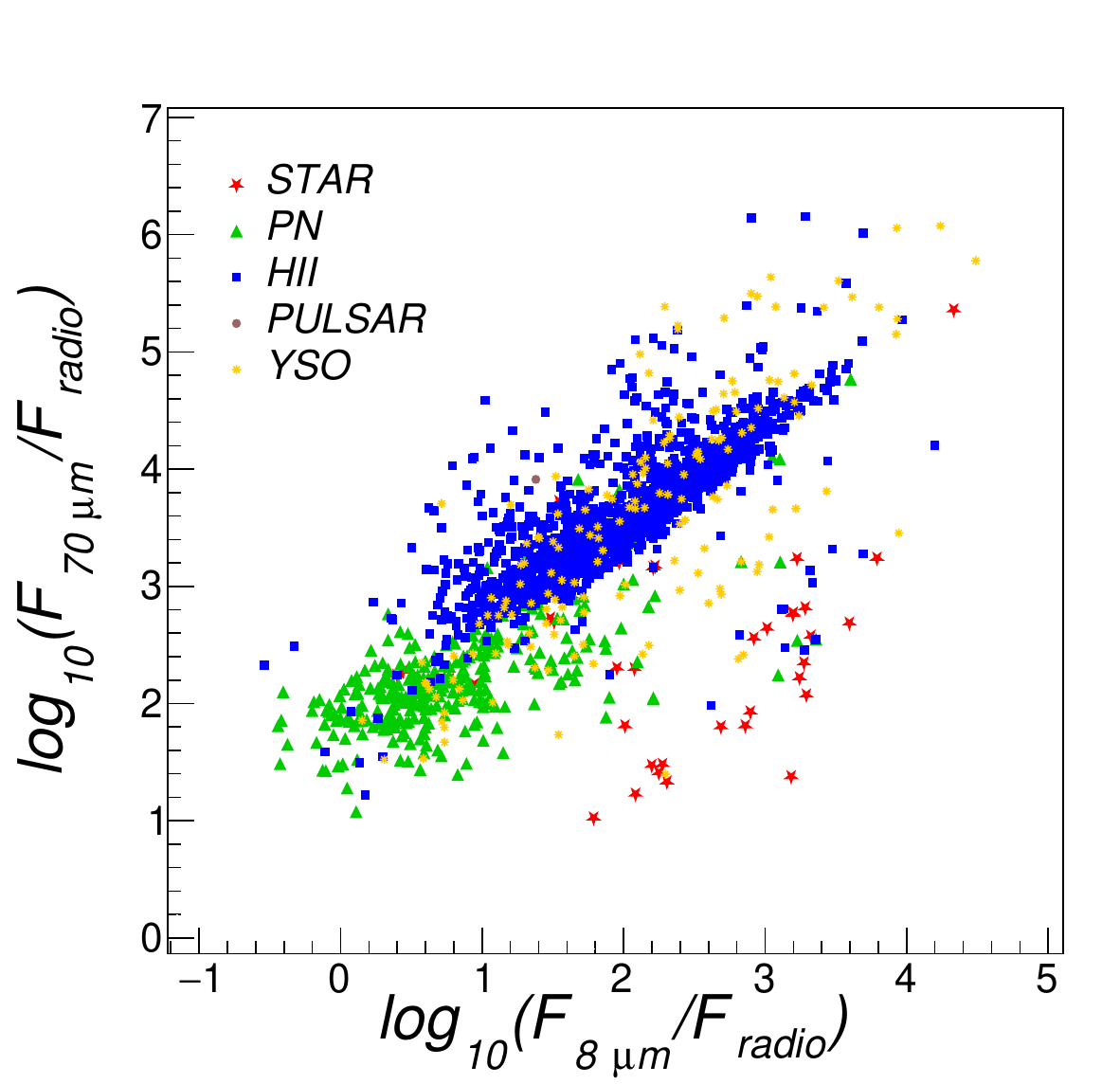}
\caption{Scatter plots of representative infrared/radio colour indices computed over the entire dataset for images with detected sources in both the radio and infrared channels (IoUs>0). Radio flux densities are obtained at different frequencies ranging from 0.912 GHz (ASKAP Early Science survey data) to 5.8 GHz (GLOSTAR). See Section~\ref{sec:observations} for details on survey frequencies.}
\label{fig:color-indices}
\end{figure}

\subsection{Infrared-radio color parameters}
\label{subsec:color-indices}
Colour indices $c_{i,j}$ are defined as the magnitude difference between measured fluxes $F_{i}$ and $F_{j}$ in band $i$ and $j$ where $\lambda_j$>$\lambda_i$ \citep{Nikutta2014}, e.g. $c_{i,j}=\log_{10}(F_j/F_i)$. We considered these radio-infrared colour indices ($c_{radio,3.4\mic}$, $c_{radio,4.6\mic}$, $c_{radio,8\mic}$, $c_{radio,12\mic}$, $c_{radio,22\mic}$, $c_{radio,70\mic}$), in which source fluxes $F$ were computed for each band as follows:
\begin{itemize}
\item Compute background level $B$ and noise rms $\sigma_{rms}$ from median and standard deviation of 3$\sigma$-clipped pixel flux distribution;
\item Find local maxima (or peaks) in image;
\item Extract sources with a flood-fill algorithm, assuming a 5$\sigma$ and 2.5$\sigma$ seed and merge detection thresholds, respectively, with respect to previously computed background. Further, require at least one peak detected inside extracted source aperture;
\item Compute flux information by standard aperture photometry, i.e. $F=\sum_{i}^{N}F_{i}-N\times B$, where $F_{i}$ and $N$ are the flux of $i$-th pixel and number of pixels in source aperture, respectively. 
\end{itemize}
Besides colour indices, we also computed these additional parameters for radio-infrared band combinations (radio,$j$ with $j$=[3.4$\mic$, 4.6$\mic$, 8$\mic$, 12$\mic$, 22$\mic$, 70$\mic$]) to quantify the likelihood of source cross-match association:
\begin{itemize}
\item $\text{\textit{IoU}}_{\text{radio},j}$: Intersection-Over-Union (IoU) between source islands detected in radio and infrared band $j$. IoU is computed as:
\begin{equation*}
\text{IoU}=\frac{n_{overlap}}{n_{union}}
\end{equation*}
where $n_{overlap}$ is the number of pixels that overlap in radio and infrared islands, while $n_{union}$ is the number of pixels of island union. IoU is set to 0 if no source is detected in band $j$;
\item $\text{\textit{SSIM}}_{\text{radio},j}$: Average Structural Similarity Index (SSIM, \citealt{Wang2004}) computed between source image in radio and infrared band $j$. SSIM metric is computed on various image windows and measures the perceptual difference between two images. For two windows $x$ and $y$ of size $K\times K$, SSIM is computed as\footnote{The SSIM implementation of the scikit-image library \citep{skimage} was used.}:
\begin{equation}
\text{SSIM}_{x,y}=\frac{(2\mu_x\mu_y+c_1)(2\sigma_{xy}+c_2)}{(\mu^2_x+\mu^2_y+c_1)(\sigma^2_x+\sigma^2_y+c_2)}
\end{equation}
where $\mu_x$/$\sigma_x$, $\mu_y$/$\sigma_y$ are the pixel sample mean/variance of $x$ and $y$, respectively, and $\sigma_{xy}$ is their covariance. $c_1$ and $c_2$ are constant values used to stabilize the ratio. SSIM index close to 1 indicates high similarity, while negative or close to zero indices denote a high discrepancy.
\end{itemize}
Overall, 12 (18) parameters are selected for classification analysis with the 5-band (7-band) dataset (see feature summary Table~\ref{tab:color-features}). In Fig.~\ref{fig:corr-coeff-5bands} and \ref{fig:corr-coeff-7bands} we explored the degree of correlation among the extracted features, reporting the Pearson correlation coefficient $r$ for each class in both the 5-band and 7-band datasets, respectively. In general, we observe a moderate correlation trend ($r$=0.5-0.7) for many variables in all classes. The strongest correlation ($r$>0.8) is found between radio-3.4$\mic$ and radio-4.6$\mic$ colors, but also among SSIM and IoU parameters computed for these infrared bands. For galactic classes, the correlation becomes more important also among 12$\mic$ and 22$\mic$ parameters. Given the computed 2-tailed p-values, we conclude that these correlations are significant at the 1\% confidence level.\\In Table~\ref{tab:dataset-match-stats} we report the fraction of sources detected in each infrared band (according to the above criteria) having a minimum overlap (IoU>0) with the radio source. These counts include possible spurious detections. On the other hand, missed counts may include IR sources failing to pass the applied detection criteria. In Fig.~\ref{fig:color-indices} we report scatter plots of ($c_{\text{radio},3.4\mic}$, $c_{\text{radio},22\mic}$), ($c_{\text{radio},8\mic}$, $c_{\text{radio},70\mic}$) colour indices obtained for sources simultaneously detected (IoU>0) in both bands over the entire dataset. As can be seen, extragalactic objects tend to cluster on the bottom left region of near- and mid-infrared colour space. Unfortunately, no data for extragalactic sources are available at 8$\mic$ and 70$\mic$ in our dataset, where a larger separation is found among classes of Galactic sources, compared to other colour parameters.

\begin{figure}[htb]
\centering%
\includegraphics[scale=0.37]{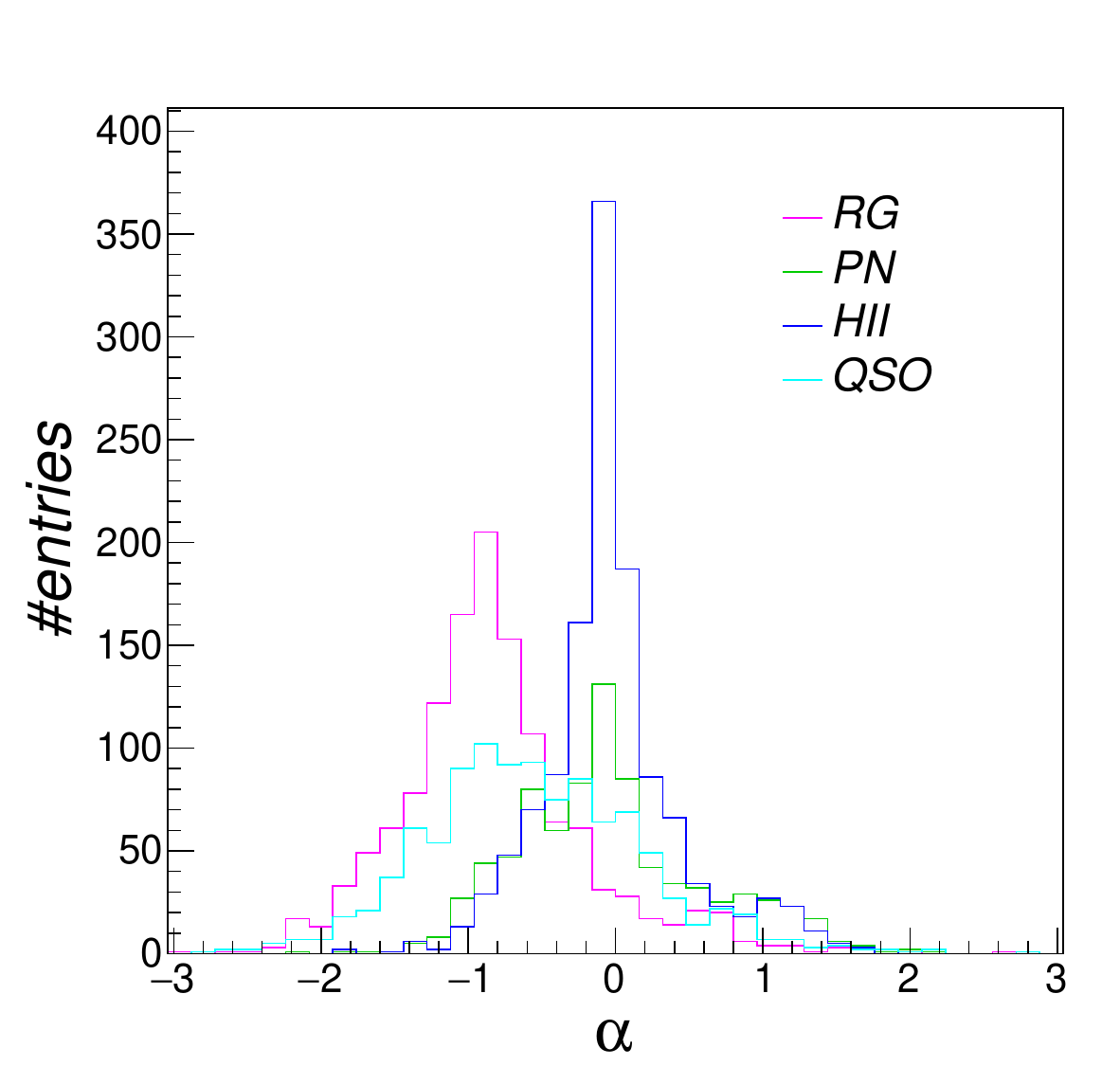}\\%
\vspace{-0.2cm}%
\includegraphics[scale=0.37]{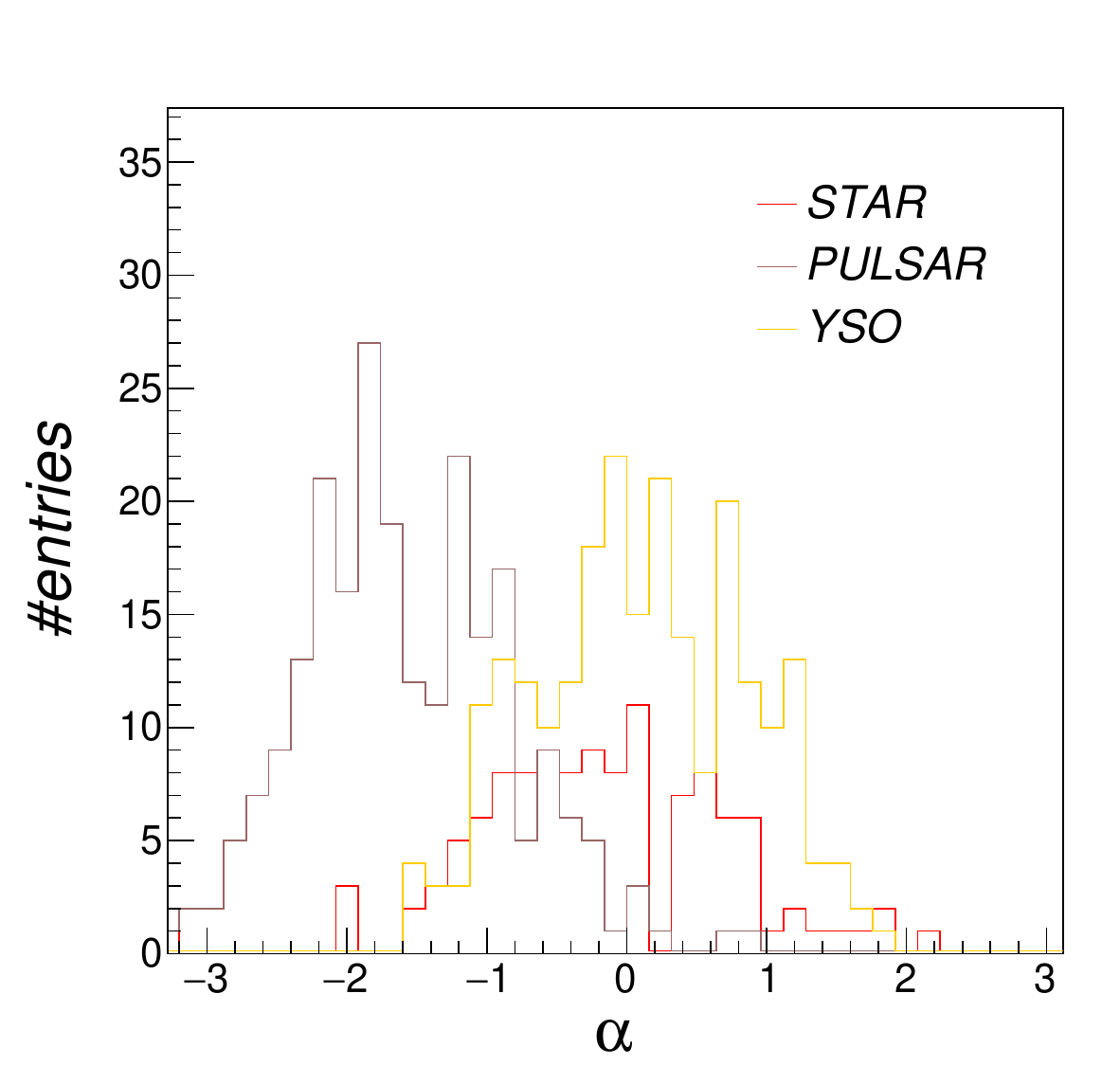}
\caption{Radio spectral indices measured for different source classes with the T-T plot method. Spectral indices for RG and QSO sources were computed using RACS-FIRST radio frequencies (887.5$-$1400 MHz). Indices for the remaining Galactic classes were computed from survey selected sub-bands (when available), i.e. 871$-$1480 MHz (ASKAP Scorpio), 1060$-$1440 MHz (THOR), 4240$-$4670 MHz (GLOSTAR).}
\label{fig:alpha}
\end{figure}

\subsection{Radio spectral indices}
\label{subsec:spectral-index}
We computed the radio spectral index $\alpha$ ($F\propto\nu^{\alpha}$) of sources in our dataset using the T-T plot method \citep{Turtle1962}, e.g. taking the slope of a linear regression of pixel flux densities for source images at two different radio frequencies. This method enables a measurement of the spectral index that is less dependent on the zero level of each image, under the hypothesis of background isotropy and constant $\alpha$. These conditions are holding since we are considering compact sources and regions of size comparable with the synthesised beam of the instrument.\\A subset of our survey data (THOR, ASKAP pilot, GLOSTAR) provide sub-band data that can be used for T-T spectral fit. For VLA FIRST data, instead, we resorted to use data from the ASKAP RACS survey to obtain an estimate of the radio spectral index. 
It is worth to note that such two-point spectral index estimate is not accurate for sources having a curved spectrum, not well described by a power-law model. Indeed, some classes of sources, such as PN \citep{Hajduk2018} or UC \hii{} regions \citep{Yang2021}, could present a turnover frequency in the frequency range (0.8-5 GHz). The frequency coverage of our in-band survey data is, however, rather limited (e.g. 0.87-1.6 GHz for ASKAP) to expect a reliable measurement of any spectral turnovers. Nevertheless, we inspected the ASKAP dataset to search for possible departures from the power-law assumption, by fitting ASKAP source SEDs with different curved spectrum models (e.g. free-free, synchrotron with free-free absorption, see \citealt{Tingay2003}). We found only 5 sources (out of 190 sources with flux measurement available in all five ASKAP sub-bands) that can be fitted ($\tilde{\chi}^{2}$<5) with a curved model.
\\In Fig.~\ref{fig:alpha} we report the obtained spectral indices for different source classes in our dataset. To select more reliable measurements, we selected sources for which the spectral regression correlation coefficient was larger than 0.9. The number of sources per class with measured spectral index (and infrared data) have been reported in Tables~\ref{tab:dataset-train} and \ref{tab:dataset-test} (columns 7-9). The obtained values follow expectations (e.g. see \ref{appendix:source-classes}) or previous measurements for some source classes. 
For example, pulsars have the steepest radio spectrum, while \hii{} regions and PNe have predominantly flatter radio spectra ($\alpha\sim$0), with a significantly smaller fraction peaking around $\alpha$=1. The observed spectral indices of radio galaxies and quasars peak around $-$0.9, in general agreement with the $-$0.95 value reported by \cite{Randall2012} (Fig. 8) in the frequency range 0.843-2.3 GHz, but slightly steeper than conventional value $ \langle\alpha\rangle$=$-$0.7 \citep{Condon2002,Best2005} or measured averages reported at different frequency ranges, e.g. $\langle\alpha\rangle$=$-$0.79 (0.147$-$1.4 GHz) \citep{DeGasperin2018} or $\langle\alpha\rangle$=$-$0.71 (1.4$-$3.0 GHz) \citep{Gordon2021}. This comparison is only indicative as the measured average spectral indices are known to steepen (from $-$0.7 to $-$1) with increasing flux densities, and vary with other parameters such as the size of the source or the flux density threshold (e.g. see \citealt{DeGasperin2018} and references therein).\\Considering the large synthesized beams, it is worth to note that for some radio star types (e.g. LBVs) we could be actually measuring a composite spectral index of the point-source (typically $\alpha\sim$0.6) and the surrounding nebula (which could be $\alpha$<0). This may represent a potential source of misclassification of radio stars when incorporating the spectral index information in the classification analysis (Section~\ref{subsec:results-alpha}).
We also note the absence of radio stars with spectral indices in the range [0.2$-$0.3], where we would expect about 4 counts. This is not understood at present and should be investigated in the future with an extended source sample.  

\begin{figure}[htb]
\centering%
\includegraphics[scale=0.45]{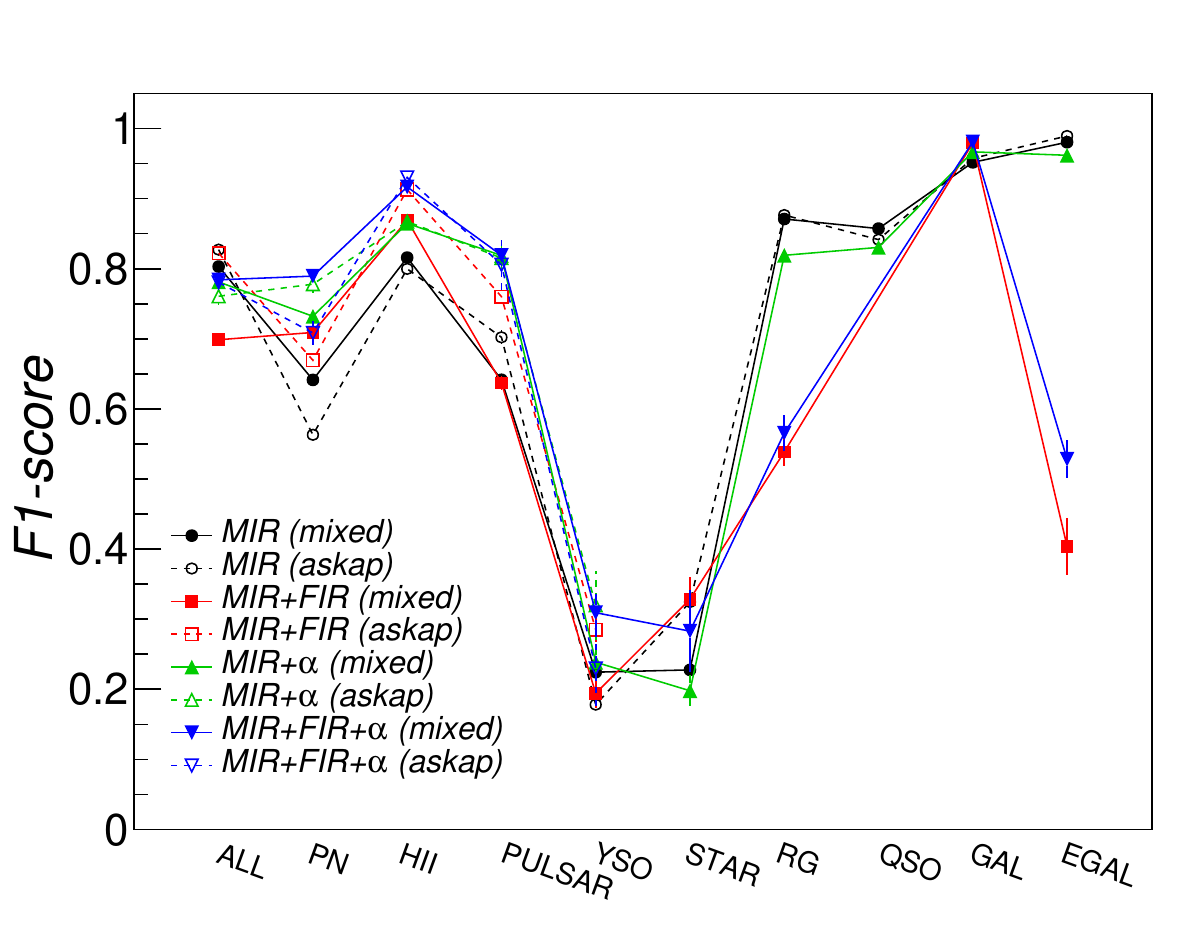}
\caption{Average F1-score metric achieved by the LightGBM trained classifier for binary classification of Galactic and Extragalactic source groups and for multiclass classification, computed over five "mixed" survey test sets (labelled as "mixed" and shown with filled markers) and pure ASKAP test sets (labelled as "askap" and shown with open markers). The error bars are the F1-score standard deviations obtained over the five test sets. Results obtained over the 5-band (radio+MIR) datasets without and with the spectral index ($\alpha$) information are respectively shown with black dots and green triangles, while results obtained over the 7-band (radio+MIR+FIR) datasets are respectively shown with red squares and blue inverted triangles.}
\label{fig:lgbm-f1score-summary}
\end{figure}

\begin{table*}[htb]
\centering%
\footnotesize%
\caption{Average F1-score metrics achieved by the LightGBM trained classifier for binary classification of Galactic and Extragalactic source groups and for multiclass classification, computed over five "mixed" survey test sets (labelled as "mixed") and pure ASKAP test sets (labelled as "askap"). Metrics were not reported if less than 10 sources are available in the test set. Column groups (2-3) and (6-7) report the results obtained over the 5-band (radio+MIR) datasets without and with the spectral index ($\alpha$) information, respectively. Results in column groups (4-5) and (8-9) are relative to the 7-band (radio+MIR+FIR) dataset. Parameters for binary (multiclass) models were set to: \texttt{num\_leaves}=2 (32), \texttt{min\_data\_in\_leaf}=20 (20), \texttt{max\_depth}=1 (5).}
\begin{tabular}{lcccccccc}
\hline%
\hline%
& \multicolumn{8}{c}{F1-score (\%)} \\%
\cmidrule{2-9}%
 & \multicolumn{2}{c}{MIR} & \multicolumn{2}{c}{MIR+FIR} & \multicolumn{2}{c}{MIR+$\alpha$} & \multicolumn{2}{c}{MIR+FIR+$\alpha$}\\%
 \cmidrule{2-9}%
  & mixed & askap & mixed & askap & mixed & askap & mixed & askap\\%
 & (2) & (3) & (4) & (5) & (6) & (7) & (8) & (9)\\%
\hline%
\textsc{gal} & 95.2$\pm$0.1 & 95.8$\pm$0.1 & 97.9$\pm$0.1 & $-$ & 96.7$\pm$0.1 & $-$ & 98.1$\pm$0.2 & $-$\\%
\textsc{egal} & 98.1$\pm$0.1 & 98.9$\pm$0.1 & 40.4$\pm$4.1 & $-$ & 96.2$\pm$0.1 & $-$ & 52.8$\pm$2.8 & $-$\\%
\rowcolor{lightgray}
\textsc{all} & 97.2$\pm$0.1 & 98.3$\pm$0.1 & 95.3$\pm$0.4 & $-$ & 96.4$\pm$0.1 & $-$ & 95.9$\pm$0.4 & $-$\\%
\hline%
\textsc{pn} & 64.1$\pm$0.5 & 56.3$\pm$0.5 & 70.9$\pm$0.6 & 66.9$\pm$0.6 & 73.2$\pm$0.8 & 77.8$\pm$1.3 & 78.9$\pm$0.5 & 70.8$\pm$2.5\\%
\textsc{\hii{}} & 81.6$\pm$0.3 & 80.0$\pm$0.4 & 86.9$\pm$0.2 & 91.3$\pm$0.3 & 86.5$\pm$0.5 & 86.7$\pm$0.7 & 91.7$\pm$0.3 & 93.0$\pm$0.6\\%
\textsc{pulsar} & 64.2$\pm$0.3 & 70.2$\pm$1.0 & 63.7$\pm$1.0 & 76.0$\pm$0.9 & 81.9$\pm$1.4 & 81.5$\pm$2.5 & 81.9$\pm$0.8 & 80.6$\pm$4.2\\%
\textsc{yso} & 22.4$\pm$0.7 & 17.8$\pm$0.6 & 19.4$\pm$2.1 & 28.5$\pm$2.9 & 23.9$\pm$1.7 & 32.0$\pm$4.9 & 30.9$\pm$2.7 & 23.0$\pm$5.3\\%
\textsc{star} & 22.8$\pm$1.4 & 32.4$\pm$2.8 & 32.8$\pm$3.1 & $-$ & 19.8$\pm$2.2 & $-$ & 28.3$\pm$5.8 & $-$\\%
\textsc{rg} & 87.1$\pm$0.3 & 87.7$\pm$0.2 & 53.9$\pm$2.1 & $-$ & 81.9$\pm$0.6 & $-$ & 56.5$\pm$2.5 & $-$ \\%
\textsc{qso} & 85.7$\pm$0.4 & 84.1$\pm$0.2 & $-$ & $-$ & 83.0$\pm$0.5 & $-$ & $-$ & $-$ \\%
\rowcolor{lightgray}
\textsc{all} & 80.3$\pm$0.3 & 82.7$\pm$0.1 & 69.9$\pm$0.5 & 82.2$\pm$0.4 & 78.1$\pm$0.4 & 76.1$\pm$1.3 & 78.4$\pm$0.7 & 78.1$\pm$0.5\\%
\hline%
\hline%
\end{tabular}
\label{tab:f1score-lgbm}
\end{table*}

\section{Source classification analysis}
\label{sec:classification-analysis}
We used the dataset described in Section~\ref{sec:dataset} to perform classification studies with supervised learning algorithms. We carried out two different analysis. The first one, reported in Section~\ref{subsec:manual-feature-analysis}, uses the set of conventional features (color indices, spectral indices) extracted from the dataset, as described in Section~\ref{sec:feat-extraction}, and gradient-boosted decision trees as classifier method. 
A second analysis, reported in Section~\ref{subsec:auto-feature-analysis}, employs convolutional neural networks for automated feature extraction and source image classification.\\The entire dataset, including data from all radio surveys, was split into three "mixed" survey subsets (train, validation, test sets), containing 55\%/15\%/30\% of the original data, respectively. Five train/validation/test splits were randomly generated to estimate the model performance uncertainties. We also produced additional data splits with exclusively ASKAP data in the test set, and previous radio surveys in train and validation sets (with a 70\%/30\% data proportion). These samples were used to estimate how the classifier performs on a specific survey, when trained on a mixture of completely different surveys.
\\In both analysis, we made use of the following metrics\footnote{More details at \scriptsize{\url{https://scikit-learn.org/stable/modules/model_evaluation.html}}.}, widely adopted in multi-class classification problems, to estimate the achieved classification performances:
\begin{itemize}
\item \emph{Recall} ($\mathcal{R}$): Fraction of sources of a given class that were correctly classified by the model out of all sources labelled in that class, computed as:
\begin{equation*}
\mathcal{R}=\frac{TP}{TP + FN}
\end{equation*}
\item \emph{Precision} ($\mathcal{P}$): Fraction of sources correctly classified as belonging to a specific class, out of all sources the model predicted to belong to that class, computed as:
\begin{equation*}
\mathcal{P}=\frac{TP}{TP+FP}
\end{equation*}
\item \emph{F1-score}: the harmonic mean of precision and recall:
\begin{equation}
\text{F1-score}=2\times\frac{\mathcal{P}\times\mathcal{R}}{\mathcal{P}+\mathcal{R}}
\end{equation}
\end{itemize}
where $TP$, $FP$, $FN$ are the number of true positives, false positives, and false negatives, respectively. These metrics were computed for each source class individually, and cumulatively over all dataset. In the latter case, individual class metrics were first weighted by the number of sources present for each class to account for class unbalance, and then averaged.

\begin{figure}[htb]
\centering%
\includegraphics[scale=0.45]{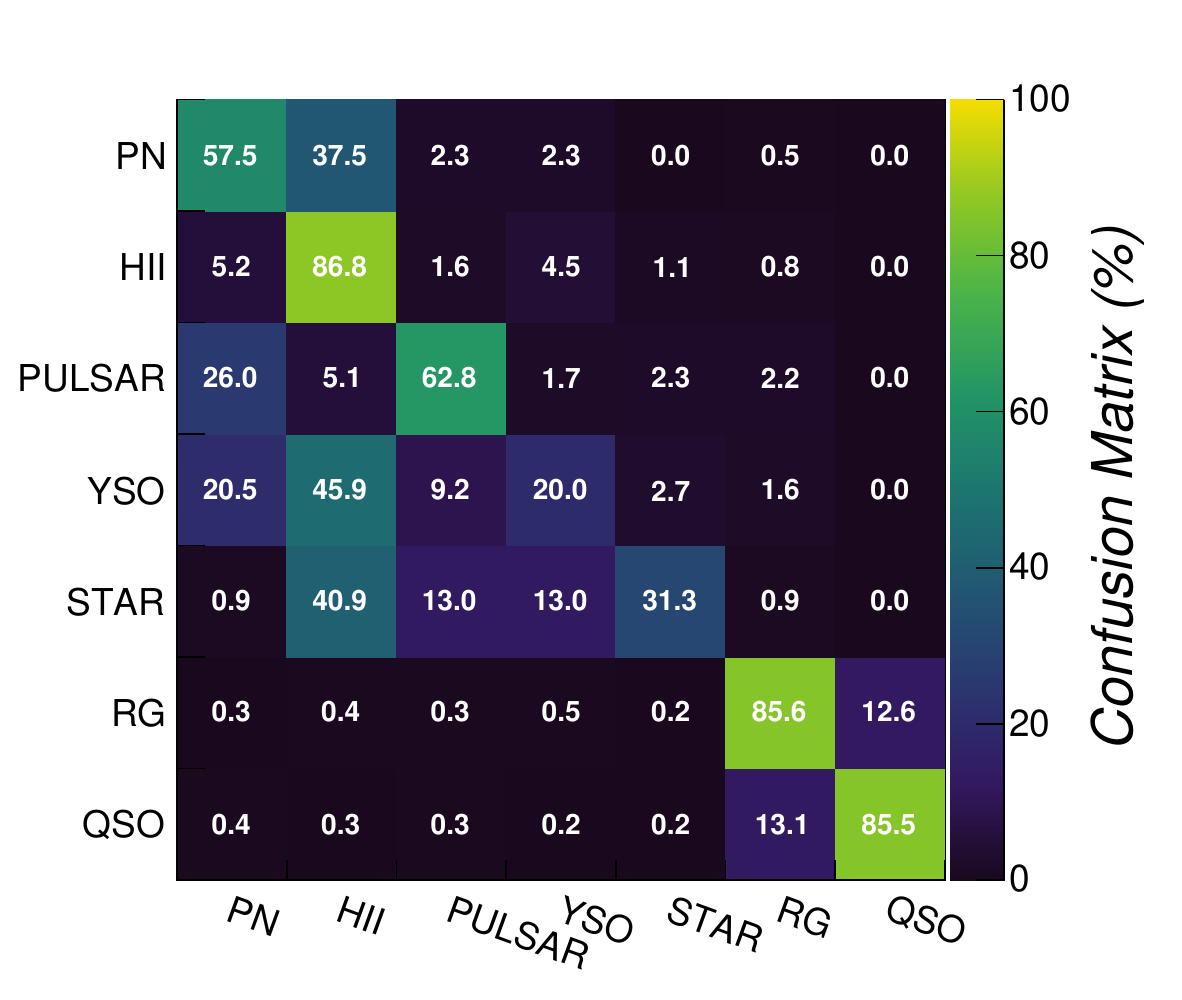}
\caption{Confusion matrix of the trained LightGBM classifier obtained over the 5-band (radio+MIR) pure ASKAP test datasets.}
\label{fig:confusion-matrix-5bands}
\end{figure}

\subsection{LightGBM Classification}
\label{subsec:manual-feature-analysis}

\subsubsection{Model training}
\label{subsec:classification}
We trained a LightGBM\footnote{\scriptsize{\url{https://lightgbm.readthedocs.io/en/latest/index.html}}} \citep{Ke2017} classifier over the produced 5-band and 7-band dataset splits ("mixed" surveys sets and non-ASKAP survey sets), using the set of feature parameters described in Section~\ref{sec:feat-extraction} as inputs. LightGBM is a distributed and high-performance gradient boosting framework based on decision tree algorithm, widely adopted for classification tasks as known to reach comparable (or even better) performances on tabular data with considerably lower training times and memory usage with respect to other popular libraries (e.g. XGBoost). 
The most important algorithm hyperparameters controlling the model accuracy and overfitting are: \texttt{max\_depth}, \texttt{num\_leaves}, \texttt{min\_data\_in\_leaf}, \texttt{num\_iterations}\footnote{
\texttt{max\_depth} is the maximum depth of each decision tree, typically chosen in the range [2,12], as very deep/shallow trees are more likely to overfit/underfit the training data. \texttt{max\_depth} has to be optimized in combination with the \texttt{num\_leaves} parameter, controlling the number of decision leaves in a single tree, with optimal \texttt{num\_leaves} values lying below the limit 2$^{\texttt{max\_depth}}$. \texttt{min\_data\_in\_leaf} specifies the minimum number of sources that fit the decision criteria in a leaf, allowing to control the model overfitting. Suitable values are typically assumed on the basis of the training sample size. Finally, the \texttt{num\_iterations} is the number of boosting iterations performed, often interpreted as the “number of trees” used.}.
\\To select suitable hyperparameter values, we performed several training runs in which we varied \texttt{max\_depth} values in the [2,12] range, and \texttt{num\_leaves}$\le$2$^{\texttt{max\_depth}}$, observing the resulting model F1-score on the test set. For each training run, we used early stopping on validation data to select the optimal \texttt{num\_iterations} parameter (typically found <100 in all performed runs). For a given tree depth choice, we also scanned different values of \texttt{min\_data\_in\_leaf} from 5 to 100.\\
Classification results achieved over the available feature subsets and dataset splits are summarized in Fig.~\ref{fig:lgbm-f1score-summary} and Table~\ref{tab:f1score-lgbm}, and discussed with more details in the following paragraphs.\\
In Figures~\ref{fig:lgbm-feat-importance-5bands}, \ref{fig:lgbm-feat-importance-5bands-alpha}, \ref{fig:lgbm-feat-importance-7bands} and \ref{fig:lgbm-feat-importance-7bands-alpha}, we inspected the relative importance of each feature provided to trained LightGBM classifiers, finding that radio-infrared colour indices are always ranked among the top most sensitive features, along with the radio spectral index, while morphological parameters (radio-infrared IoUs) are ranked last.

\begin{figure}[htb]
\centering%
\includegraphics[scale=0.37]{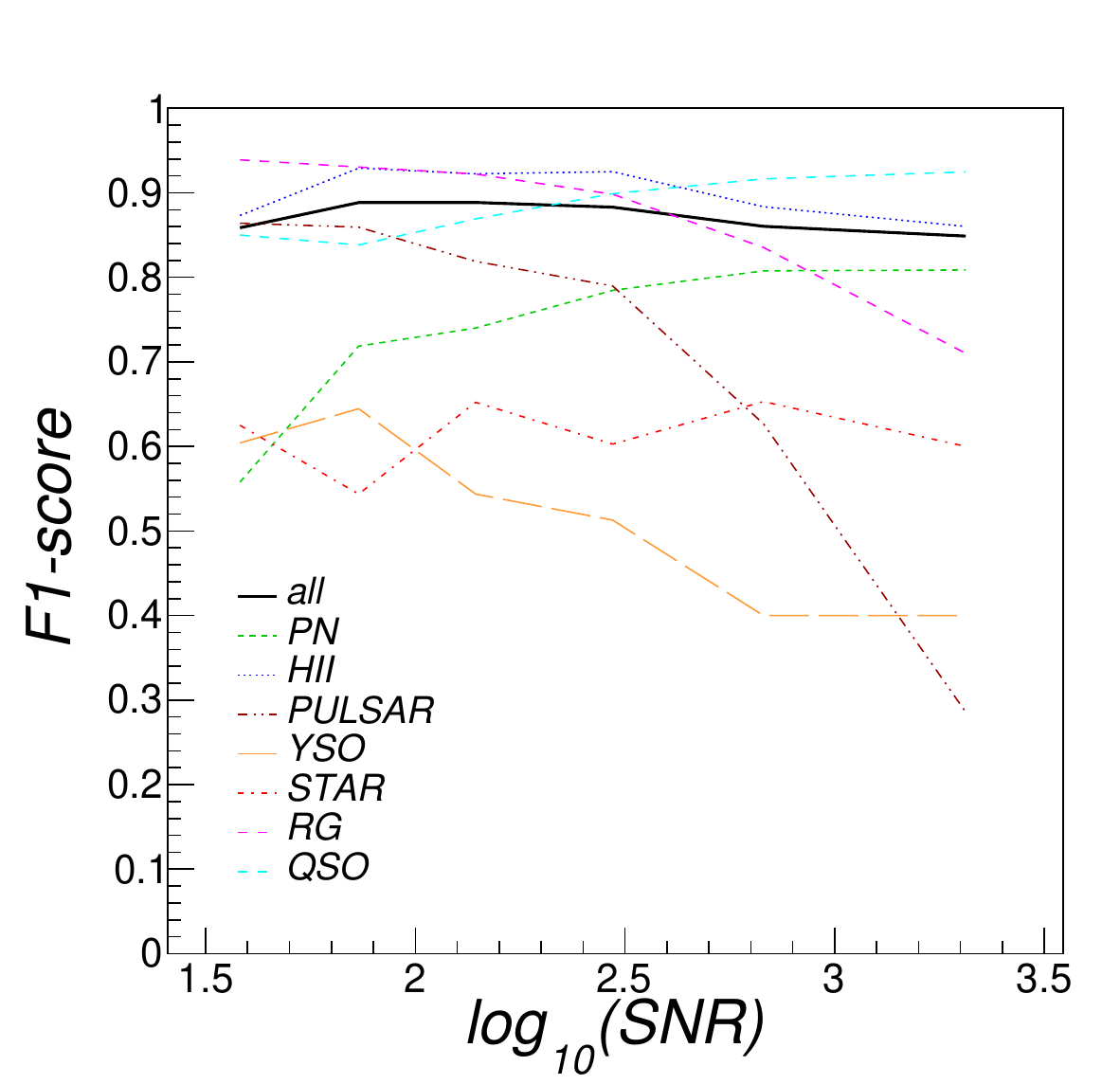}\\
\vspace{-0.1cm}
\includegraphics[scale=0.37]{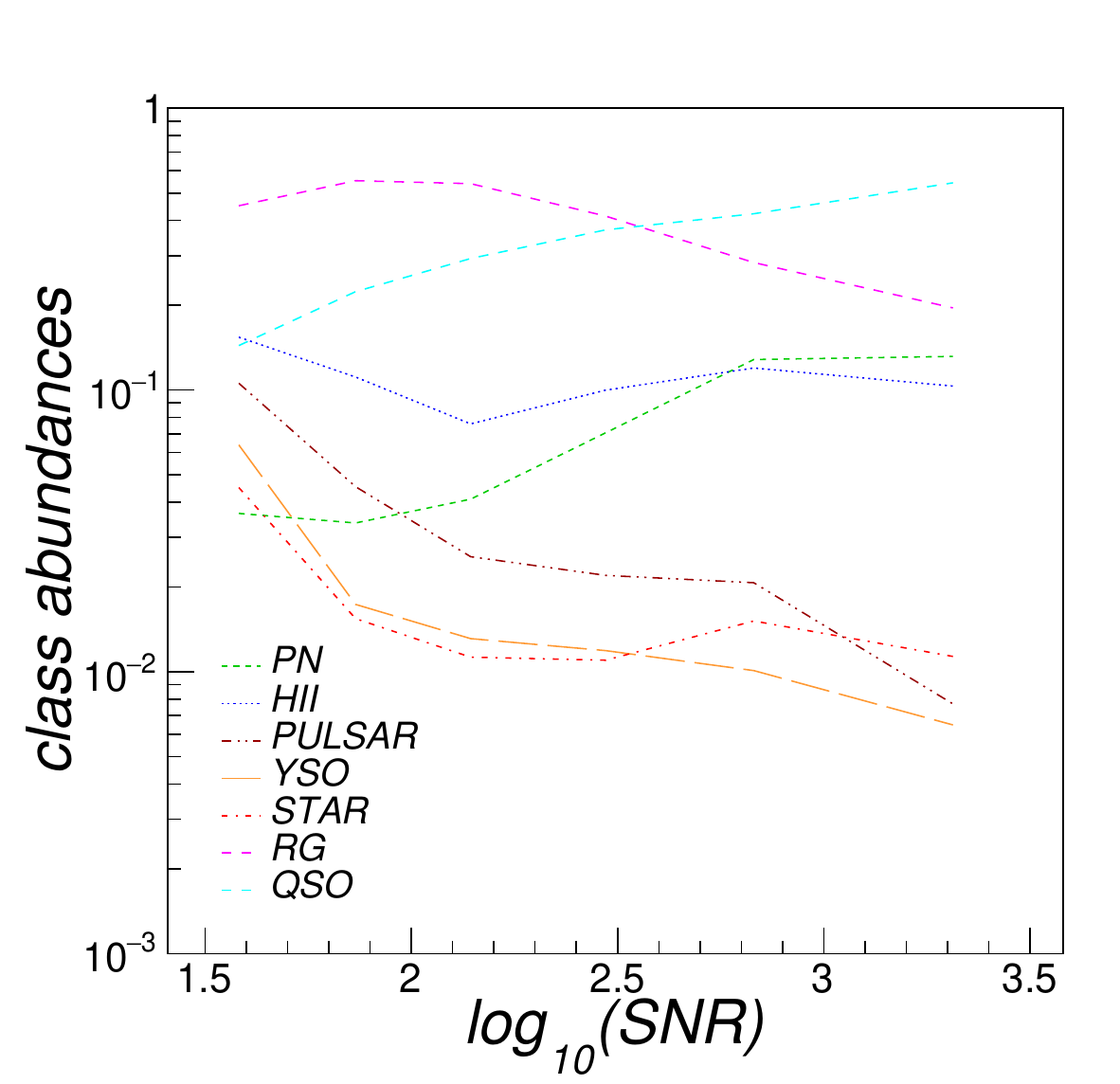}
\caption{Top: F1-score of the LightGBM classifier as a function of radio source signal-to-noise (SNR) obtained over the 5-band (radio+MIR) dataset. Bottom: Fraction of source images available in the 5-band dataset as a function of SNR.}
\label{fig:f1score-vs-snr}
\end{figure}

\subsubsection{Results on radio+MIR data}
\label{subsec:lgbm-5bands-results}
In Table~\ref{tab:f1score-lgbm} (rows 1-3, columns 2, 3), we report the F1-score metric of the trained LightGBM model, obtained on the 5-band "mixed" and "askap" test datasets, for classifying sources into two groups: Galactic (i.e. including target object classes of interest for Galactic science studies, such as PNe, \hii{} regions, pulsars, YSO, and stars), and Extragalactic (i.e. including radio galaxies and quasars). The model is able to identify sources belonging to the two groups with very high accuracy (above 90\%), with a relatively shallow tree configuration (\texttt{max\_depth}=1 or 2), even when presented with data observed with a completely different survey (ASKAP) with respect to those used in the training sample. 
As the Galactic-Extragalactic discrimination analysis can only be done using this dataset, due to the existing survey coverage and catalogue availability, this is a remarkable and encouraging result (e.g. there is no strong need for additional multi-wavelength data).\\Discrimination of individual source classes was also studied. A deeper model (\texttt{max\_depth}=5) was found to provide the best performances in the parameter scan. 
Classification metrics obtained over both "mixed" and "askap" test set are reported in Table~\ref{tab:f1score-lgbm} (rows 4-11, columns 2, 3), while the source confusion matrix obtained over the "mixed" survey test sets is plotted in Fig.~\ref{fig:confusion-matrix-5bands}. In this case, extragalactic sources (radio galaxies, QSO) can be identified with $\sim$85\% accuracy, with a rate of misclassified sources of the order of 15\%, almost entirely in the direction of the other extragalactic source category (e.g. QSO$\rightarrow$galaxy, and viceversa). PNe, \hii{} regions and pulsars are the best classified sources within the Galactic group. Lowest misclassification rates towards other classes are obtained for \hii{} regions, found below 15\%. PNe are more likely (38\%) to be misclassified as \hii{} regions. As reported in previous studies \citep{Anderson2012}, we expect that a better discrimination power between the two types can be achieved by employing far-infrared and 8$\mic$ data (see next paragraph). Poor classification results are obtained on the radio stars and YSO samples, with F1-scores ranging from 20\% to 30\%. YSOs are largely ($\sim$66\%) misclassified as \hii{} regions or PNe. This is somewhat expected, as a fraction of SIMBAD objects classified as YSO (used as a reference for building the training sample) were also found listed in the WISE \hii{} region and HASH PN catalogues. Future data releases shall therefore focus on assessing the reliability of our YSO candidates, removing the identification ambiguities before repeating the classification analysis.
The same labelling issue is also potentially affecting the radio star classification. 
Poor results on some Galactic class may be therefore not only due to the limited training sample, but also ascribed to the reliability of original source classification present in the literature.\\In Fig.~\ref{fig:f1score-vs-snr} (top panel) we reported the F1 classification score for all classes in the 5-band ASKAP test dataset as a function of the computed radio source signal-to-noise ratio (SNR). The overall classification performance is mostly flat over the SNR range, while individual classes do show some dependency on the SNR, e.g. F1-score is increasing with SNR for PNe/QSOs and decreasing for pulsars/radio galaxies. As shown in Fig.~\ref{fig:f1score-vs-snr} (bottom panel), the observed trends for each class seem to correlate with the number of corresponding images available in each SNR bin.

\begin{figure}[htb]
\centering%
\includegraphics[scale=0.45]{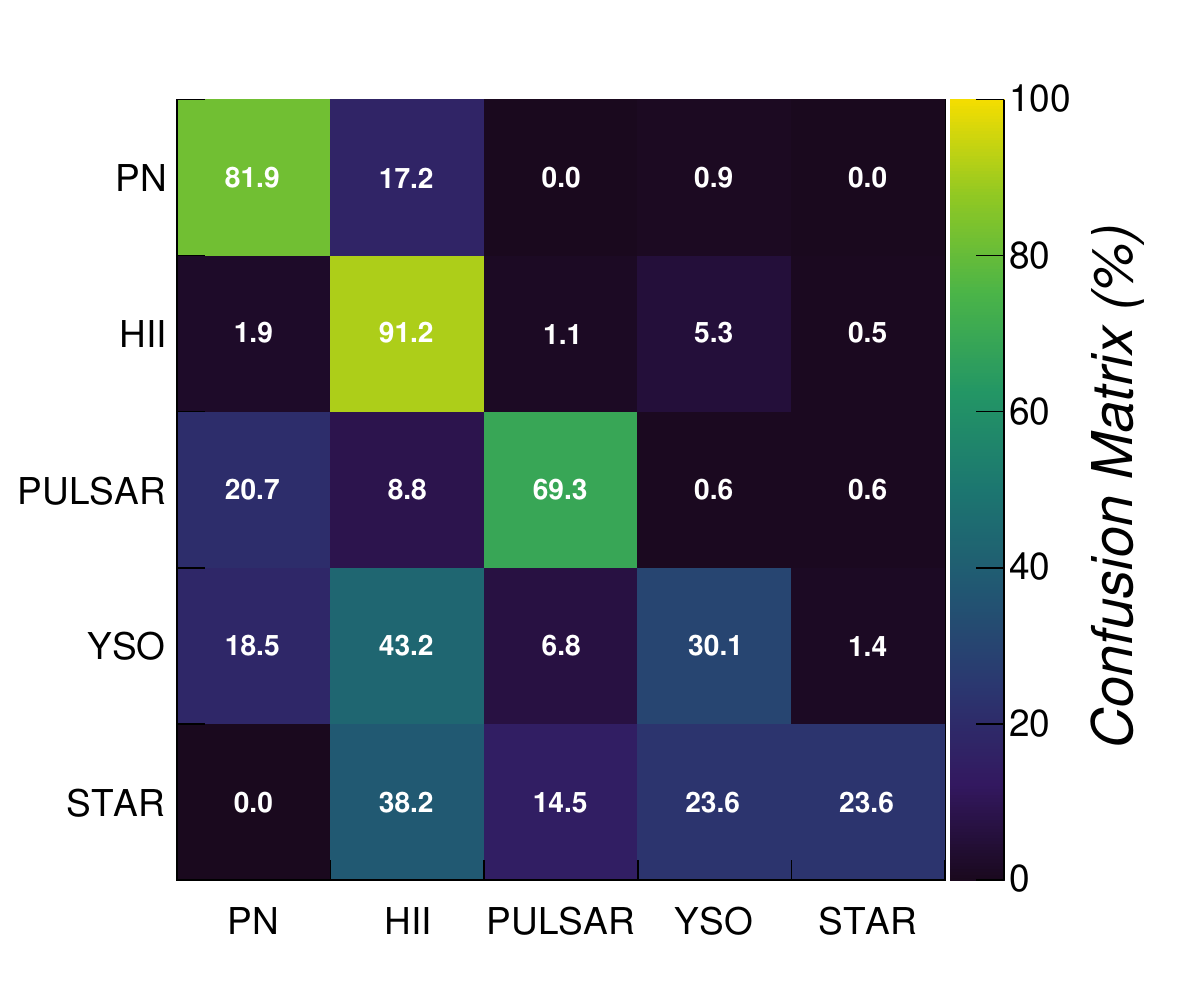}
\caption{Confusion matrix of the trained LightGBM classifier obtained over 7-band (radio+MIR+FIR) pure ASKAP test datasets.}
\label{fig:confusion-matrix-7bands}
\end{figure}

\subsubsection{Results on radio+MIR+FIR data}
In Table~\ref{tab:f1score-lgbm} (columns 4, 5) we report the F1-score metric of the trained LightGBM model, obtained on the 7-band "mixed" and "askap" test datasets. Only 5 Galactic classes are available in the latter case, but we did not report classification metrics for the "STAR" class, as less than 10 sources are available in the test set. Inclusion of 8$\mic$ and 70$\mic$ data lead to a slight improvement (5-10\%) in classification for most classes, except for pulsars that are infrared-quiet at these bands. Misclassification rates, shown in Fig.~\ref{fig:confusion-matrix-7bands}, also improved considerably for \hii{} regions and PNe, e.g. the fraction of PNe misclassified as \hii{} regions decreased by $\sim$20\% compared to the 5-band analysis, highlighting how the far-infrared information is crucial for separation of certain Galactic classes. Although a slight improvement is also seen on radio star and YSO identification, the limitations highlighted in the previous paragraph prevent to eventually obtain an effective classification of both types.

\subsubsection{Results with radio spectral index information}
\label{subsec:results-alpha}
In Table~\ref{tab:f1score-lgbm} (columns 6-9) we reported the classification results obtained on the 5-band and 7-band "mixed survey and pure ASKAP test datasets, after including the radio spectral index $\alpha$ as an additional input feature. A clear increase in performance was obtained for PNe, \hii{} regions, and pulsars, while no sensible improvements were observed on the remaining classes. Unfortunately, the training and test samples are very limited in size for some classes, e.g. less than 70 (40) radio stars in the 5-band (7-band) datasets, and therefore their corresponding metrics may not be precisely estimated. 

\subsection{CNN Classification}
\label{subsec:auto-feature-analysis}
In this section we explored the capabilities of supervised classification models, such as CNNs, that automatically extract features directly from images, e.g. they do not require the extra image processing applied in Section~\ref{subsec:color-indices}. More importantly, contrarily to the previous analysis, a CNN classifier is less tied to the source compact morphology assumption, and would be thus also potentially suited for extended source classification.

\subsubsection{Model training}
We considered two alternative CNN architectures: a custom shallow network with only two convolutional layer blocks, and a standard deep ResNet18 architecture. 
Network configurations are reported in Table~\ref{tab:cnn-models}. 
We trained six custom model configurations (denoted as \texttt{custom\_v1}, \texttt{custom\_v2}, \dots, \texttt{custom\_v6}) on our data, varying the convolutional or dense layer structure (e.g. number of filters, kernel or stride size, etc). Columns (2) and (3) report the network backbone and classification head structure, following this notation:
\begin{itemize}
\item \texttt{16C3BnP2-32C3BnP2-32-16}: indicate a network with two convolutional layer blocks and two dense layers with 32 and 16 neurons, respectively. Convolution blocks (C) have 16 and 32 3$\times$3 filters respectively, each followed by batch normalization (Bn) \footnote{Batch normalization layers normalize their inputs by subtracting the batch mean and dividing by the batch standard deviation. They are often inserted in CNN architectures to reduce the internal covariate shift and improve network stability during training.} and max pooling (P) layers\footnote{Pooling layers reduce the spatial dimension of the inputs, by applying a pooling operation (e.g. maximum or average) to a set of values in a small region of the input volume. They are commonly used to increase the receptive field of the network, reduce its computational cost, and improve its performance.} using 2$\times$2 filter and stride 2;
\item \texttt{16C3-32C5S2-16}: indicate a network with two convolutional layer blocks and a single dense layer with 16 neurons. The first convolution block has 16 3$\times$3 filters (no max pooling layer), while the second one has 32 5$\times$5 filters using stride 2.
\end{itemize}
All configurations were trained (\emph{Adam} optimizer, learning rate $\eta$=5$\times$10$^{-4}$, batch\_size=64) over five multiple train/validation/test dataset splits until overfitting is detected on the validation set (typically after 300 epochs). Classification metrics are finally computed over the test sets. In Fig.~\ref{fig:f1scores-vs-epoch} we report the classification F1-score obtained as a function of the training epoch with a representative model (\texttt{custom\_v1}) over train (blue graph) and validation (red (graph) 5-band datasets. Shaded areas correspond to the minimum and maximum F1-scores found in different training runs.\\
To avoid learning features from other nearby sources, we masked pixels not belonging to the source in all input images. Masks for each source were obtained from the radio channel in an automated way using \caesar{} source finder \citep{Riggi2016,Riggi2019}, refined manually (if not accurate enough), and eventually enlarged using a morphological dilation transform\footnote{As we expect the environment surrounding the source can provide valuable information for classification purposes, the source mask was enlarged using a morphological dilation transform with configurable kernel size (21 pixels by default).}. The resulting masks were finally applied to radio and infrared channels to produce masked image data\footnote{Masked input data are included in the dataset under version control along with unmasked images.} that are provided as CNN inputs.\\Different image pre-processing stages were applied to the masked input data during the training and inference stages:
\begin{enumerate}
\item \emph{Channel max scaling}: For each source, we scaled each channel by the maximum pixel value among all channels for that source. This step is introduced to preserve the original radio/infrared flux ratios (very sensitive to the source type) and remove the flux density degeneracy, e.g. two identical sources (e.g. same class and radio/infrared ratios) with just an absolute flux density offset will be treated as the same input by the classifier.
\item \emph{Augmentation}: we randomly applied a series of transformations to input cutout channels, including horizontal and vertical flipping, and [-90$^{\circ}$,90$^{\circ}$] rotation. This step is only applied during training to improve the model generalization capabilities; 
\item \emph{Resizing}: Finally, we resized all image cutouts in the dataset to the same size in pixels (64$\times$64 pixels by default), as the first convolutional layer of the network requires tensor of the same shape in input;
\end{enumerate}
Results are reported in the following paragraphs only for the 5-band dataset, as the 7-band and spectral index datasets are too limited in size for training a deep network.

\begin{table}[htb]
\centering%
\footnotesize%
\caption{Average F1-score metrics achieved by trained CNN models for multiclass classification, computed over five "mixed" survey 5-band (radio+MIR) test sets.}
\begin{tabular}{lllc}
\hline%
\hline%
Model & Backbone & Head & F1-score (\%)\\%
\hline%
\texttt{custom\_v1} & 16C3P2-32C3P2 & 16 & 80.7$\pm$0.3\\%
\texttt{custom\_v2} & 16C3-32C5S2 & 16 & 80.5$\pm$0.8\\%
\texttt{custom\_v3} & 16C5P2-32C5P2 & 16 & \textbf{81.3$\pm$0.2}\\%
\texttt{custom\_v4} & 32C3P2-64C3P2 & 32 & 80.2$\pm$0.5\\%
\texttt{custom\_v5} & 16C3BnP2-32C3BnP2 & 16 & 78.8$\pm$0.1\\%
\texttt{custom\_v6} & 16C3P2-32C3P2 & 32-16 & \textbf{81.3$\pm$0.4}\\%
\hline%
\texttt{resnet18} & ResNet18 & 16 & 80.1$\pm$1.1\\%
\hline%
\hline%
\end{tabular}
\label{tab:cnn-models}
\end{table}

\begin{table}[htb]
\centering%
\footnotesize%
\caption{Average F1-score metrics achieved by trained shallow and deep CNN models for source multiclass classification, computed over five "mixed" survey 5-band (radio+MIR) test sets (labelled as "mixed") and pure ASKAP 5-band (radio+MIR) test sets (labelled as "askap").}
\begin{tabular}{lcc|cc}
\hline%
\hline%
& \multicolumn{4}{c}{F1-score (\%)} \\%
\cmidrule{2-5}%
 & \multicolumn{2}{c|}{\texttt{custom\_v1}} & \multicolumn{2}{c}{\texttt{resnet18}}\\%
\cmidrule{2-5}%
& mixed & askap & mixed & askap\\%
 & (2) & (3) & (4) & (5) \\%
\hline%
\textsc{pn} & 68.3$\pm$2.3& 67.2$\pm$2.0 & 74.6$\pm$0.8 & 66.9$\pm$0.1 \\%
\textsc{\hii{}} & 86.2$\pm$0.9 & 90.5$\pm$0.8 & 88.8$\pm$0.6 & 92.6$\pm$1.1 \\%
\textsc{pulsar} & 64.0$\pm$1.4 & 63.1$\pm$1.2 & 71.6$\pm$0.6 & 52.4$\pm$2.4 \\%
\textsc{yso} & 41.0$\pm$2.2 & 40.6$\pm$2.4 & 52.1$\pm$1.3 & 49.0$\pm$2.1 \\%
\textsc{star} & 38.8$\pm$2.7 & 52.4$\pm$1.4 & 41.0$\pm$1.3 & 37.5$\pm$3.2 \\%
\textsc{rg} & 85.6$\pm$0.3 & 84.1$\pm$1.2 & 84.2$\pm$0.4 & 85.3$\pm$1.2 \\%
\textsc{qso} & 83.3$\pm$0.8 & 77.8$\pm$1.8 & 78.6$\pm$2.7 & 81.9$\pm$0.9 \\%
\rowcolor{lightgray}
\textsc{all} & 80.7$\pm$0.3 & 80.4$\pm$1.0 & 80.1$\pm$1.1 & 82.0$\pm$0.8 \\%
\hline%
\hline%
\end{tabular}
\label{tab:f1score-cnn}
\end{table}

\subsubsection{Results on radio+MIR data}
\begin{figure}[htb]
\centering%
\includegraphics[scale=0.45]{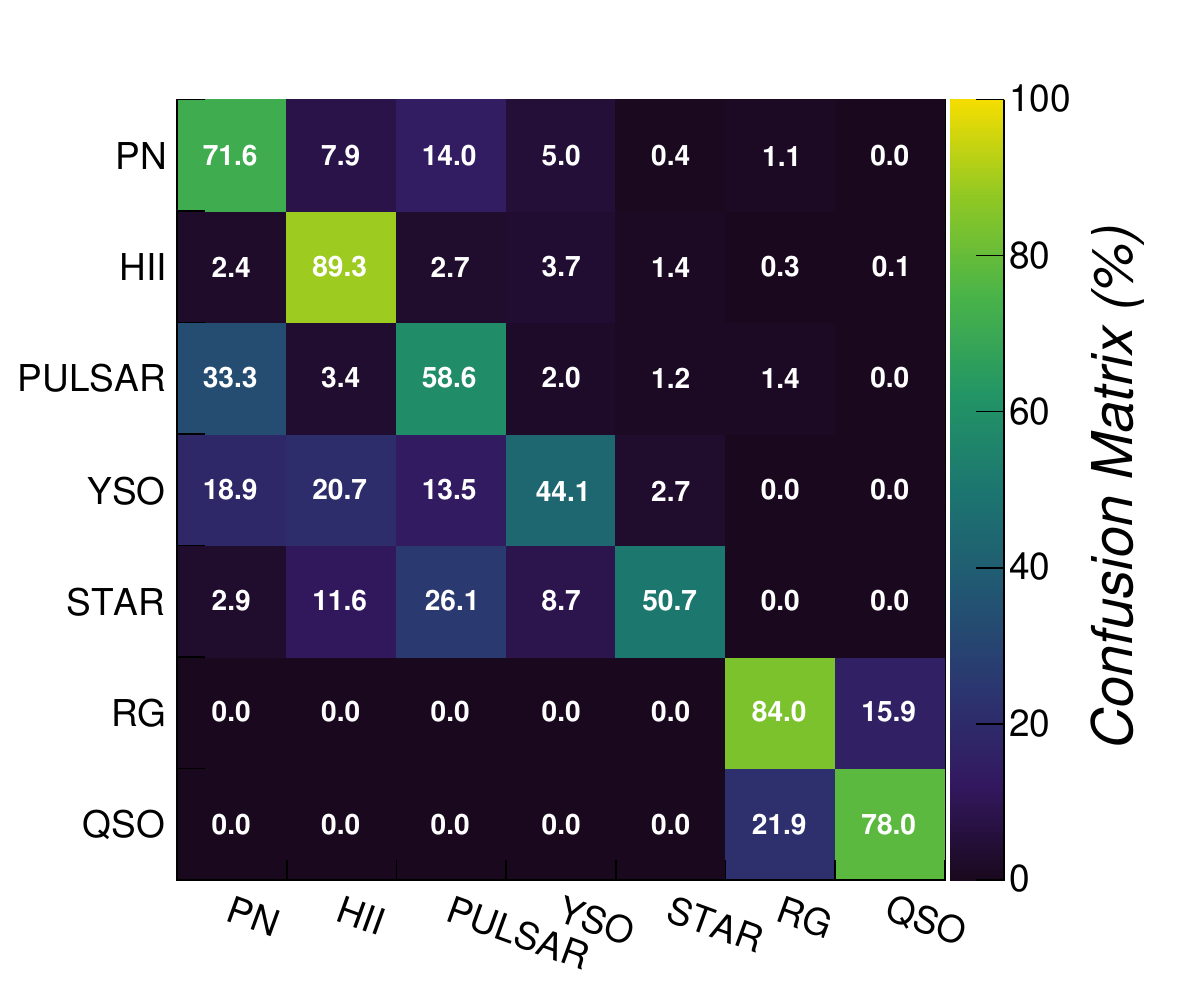}
\caption{Confusion matrix of the trained CNN \texttt{custom\_v1} classifier obtained over 5-band (radio+MIR) pure ASKAP test datasets.}
\label{fig:confusion-matrix-5bands-cnn}
\end{figure}

Classifications scores obtained by trained CNN classifiers on "mixed" survey test datasets, reported in Table~\ref{tab:cnn-models} (column 4), are rather comparable (within 1\%) across shallow and deep model configurations and training runs. A larger kernel size (5$\times$5 pixels, \texttt{custom\_v3} model) slightly improved the results, while batch normalization layers (\texttt{custom\_v5} model) produce a $\sim$2\% decrease in performance. In Table~\ref{tab:f1score-cnn} we report the classification scores obtained with the \texttt{resnet18} model and a representative shallow model (\texttt{custom\_v1}) trained on "mixed" survey data over both "mixed" survey and pure ASKAP test sets. Misclassification rates obtained on pure ASKAP test sets are reported in Fig.~\ref{fig:confusion-matrix-5bands-cnn} for the \texttt{custom\_v1} model. Overall, we conclude that the achieved metrics are comparable to those found with the LightGBM classifier (Table~\ref{tab:f1score-lgbm}, columns 2, 3). We also observe that, with regard to the individual classes, the CNN classifiers tend to better classify Galactic sources ($\sim$10\% improvement in scores and misclassification rates for some classes) with a corresponding performance drop on the extragalactic source group. Despite the already noted dataset limitations, we believe that this analysis represent a first valuable baseline for future studies aiming to explore other image-based classifiers and optimized normalization strategies for multi-wavelength data.

\section{\emph{sclassifier}: a radio source classifier tool}
\label{sec:method}
We developed a tool, dubbed \emph{sclassifier}\footnote{\scriptsize{\url{https://github.com/SKA-INAF/sclassifier}}}, for performing radio source classification using the dataset and the methods adopted in this work. An end-to-end pipeline was implemented, allowing users to obtain source classification information (e.g. predicted class labels and probabilities) and supplementary products (source image cutouts, feature data tables) from a radio continuum 2D map (FITS format) and a source catalogue (DS9 polygon regions) supplied as inputs. Additional algorithms and models (e.g. convolutional autoencoders, outlier finder, clustering, etc) were also implemented and will be presented in a future work focusing on an unsupervised analysis of the dataset.\\ 
\emph{sclassifier} is developed in python (3.x), and based on several libraries for astronomical data analysis and image processing - \texttt{Astropy} \citep{astropy1,astropy2,astropy3}, \texttt{Montage} \citep{montage}, \texttt{OpenCV} \citep{opencv} - and machine learning - \texttt{TensorFlow} \citep{tensorflow}, \texttt{Keras} \citep{keras}, \texttt{scikit-learn} \citep{sklearn}. As some stages, e.g. source cutout provision, regridding/reprojection, are quite computationally intensive for large catalogues, we parallelized them using the \texttt{mpi4py} library \citep{mpi4py}, splitting the computation for all sources across multiple computing nodes.

\section{Summary}
\label{sec:summary}
In this work, we carried out a supervised classification analysis of compact radio sources
over a large annotated dataset of $\sim$20,000 Galactic and Extragalactic objects, extracted from novel ASKAP radio observations and previous radio and infrared surveys. We trained two different classifiers on the produced data. The first uses the LightGBM gradient-boosting framework and is trained on a set of pre-computed features derived from the multi-wavelength data, including the radio-infrared colour indices and the radio spectral index. The second model uses convolutional neural networks and is trained directly on multi-channel images.\\
The LightGBM classifier achieved very high performances (above 90\%) for the identification of Galactic objects against sources belonging to the extragalactic group, using only radio and mid-infrared data. Classification metrics largely vary among individual source classes. Extragalactic objects (radio galaxies, QSO) are best classified, with F1-scores exceeding 85\%. PNe, \hii{} regions, and pulsars are the second group of best classified objects, with F1-scores ranging from 60\% to 75\%. Poor performances are obtained on radio star group and YSOs, due to the limited sample size, object spectral type heterogeneity, and unreliable classification information reported in the reference catalogues. We also tested how the classification performances changed for Galactic objects when including additional infrared band data (8$\mic$, 70$\mic$) and the radio spectral index information in the analysis. We obtained a significant boost in performance ($\sim$10\%) for PNe, \hii{} regions, and pulsars.\\CNN classifier was only trained on 5-band (radio+MIR) data due to the limited number of images available in the FIR band. The classification metrics achieved by trained shallow and deep network architectures are overall comparable to LightGBM, with better classifications observed on the Galactic source group at the expense of the extragalactic source group.\\
The obtained results motivate further analysis to be done to improve overall source classification results and tackle some reported limitations, before applying the method on unclassified ASKAP sources. Firstly, test data sample size can be slightly increased once new ASKAP observations towards the Galactic plane will be completed. This would increase the reliability of the reported classification metrics for some classes. Secondly, analysis should be repeated with a revised YSO and star reference catalogue, as the ones used in this work may contain spurious association to \hii{} regions or extragalactic objects, that could partly explain the misclassification rates obtained. In this context, we foresee to carry out a completely unsupervised analysis of the dataset to detect possible label anomalies and perform new classification studies.\\As commented in Section~\ref{subsec:compact-sources}, our training set does not contain star-forming galaxies, expected to contribute with a non-negligible fraction ($\sim$25\%) in EMU survey data, and therefore our current classifier could potentially misclassify them as Galactic sources, if their radio-infrared colors are similar. Luckily, there are ongoing studies within EMU, aiming to produce a curated sample of SFGs from EMU pilot observations, that would allow us to extend the training dataset, study SFG color parameter distribution and re-train our classifiers.\\
On a longer term, we also would like to extend our dataset with additional complementary data (e.g. optical, H$\alpha$ or radio polarization information), that could potentially lead to improved classification results.\\In this work, we produced a python-based tool, enabling users to run source classification on their new data. The implemented methods are rather general-purpose, allowing for the future to include additional image wavelength data, or to perform a similar analysis on extended sources. We plan to integrate it in the list of source analysis applications supported in the \emph{caesar-rest} service\footnote{\scriptsize{\url{https://github.com/SKA-INAF/caesar-rest}}}, developed within the CIRASA (\emph{Collaborative and Integrated platform for Radio Astronomical Source Analysis}) project \citep{Riggi2021b} to enable SKA Galactic science teams or high-level service API to run source analysis tasks (source extraction, classification, cross-matching, etc) over an http interface. This service is currently deployed on the European Open Science Cloud (EOSC) prototype, setup for the H2020 NEANIAS (\emph{Novel EOSC Services for Emerging Atmosphere, Underwater \& Space Challenges}) project\footnote{\scriptsize{\url{https://www.neanias.eu/}}} \citep{Sciacca2021}.

\section*{Acknowledgements}
\label{sec:acknowledgements}
\small{
The Australian SKA Pathfinder is part of the Australia Telescope National Facility which is managed by CSIRO. Operation of ASKAP is funded by the Australian Government with support from the National Collaborative Research Infrastructure Strategy. Establishment of the Murchison Radio-astronomy Observatory was funded by the Australian Government and the Government of Western Australia. This work was supported by resources provided by the Pawsey Supercomputing Centre with funding from the Australian Government and the Government of Western Australia. We acknowledge the Wajarri Yamatji people as the traditional owners of the Observatory site.

This research has made use of the HASH PN database at hashpn.space. This publication makes use of data products from the Wide-field Infrared Survey Explorer, which is a joint project of the University of California, Los Angeles, and the Jet Propulsion Laboratory/California Institute of Technology, funded by the National Aeronautics and Space Administration. Additionally, this research has made use of the SIMBAD database, operated at CDS, Strasbourg, France \citep{Wenger2000}. 
}%

\section*{Funding Statement}
\small{
This research was supported by the INAF CIRASA \& SCIARADA grants. C.B. acknowledges support from European Commission Horizon 2020 research and innovation programme under the grant agreement No. 863448 (NEANIAS).
}%

\section*{Data Availability}
\label{sec:data-availability}
\small{ 
The software code used in this work is publicly available under the GNU General Public License v3.0\footnote{\scriptsize{\url{https://www.gnu.org/licenses/gpl-3.0.html}}} on the GitHub repository \url{https://github.com/SKA-INAF/sclassifier/}. The trained model weights have been made available on Zenodo repository at \url{https://doi.org/10.5281/zenodo.10477860}.
}




\newpage%

\appendix%
\renewcommand{\thesection}{\Alph{section}}

\section{Source class physical properties}
\label{appendix:source-classes}

\subsection{Young Stellar Objects (YSOs)}
\label{subsec:yso}
Young Stellar Objects (YSOs) denote the early stages of star development, e.g. protostars and pre-main sequence stars. They have been classified into different classes (0, I, II, III), depending on their evolution phase \citep{Gomez2013}. Class 0 objects are characterized by an embedded central core, surrounded by a larger (not visible yet) accreting envelope, typically observed through far-infrared and millimetre wavelength emission from the dust. Class I objects denote the late mass accretion phase, in which the central core grows, a flattened circumstellar accretion disk develops, and the protostar also expels matter via bipolar jets and outflows. Typically, their SEDs rise in the far- and mid-infrared range ($\alpha_{IR}>$0.3). In Class II objects, the majority of the circumstellar material is found in a disk of gas and dust.
Flatter infrared spectral indices are typically observed in this stage. In Class III objects, the gas has been cleared out, the young planetary disk is formed, and the stellar atmosphere is recognizable. The emission from the disk becomes now negligible, and the SED is dominated by the pure stellar photosphere contribution ($\alpha_{IR}<-$1.6).\\YSOs are observed in the radio continuum, mainly through thermal free-free emission from ionized regions in their components (disks, winds, coronae, and jets), specially in massive YSOs. For low mass YSOs, however, the emission is thought to be driven by outflow processes that shock the surrounding material causing the required gas ionization \citep{Anglada1998}. The radio spectral indices $\alpha$ are expected in the range $-$0.1<$\alpha$<1.1, depending on the evolution phase and emission mechanism \citep{Scaife2012}. For example, in collimated outflows, typical of early protostar stages (Class 0, I), a radio spectral index $\alpha\sim$0.25 is favoured, while, for standard conical jets, spectral indices around 0.6 are expected \citep{Reynolds1986,Anglada1998}. Non-thermal emission can be found around more developed pre-main sequence stages (Class II, III), like T Tauri stars, and may partially contribute to the very negative indices observed in some YSOs \citep{Ainsworth2012}.

\subsection{Radio stars}
\label{subsec:stars}
Thermal and non-thermal radio emission has been detected so far from stars of different types and evolution stages across the entire Hertzsprung-Russell (H-R) diagram, among them magnetic stars, early-type stars (e.g. O-B) with winds and strong mass loss, later stages like Wolf-Rayet (WR) stars and LBVs, binaries (also bright in X-rays), and ultra cool-dwarfs \citep{Gudel2002,Matthews2013}. Thermal free-free emission, as in the stellar wind emission, originates from stellar outflows and chromospheres. Stars with spherically symmetric, isothermal, and stationary outflows are expected to have a radio spectrum S$_{\nu}\propto\nu^{\alpha}$ ($\alpha$=0.6) \citep{Panagia1975}, although spectral indices deviating from the canonic value may be obtained when variating the wind parameters (e.g. electron densities, velocity gradients, mass loss rate). Non-thermal gyrosynchrotron and synchrotron emission is generated in flares and also found in stars with magnetic activity and colliding winds in binaries. Negative spectral indices ($\alpha<$0) are expected in this case \citep{Umana2015b}. 
The stable non-thermal radio emission from the MCP stars is instead expected with an almost flat spectrum \citep{Leto2021} and partially polarized, with the circularly polarized emission increasing as the radio frequency increases \citep{Leto2020}.
In stars with high radio brightness temperatures, coherent emission mechanisms, like plasma radiation or electron cyclotron maser emission \citep{Trigilio2000}, are also operating.

\subsection{\hii{} regions}
\label{subsec:hii}
\hii{} regions are discrete ionized clouds surrounding young and massive hot stars (type O-B), thus enabling to trace massive star formation across the entire Galaxy, particularly in their youngest and compact stages, known as hyper-compact \hii{} (HC\hii{}, diameter$\le$0.05 pc) and ultra-compact \hii{} (UC\hii{}, diameter$\le$0.1 pc) regions \citep{Kurtz2005}. They are detected through their bright radio and infrared emission. The observed radio continuum spectrum can be described by a standard thermal free-free model assuming an optically thin regime above $\sim$1 GHz, with spectral indices $\alpha\sim-$0.1, and an optically thick scenario at lower frequencies (e.g. below a turnover frequency), leading to increased spectral indices ($\alpha\sim$2) due to self-absorption mechanisms. Younger \hii{} regions with
higher density typically remain optically thick at higher frequencies, with a positive spectral index and turnover at $\nu\sim$5 GHz for UC\hii{} and $\nu\sim10\div100$ GHz for HC\hii{} \citep{Yang2019,Yang2021}.\\The infrared emission comes from
different dust populations \citep{Robitaille2012}, mostly located in the photodissociation
regions, e.g. polycyclic aromatic hydrocarbons
(PAHs) at 8$\mic$ and 12$\mic$, small and large grains at 22-24$\mic$ and above 70$\mic$, respectively.

\subsection{Planetary Nebulae}
\label{subsec:pn}
Planetary nebulae are shells of ionized gas, ejected from central hot stars of low to intermediate mass ($\sim$1$-$8 \(\textup{M}_\odot\)) at the end of their asymptotic giant branch (AGB) phase. A more precise definition, and a summary of their observational characteristics, were presented in \cite{Frew2010}.\\
Radio continuum radiation (thermal free–free) is observed from the nebula shell, due to the gas ionized by the ultra-violet radiation produced by the central star \citep{Kwok2000}. Typical radio spectral indices observed range from $\sim -$0.1 in an optically thin regime, to positive indices, up to $\sim$2 \citep{Pottasch1984}, for optically thick PNe. Infrared emission is due to cool dust (T$\sim$100$-$200 K) material surrounding the ionized region, peaking at $\sim$20$\mic$ for most PNe. Polycyclic aromatic hydrocarbon (PAH) emission at 8$\mic$ from the photodissociation region (PDR) surrounding the ionized gas, can also be present in more compact objects.

\subsection{Pulsars}
\label{subsec:pulsars}
Pulsars are highly magnetized rotating neutron stars, emitting beams of radiation from their magnetic poles, observed as they point towards Earth, with periods ranging from milliseconds to seconds. The radio emission shows a high degree of linear polarization, and a small fraction of circular polarization. Pulsars are known to have steep flux-density spectra, with observed average spectral indices around $-$1.8$\pm$0.2 \citep{Maron2000}, in some cases ($\sim$10\%) described by double power-laws with spectral breaks around 1 GHz. The emission is thought to be due to coherent processes, but its origin and generation mechanism is still debated \citep{Beskin2015,Beskin2018,Melrose2021}.

\subsection{Active Galactic Nuclei}
\label{subsec:extragal-sources}
Active Galactic Nuclei (AGN) of different radio-loud (RL) and radio-quiet (RQ) types (e.g. RL/RQ quasars, FR I/FR II/Seyfert radio galaxies, blazars) dominate the observed counts of continuum radio sources above the mJy level, the latter type representing $\sim$90\% of all AGNs. In the unification schema \citep{Padovani1995}, much of their observational properties (e.g. radio components, multi-wavelength spectral features) arise from the orientation of the accretion disk and the observer's line-of-sight.\\The radio emission at cm wavelengths, explained as synchrotron radiation from GeV electrons, has a relatively steep spectrum for extended regions (jets, lobes) with $\alpha\sim -$0.7, while compact regions have flatter or inverted spectra ($\alpha\ge -$0.5), resulting from the superposition of multiple self-absorbed components.
The latter scenario is observed in core-dominated sources, such as blazars or FSRQs. On the other hand, Gigahertz-Peaked Spectrum (GPS) and Compact Steep Spectrum (CSS) compact sources, with their observed steep spectra and well-defined spectral turnovers (around 1 GHz for GPS sources, and 100 MHz for CSS), are a notable exception. They are believed to be young objects eventually evolving into more extended radio objects of type FR I/II, and overall they represent a considerable fraction (around 10\% for GPS, and 30\% for CSS) of the bright compact radio source population \citep{ODea1998,ODea2021,Sadler2016}. A high degree of linear polarization (up to 30\%) is also observed, particularly in extended components.

\newpage%
\onecolumn%

\renewcommand\thefigure{\thesection.\arabic{figure}} 

\section{Supplementary plots}
\label{appendix:addon_plots}
\setcounter{figure}{0}


\begin{figure*}[htb]
\centering%
\subtable[RG]{\includegraphics[scale=0.4]{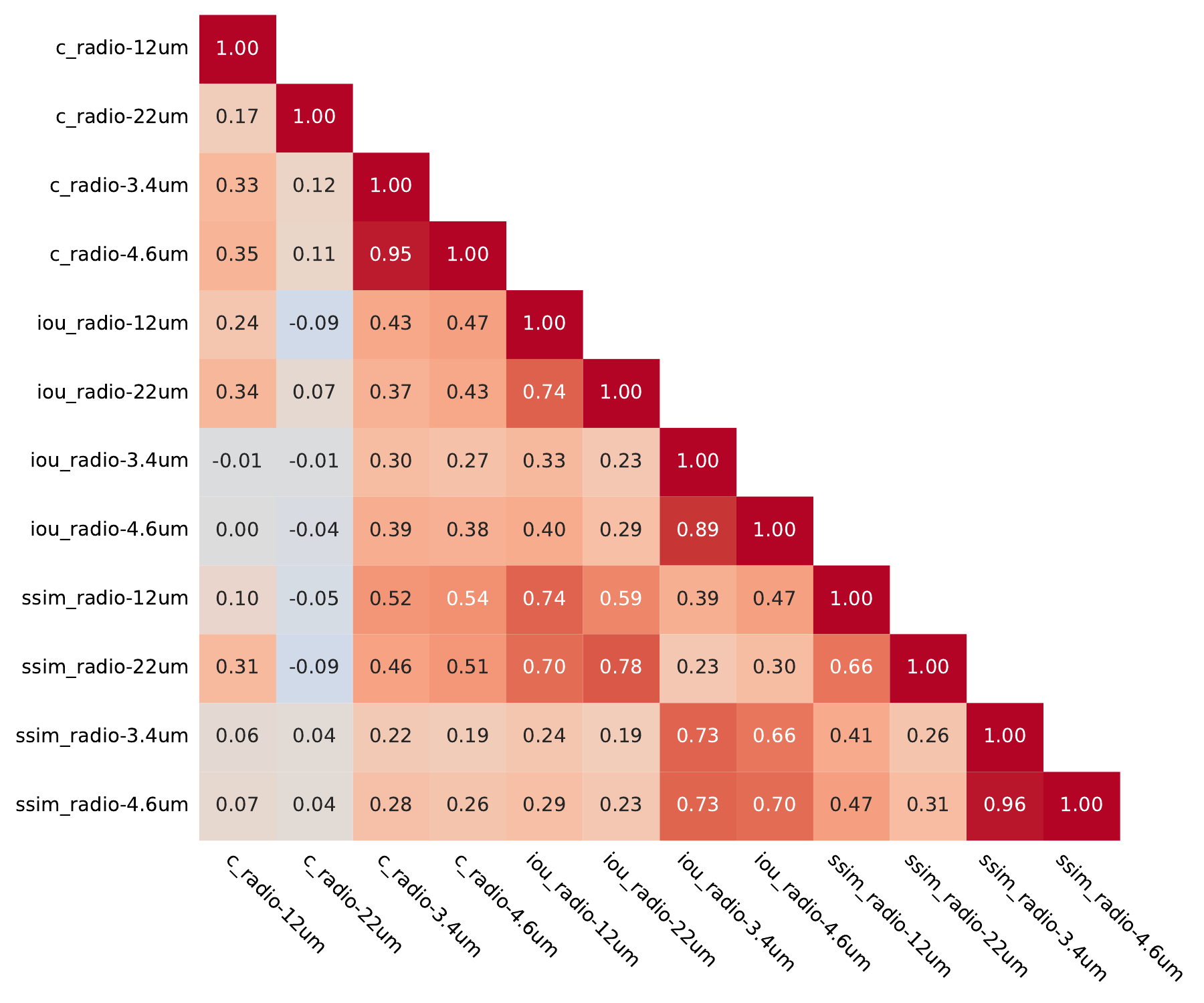}}%
\vspace{-0.2cm}
\caption{Pearson correlation coefficient matrix computed over for the 5-band color feature sets (see Table~\ref{tab:color-features}) \emph{(continued on next page)}}%
\label{fig:corr-coeff-5bands}
\end{figure*}

\newpage%

\begin{figure*}[htb]
\ContinuedFloat%
\centering%
\addtocounter{subfigure}{1}%
\subtable[QSO]{\includegraphics[scale=0.4]{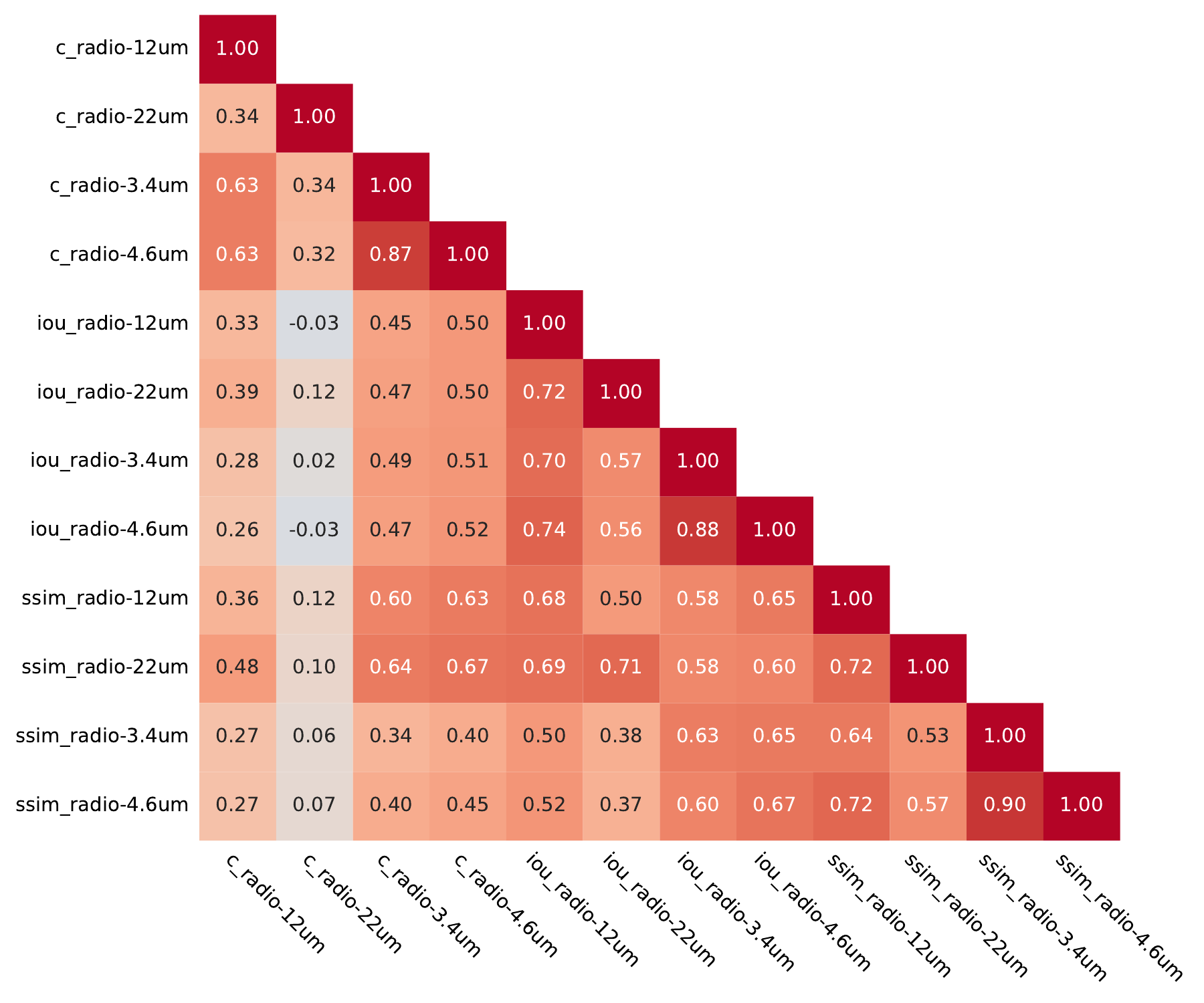}}\\%
\vspace{-0.4cm}%
\subtable[PN]{\includegraphics[scale=0.4]{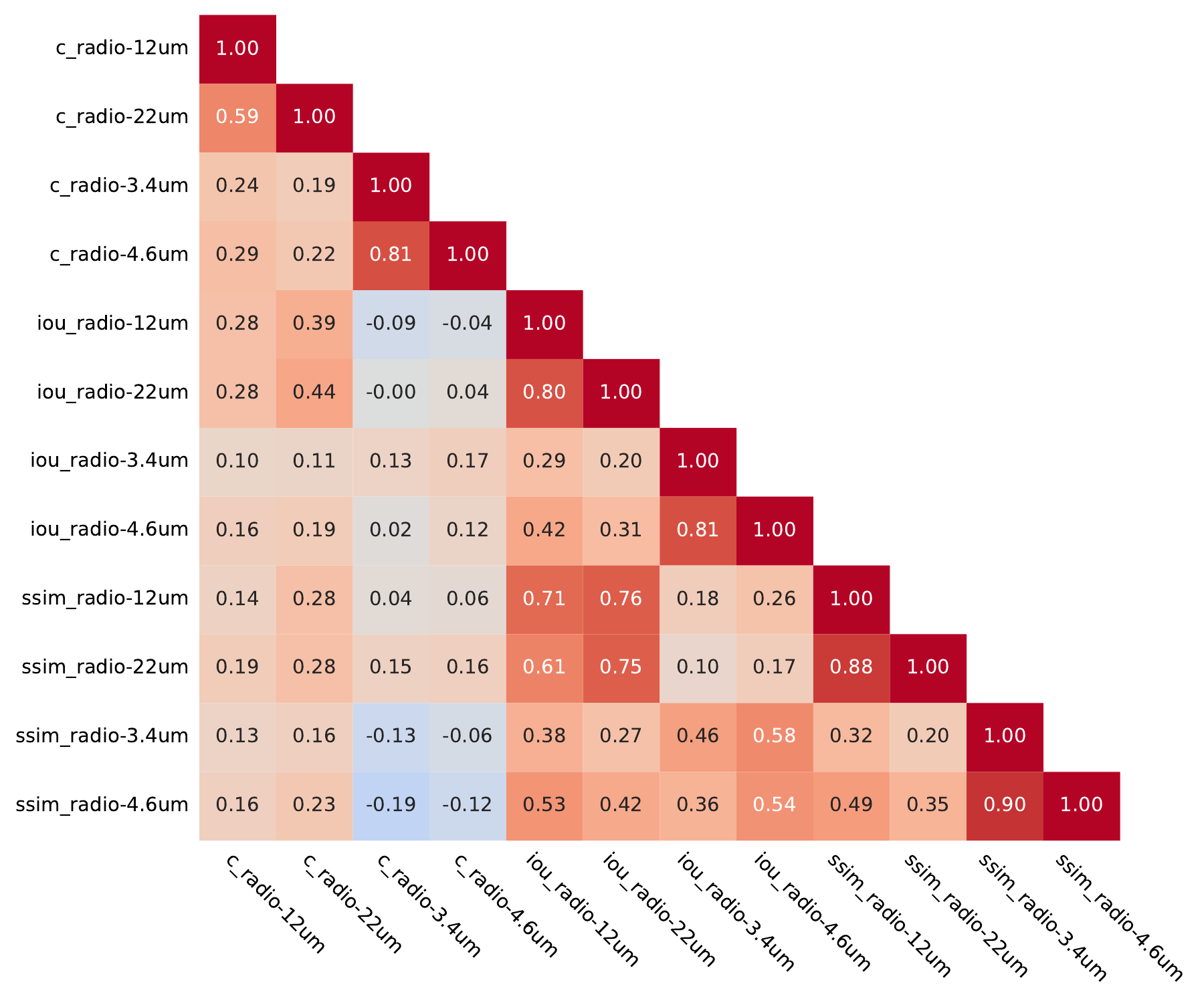}}%
\vspace{-0.2cm}
\caption{Pearson correlation coefficient matrix computed over for the 5-band color feature sets (see Table~\ref{tab:color-features}) \emph{(continued on next page)}}%
\label{fig:corr-coeff-5bands_2}
\end{figure*}

\newpage%

\begin{figure*}[htb]
\ContinuedFloat%
\centering%
\subtable[\hii{}]{\includegraphics[scale=0.4]{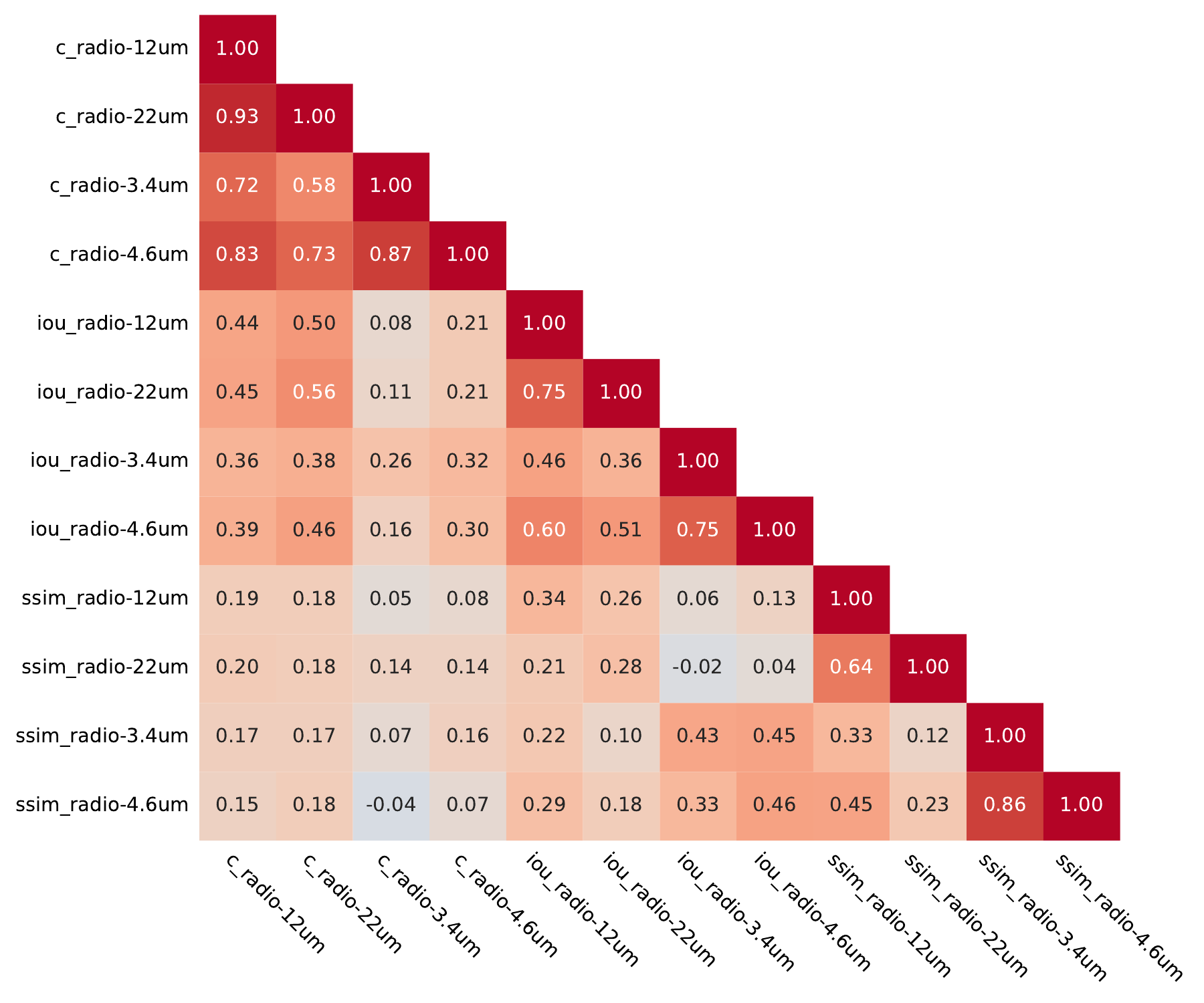}}\\%
\vspace{-0.4cm}%
\subtable[STAR]{\includegraphics[scale=0.4]{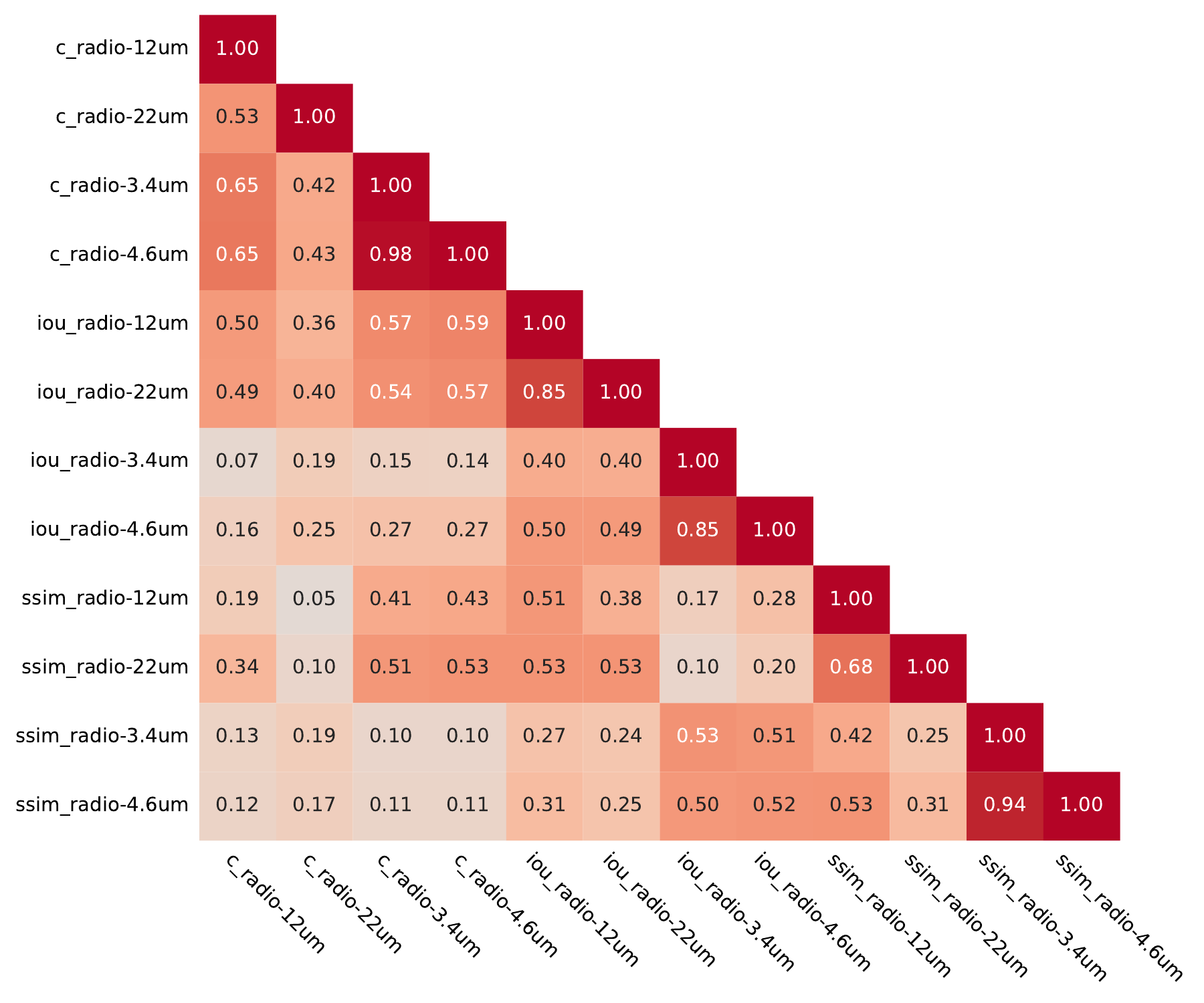}}%
\vspace{-0.2cm}
\caption{Pearson correlation coefficient matrix computed over for the 5-band color feature sets (see Table~\ref{tab:color-features}) \emph{(continued on next page)}}%
\label{fig:corr-coeff-5bands_3}
\end{figure*}

\newpage%

\begin{figure*}[htb]
\ContinuedFloat%
\centering%
\subtable[PULSAR]{\includegraphics[scale=0.4]{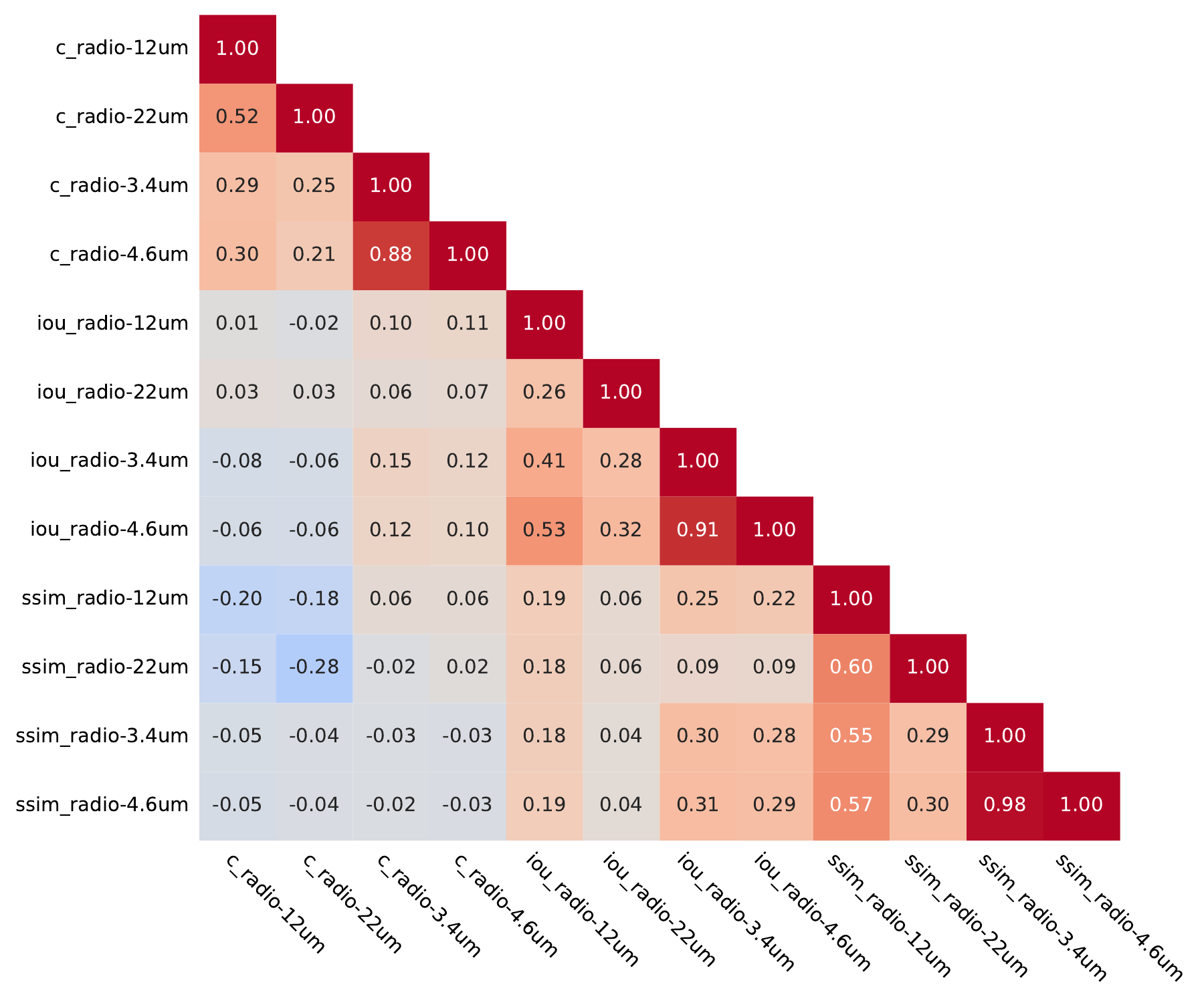}}\\%
\vspace{-0.4cm}%
\subtable[YSO]{\includegraphics[scale=0.4]{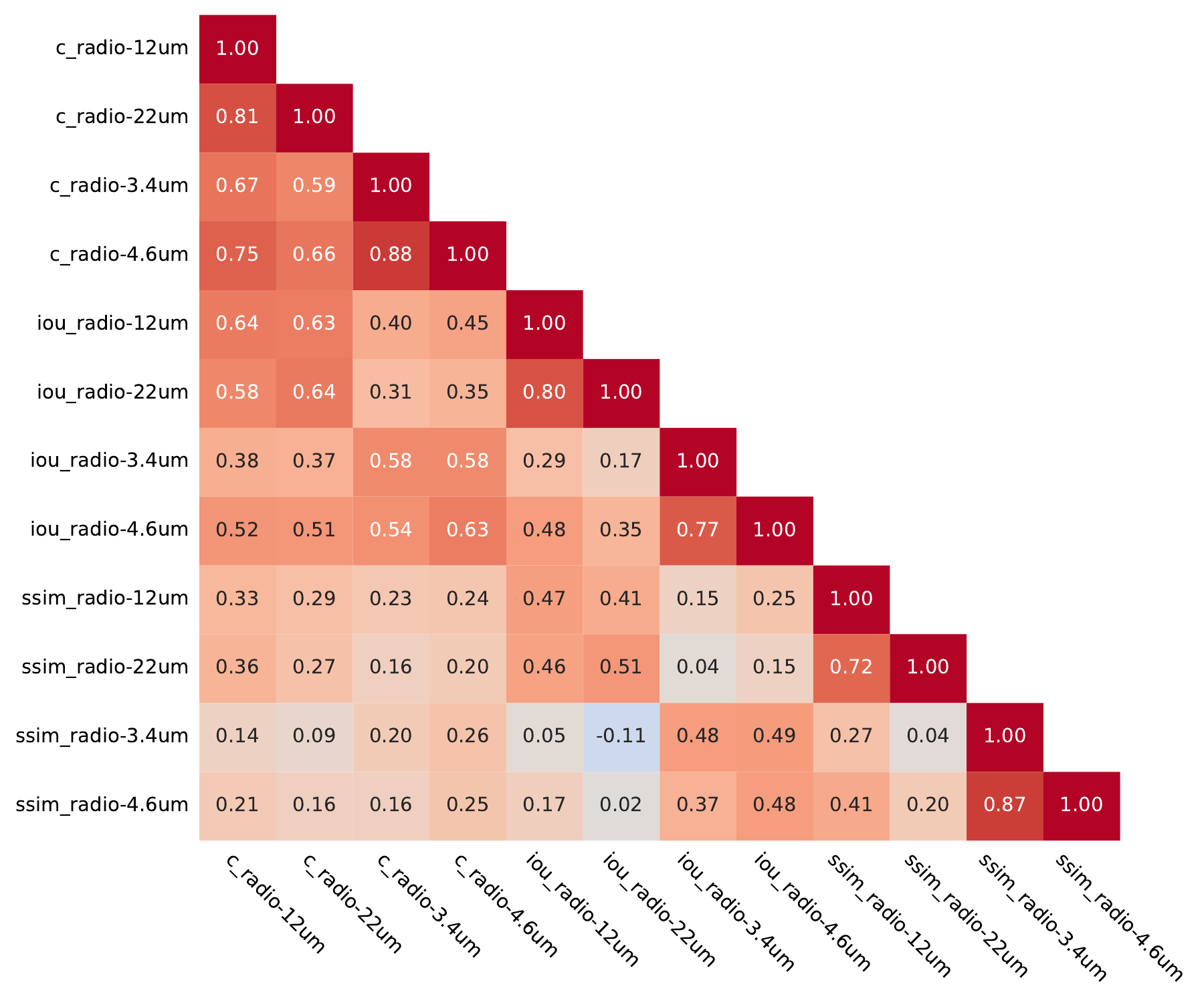}}%
\vspace{-0.2cm}
\caption{Pearson correlation coefficient matrix computed over for the 5-band color feature set (see Table~\ref{tab:color-features}).}%
\label{fig:corr-coeff-5bands_4}
\end{figure*}

\newpage%

\begin{figure*}[htb]
\centering%
\addtocounter{subfigure}{-7}%
\subtable[PN]{\includegraphics[scale=0.42]{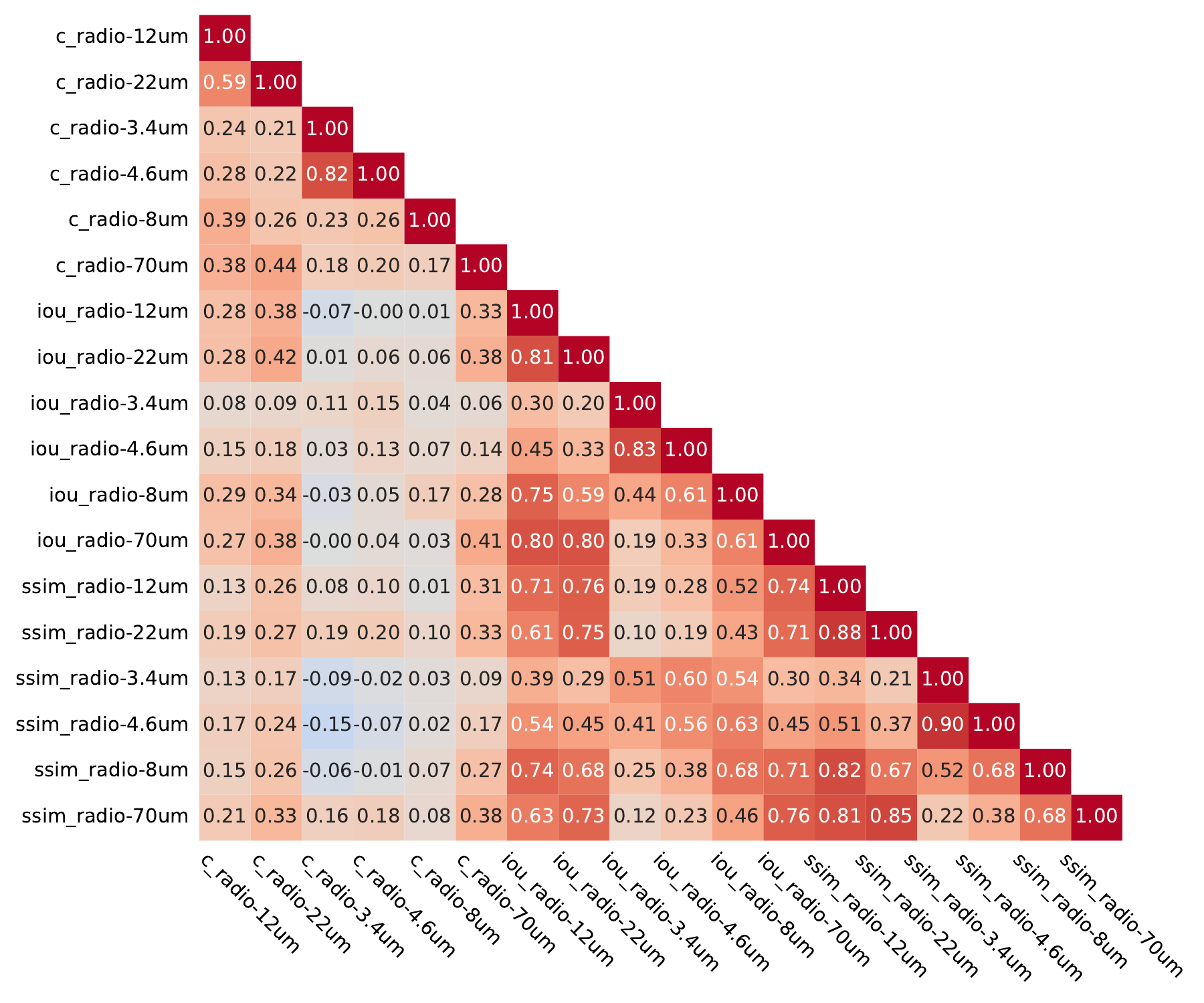}}\\%
\vspace{-0.4cm}%
\subtable[\hii{}]{\includegraphics[scale=0.42]{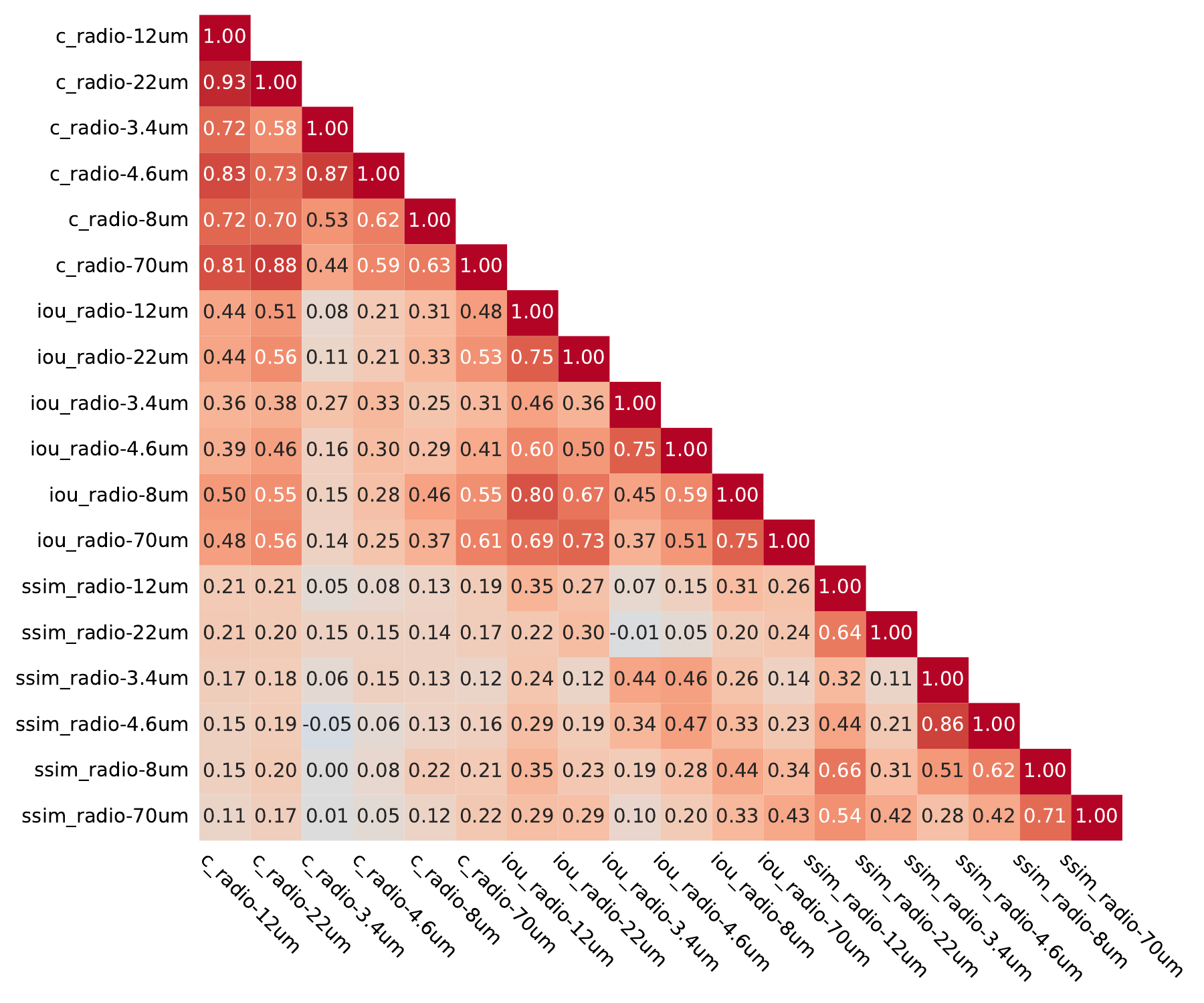}}%
\vspace{-0.2cm}
\caption{Pearson correlation coefficient matrix computed over for the 7-band color feature set (see Table~\ref{tab:color-features}) \emph{(continued on next page)}}%
\label{fig:corr-coeff-7bands}
\end{figure*}

\newpage%

\begin{figure*}[htb]
\ContinuedFloat%
\addtocounter{subfigure}{2}%
\centering%
\subtable[STAR]{\includegraphics[scale=0.42]{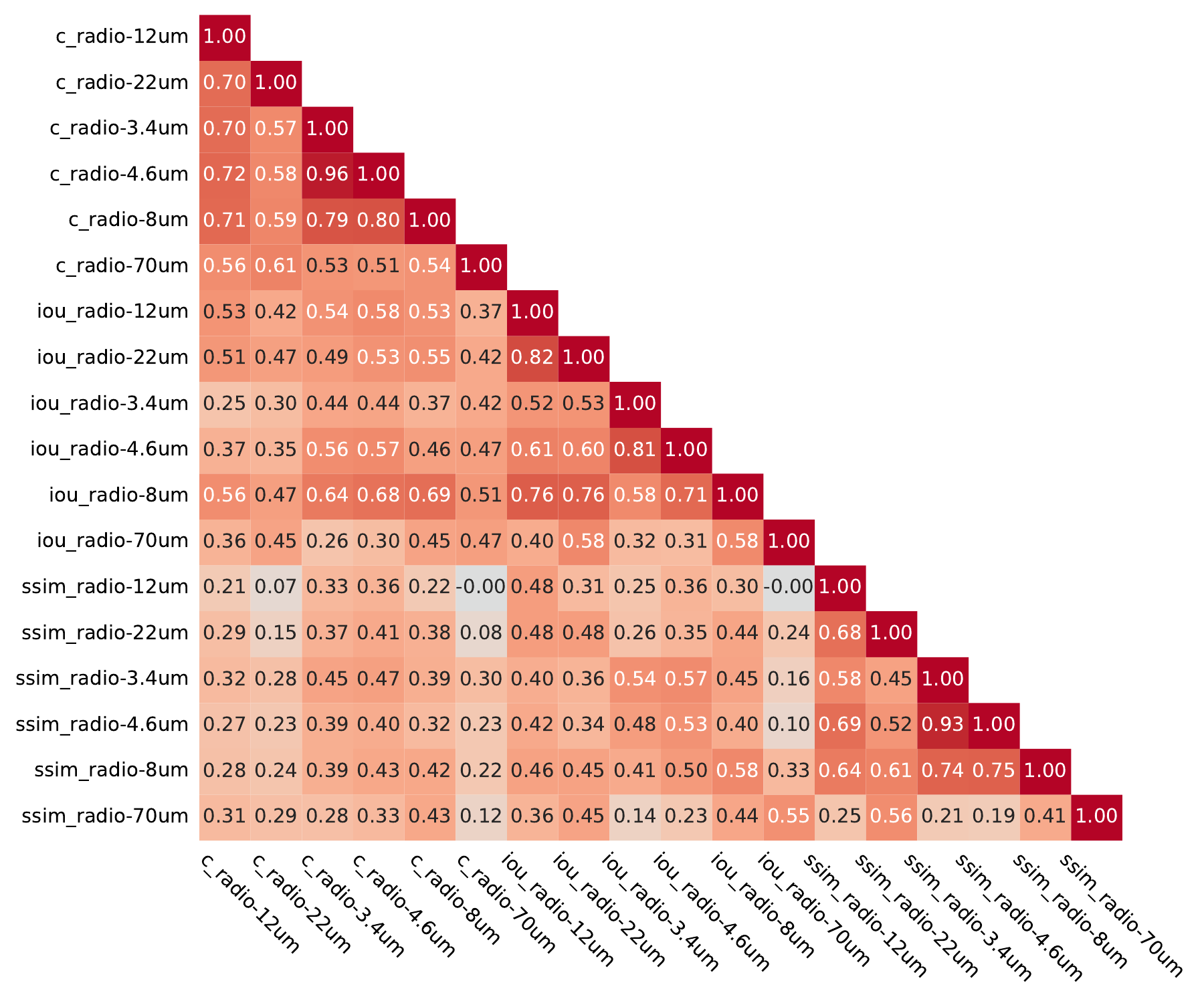}}\\%
\subtable[PULSAR]{\includegraphics[scale=0.42]{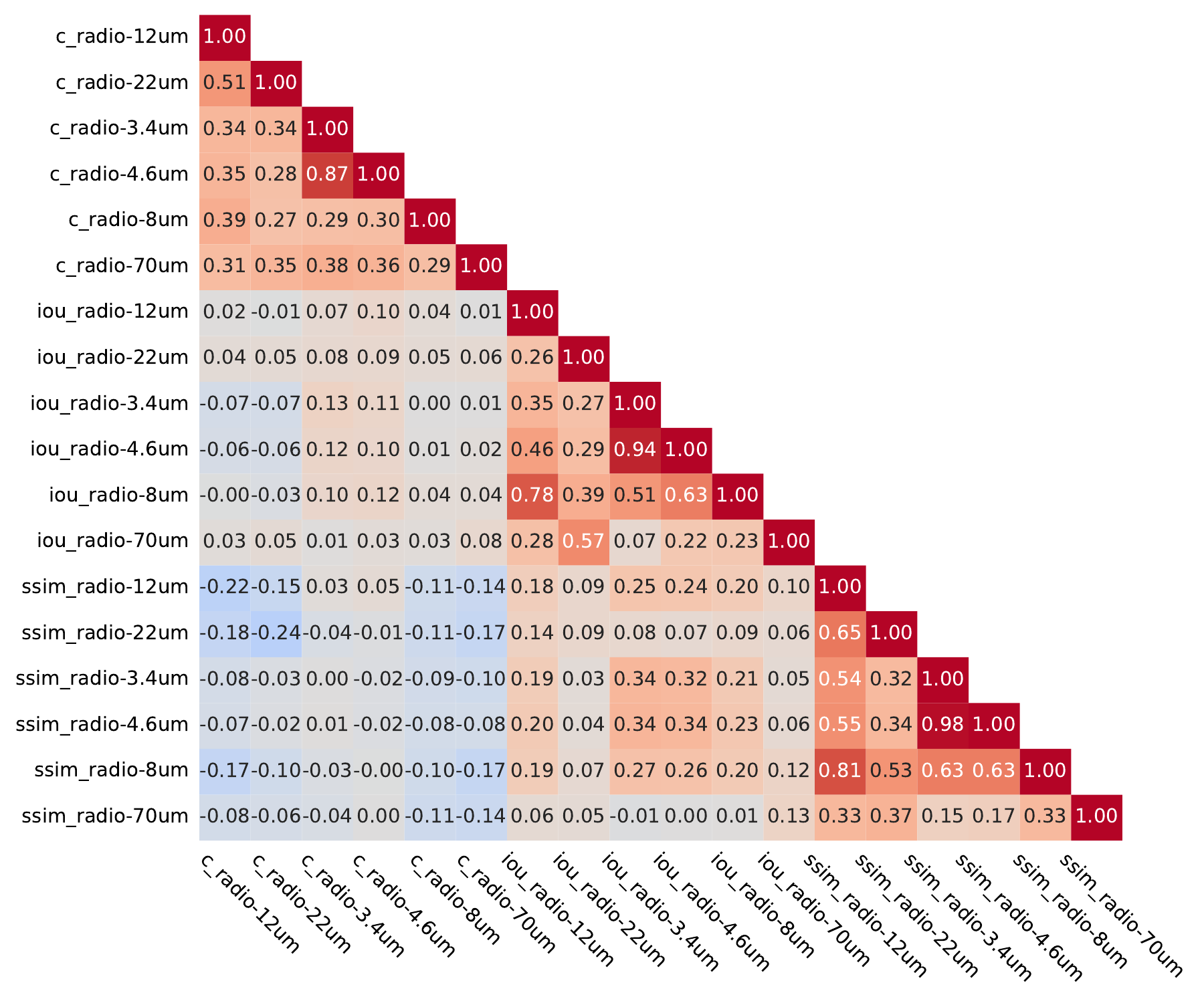}}
\vspace{-0.2cm}
\caption{Pearson correlation coefficient matrix computed over for the 7-band color feature set (see Table~\ref{tab:color-features}) \emph{(continued on next page)}}%
\label{fig:corr-coeff-7bands_2}
\end{figure*}


\begin{figure*}[htb]
\ContinuedFloat%
\centering%
\subtable[YSO]{\includegraphics[scale=0.42]{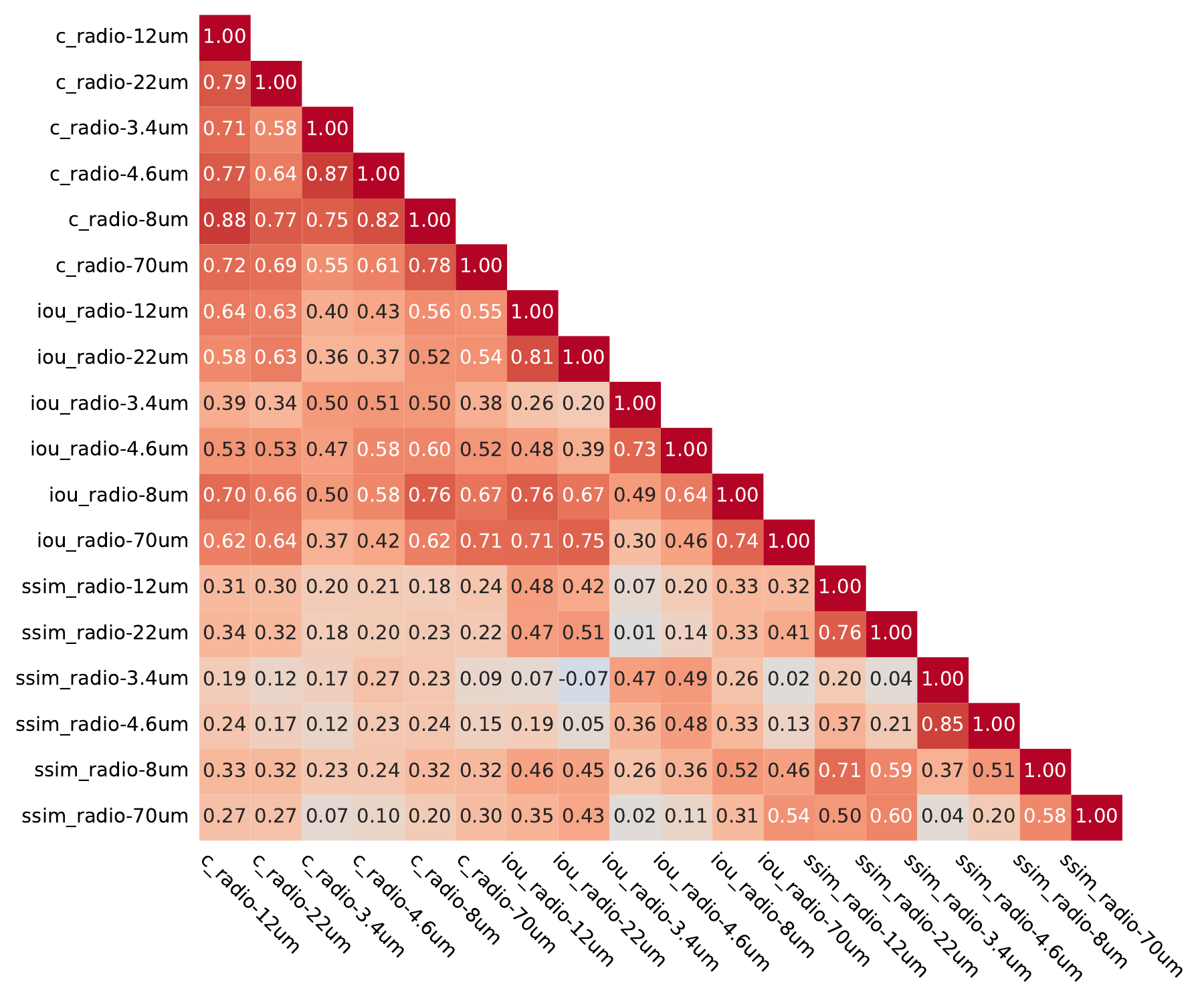}}%
\vspace{-0.2cm}
\caption{Pearson correlation coefficient matrix computed over for the 7-band color feature set (see Table~\ref{tab:color-features})}%
\label{fig:corr-coeff-7bands_3}
\end{figure*}


\begin{figure*}[htb]
\centering%
\includegraphics[scale=0.63]{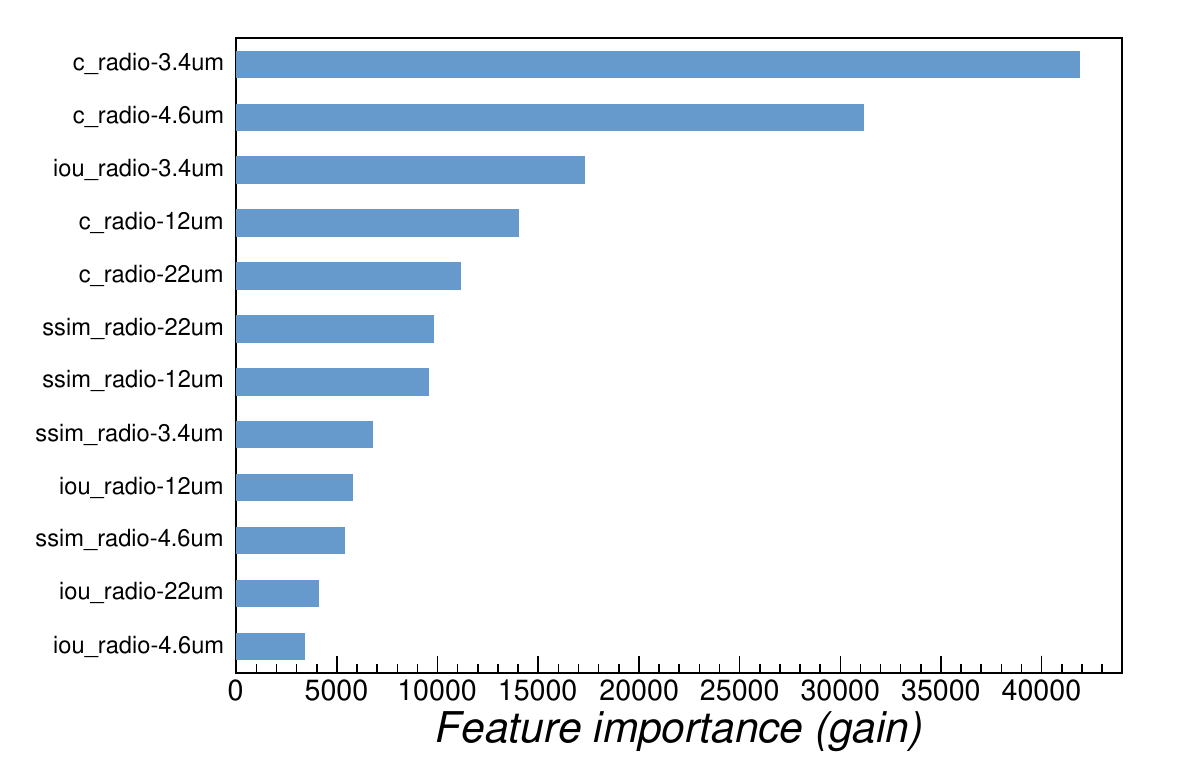}%
\caption{Feature importance for LightGBM classifier trained on 5 bands (MIR+FIR)  data.}%
\label{fig:lgbm-feat-importance-5bands}
\end{figure*}

\vspace{-0.5cm}%

\begin{figure*}[htb]
\centering%
\includegraphics[scale=0.63]{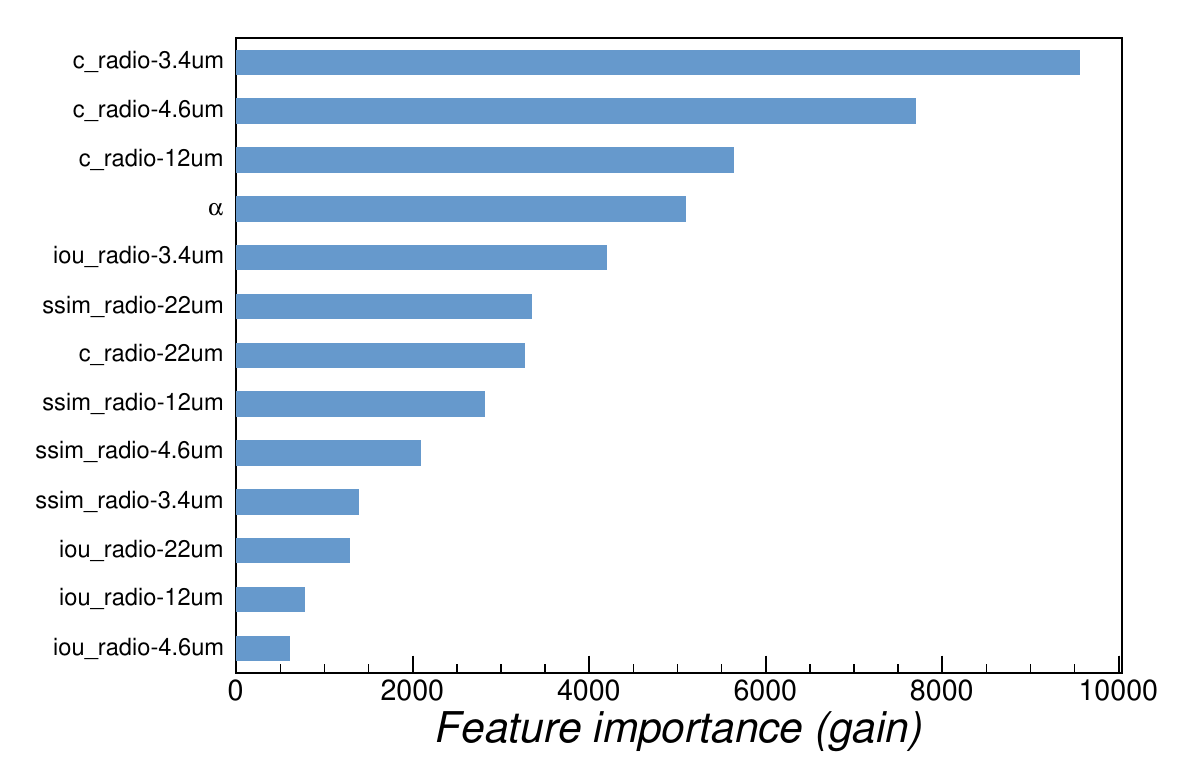}%
\caption{Feature importance for LightGBM classifier trained on 5 bands (MIR+FIR) + $\alpha$ data.}%
\label{fig:lgbm-feat-importance-5bands-alpha}
\end{figure*}


\begin{figure*}[htb]
\centering%
\includegraphics[scale=0.7]{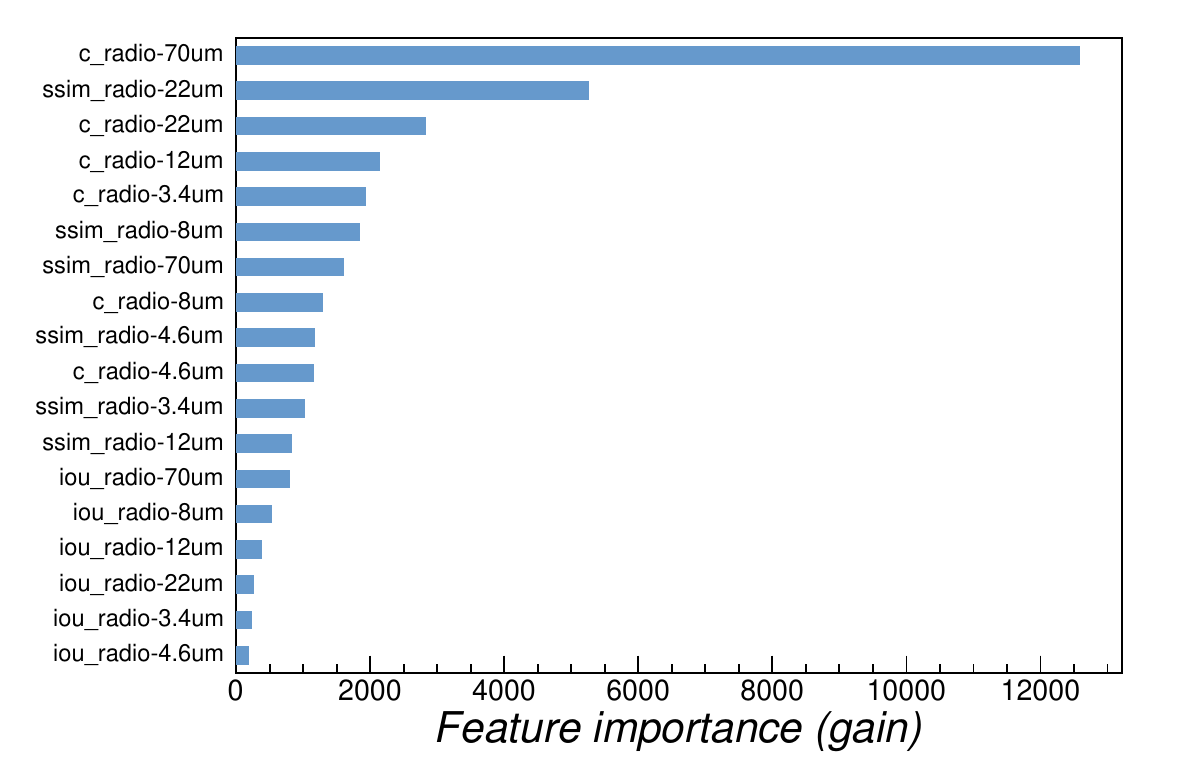}%
\caption{Feature importance for LightGBM classifier trained on 7 bands (MIR+FIR) data.}%
\label{fig:lgbm-feat-importance-7bands}
\end{figure*}

\vspace{-0.5cm}%

\begin{figure*}[htb]
\centering%
\includegraphics[scale=0.7]{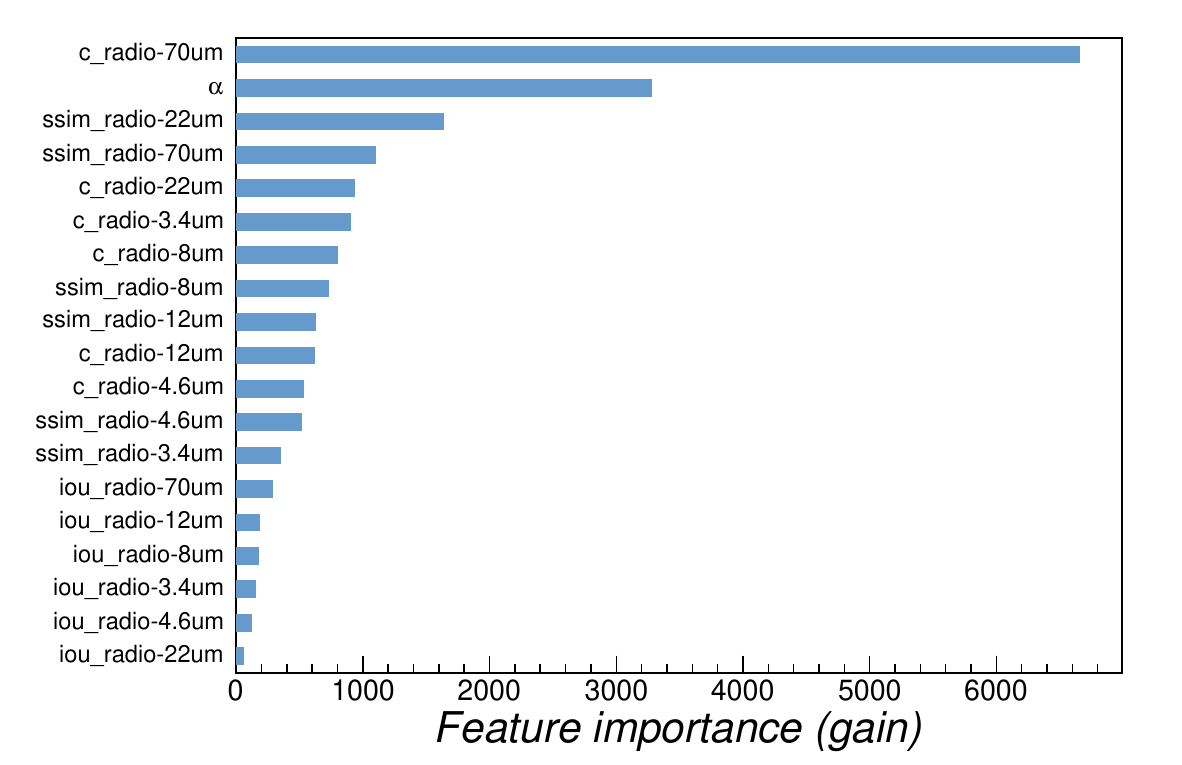}%
\caption{Feature importance for LightGBM classifier trained on 7 bands (MIR+FIR) + $\alpha$ data.}%
\label{fig:lgbm-feat-importance-7bands-alpha}
\end{figure*}

\begin{figure*}[htb]
\centering%
\includegraphics[scale=0.6]{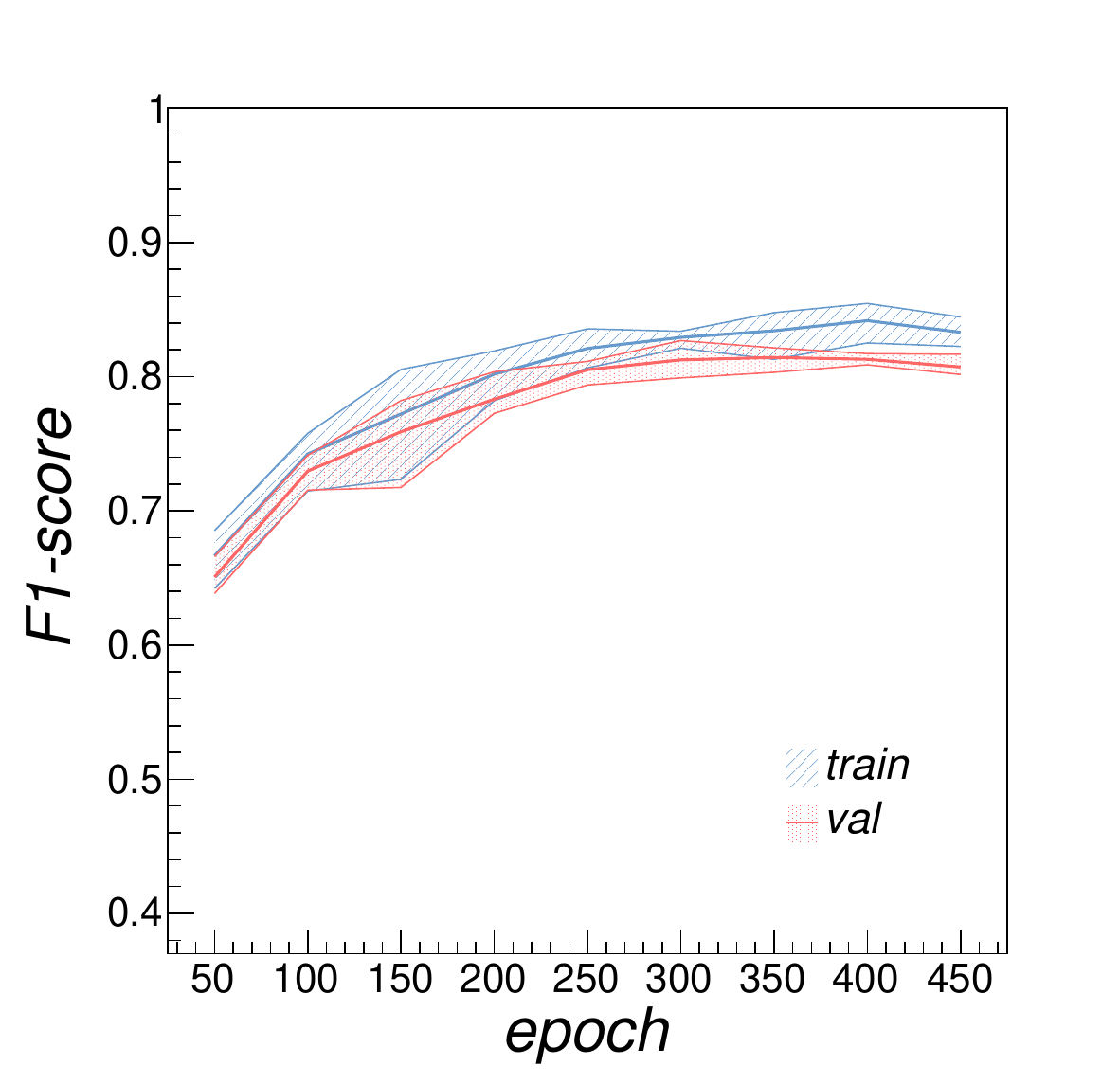}%
\caption{F1-score of CNN classifier (\texttt{custom\_v1} model) computed as a function of the training epoch over five "mixed" survey 5-bands (MIR+FIR) train (blue graph) and validation (red graph) datasets. Shaded areas correspond to the range of minimum and maximum F1-scores obtained in different training runs, each with different train/validation/test data splits.}%
\label{fig:f1scores-vs-epoch}
\end{figure*}

\end{document}